\documentclass[superscriptaddress,amsmath,amssymb,aps,twocolumn,nofootinbib,twoside,floatfix,nofootinbib]{revtex4}
\usepackage{graphicx}
\usepackage[usenames,dvipsnames]{color}
\usepackage{amsmath}
\usepackage{amssymb}
\usepackage{amsfonts}

\usepackage{mathrsfs}
\usepackage{bm}
\usepackage{hyperref}
\usepackage[percent]{overpic}
\usepackage{overpic}
\usepackage{bbold}

\usepackage{ulem}
\newcommand{\beq}{\begin{equation}}
\newcommand{\eeq}{\end{equation}}
\newcommand{\bea}{\begin{eqnarray}}
\newcommand{\eea}{\end{eqnarray}}

\newcommand{\ea}{\end{align}}

\newcommand{\bma}{\begin{pmatrix}}
\newcommand{\ema}{\end{pmatrix}}

\newcommand{\CITA}{\affiliation{\mbox{Canadian Institute for Theoretical Astrophysics, University of Toronto, ON M5S 3H8, Canada}}}
\newcommand{\CIAR}{\affiliation{\mbox{Canadian Institute for Advanced Research, Toronto, ON M5G 1Z8, Canada}}}
\newcommand{\WVU}{\affiliation{\mbox{Department of Physics and Astronomy, West Virginia University, PO Box 6315, Morgantown, WV 26506, USA}}}
\newcommand{\BNU}{\affiliation{Center for Cosmology and Gravitational Wave, Department of Astronomy, Beijing Normal University, Beijing 100875, China}}

\begin{document}
\title{Stability of exact force-free electrodynamic solutions \\and scattering from spacetime curvature}

\author{Fan Zhang} \BNU \WVU 
\author{Sean T. McWilliams} \WVU
\author{Harald P. Pfeiffer} \CITA \CIAR
\date{\today}

\begin{abstract}
Recently, a family of exact force-free electrodynamic (FFE) solutions was given by Brennan, Gralla and Jacobson, which generalizes earlier solutions by Michel, Menon and Dermer, and other authors. These solutions have been 
proposed as useful models for describing the outer magnetosphere of conducting stars. As with any exact analytical solution that aspires to describe actual physical systems, it is vitally important that the solution possess 
the necessary stability. In this paper, we show via fully nonlinear numerical simulations that the aforementioned FFE solutions, despite being highly special in their properties, are nonetheless stable under small perturbations. 
Through this study, we also introduce a three dimensional pseudospectral relativistic FFE code that achieves exponential convergence for smooth test cases, as well as two additional well posed FFE evolution systems in the 
appendix that have desirable mathematical properties. Furthermore, we provide an explicit analysis that demonstrates how propagation along degenerate principal null directions of the spacetime curvature tensor simplifies scattering, 
thereby providing an intuitive understanding of why these exact solutions are tractable; i.e. why they are not backscattered by spacetime curvature. 
\end{abstract}

\maketitle
\section{Introduction}

Force-free electrodynamics \cite{1977MNRAS.179..433B,Goldreich:1969sb,Gruzinov:2007se} (FFE) is a simplification to the joint electromagnetic and plasma dynamics that is applicable in the limit of magnetic domination. Within FFE, the inertia of the plasma is neglected, so that the equation of motion for the plasma is not required to close the set of evolution equations.  This leads to a significant reduction in computational complexity, with the presence of the plasma becoming a nonlinear modification to the vacuum Maxwell equations. Astrophysically, FFE is recognized as the appropriate limit for describing the magnetospheres of  black holes \cite{1977MNRAS.179..433B} and particularly neutron stars \cite{Goldreich:1969sb}, where one can find intense magnetic fields of $10^{8}$ -- $10^{15}$ gauss accompanied by charged particles supplied by electron-positron pair-production \cite{Ruderman:1975ju,1977MNRAS.179..433B}. FFE is an integral part of most proposed mechanisms for extracting rotational energy from neutron stars \cite{Goldreich:1969sb} or black holes \cite{1977MNRAS.179..433B,MacDonald:1982zz}, and electromagnetic dominance (over gas dynamics) is argued to be valid in all ultra-relativistic outflows \cite{Blandford2002}. For instance, the jets in quasars and active galactic nuclei \cite{1976MNRAS.176..465B} or gamma-ray bursts \cite{Meszaros:1996ww} are generally simulated using the FFE approximation.  
To understand these astrophysical phenomena, it is therefore important to analytically (e.g. \cite{Menon:2011zu,Michel1973,1976MNRAS.176..465B,Lyutikov:2011tq,Brennan:2013jla}) and numerically (e.g. \cite{Spitkovsky:2006np,Kalapotharakos:2008zc,Palenzuela:2010xn,
Petri:2012cs,Parfrey:2011ta,Komissarov:2002my,Asano:2005di,Cho:2004nn,
McKinney:2006sc,Yu:2010bp,Contopoulos:1999ga,Uzdensky:2004qu,
Timokhin:2006ur,Palenzuela:2010nf,Alic:2012df,Kalapotharakos:2011db})
study the solutions to the FFE equations. 

One step in this direction was achieved recently with the presentation of a family of analytical solutions in Kerr spacetime by Brennan, Gralla and Jacobson \cite{Brennan:2013jla}, which combines and generalizes some earlier solutions by Michel \cite{1969PhRvL..23..247M,Michel1973}, Menon and Dermer \cite{Menon:2011zu}, and puts them in a language more accessible to relativists. It has been suggested 
that these solutions (containing nonlinear ingoing or outgoing waves) can describe astronomical systems such as the outer magnetosphere of a pulsar \cite{Brennan:2013jla,Brennan:2013ppa,Gralla:2014yja}, describing the mechanism for transporting energy extracted from the interior regions towards infinity. 
Because the FFE equations are highly nonlinear, analytical solutions are relatively rare (see \cite{Lupsasca:2014,Lupsasca:2014hua,Zhang:2014pla,Li:2014bta} for some additional solutions in extremal Kerr spacetime), so it is worthwhile to examine these solutions in greater details, especially those aspects related to their applicability to real astronomical problems. For the benefit of finding further exact solutions, it is also interesting to study the properties that make these known solutions tractable. 

A remarkable feature of these solutions is that their wave contents are not backscattered\footnote{In this paper, we will refer to ``backscattering'' as the scattering capable of altering the propagation direction of the waves, or introducing a Coulomb component where initially there was none.} 
by the spacetime curvature, a fact that significantly simplifies the analysis, and in no small part contributes to the possibility of expressing these solutions in closed form. 
Such scattering-avoidance behaviour is not typical among waves travelling in a curved spacetime; generically the waves will scatter against the spacetime curvature and travel inside (as well as on) the null cones. In addition to not being scattered by spacetime curvature, these solutions are also not backscattered by nonlinear electromagnetic interactions, which makes them particularly efficient channels of energy transfer. A question then naturally arises: does the physical specialness and mathematical simplicity of these solutions equate to fragility? In other words, are these solutions stable under initial perturbations?  
The answer to this question is critical, as a negative answer would mean that these solutions do not describe realistic astronomical systems that are always subject to perturbing influences from their surrounding environments
If these solutions represent ``repellers'' in the FFE solution space, then one would expect that a small perturbation in the initial data will quickly drive the physical system away from them and perhaps establish alternative, physically less special and less efficient energy transportation channels. 
Even worse, if uncontrolled growth in the magnitude of the perturbations appears due to the presence of unstable modes, it would be a sign that the mathematical model may be intrinsically inadequate for describing real physics. 

On the topic of stability of force-free and magnetohydrodynamical configurations, there is a rich and impressive list of literature on the structure (see e.g. \cite{1996ASPC..100..129A,1996ApL&C..34..395A,
1993A&A...270...71A,1993A&A...274..699A,
1998A&A...337..603L,1999A&A...347.1055L}) and stability of jets. Researchers have examined in detail the various instability types that could alter the jet structure or even disrupt it. For example, Refs.~\cite{1993ApJ...419..111E,1994MNRAS.267..629I,
1996MNRAS.281....1I,1996AGAb...12...63A,
1997MNRAS.288..333S,1998ApJ...493..291B,
1999MNRAS.308.1006L,2001PhRvD..64l3003T,
2000A&A...355..818A,2000ApJ...533..897L, 2000A&A...355.1201L,2014IJMPS..2860201M,
2012ApJ...757...16M,2007ApJ...662..835M,
2009ApJ...697.1681N} examined various types of current driven instabilities, providing important observations such as that instabilities can change the current density in the jets \cite{2000A&A...355..818A, 2000A&A...355.1201L}, forming structures that can heat and accelerate particles, and that increased magnetization tends to have a stabilizing effect \cite{2015MNRAS.450..982K}. Other instability types in jets such as Kelvin-Helmholtz and pressure driven modes have also been the subject of intense studies (see for example \cite{1979ApJ...234...47H,1982ApJ...257..509H,1983ApJ...269..500C,1985ApJ...291..655P,1992A&A...256..354A,1996A&A...314..995A,1997ApJ...485..533H,1998A&A...333.1117B,
1998A&A...333.1001M,2014arXiv1408.3318P}). Currently, a consensus on the reason behind the remarkable stability of observed and numerically simulated (see e.g. \cite{
Palenzuela:2011es,Palenzuela:2010xn,Lehner:2011aa,
Komissarov:2007rc}) jets is still lacking \cite{2014HEAD...1410622K}, and will continue to be a fascinating area of research. 

In this paper, we tackle a rather different problem. (1) We do not examine collimated jets, but rather more isotropic radiation which appears to contribute to a significant \cite{Palenzuela:2010nf} or even dominant \cite{Alic:2012df} portion of the energy budget in the radiation emitted by e.g. a binary black hole system inside a common magnetosphere. This type of radiation has received less attention previously, but the particular scatter-avoiding solutions we examine, if stable (their stability has not been examined before), may prove to be a preferred (as it is the most efficient, without backscattering of energy flux) channel through which isotropic energy flux escapes through the magnetosphere, thus providing us with a perfect entry point for further research. Because a large pulse of isotropic radiation is emitted during the merger phase of a black hole binary \cite{Palenzuela:2010nf}, which can potentially be picked up by observers on Earth, such research should have relevance for multimessenger (gravitational and electromagnetic waves) astronomy. (2) Because these solutions are envisaged to be the couriers that carry energy across magnetospheres of black holes or neutron stars, relativistic effects would become important, so our analysis will necessarily have to take spacetime curvature into account. In contrast, most of the previous studies on jet stability assume flat spacetime, as jet disruption does not occur until very far from the central compact object. 
(3) We are examining global (while jet stability studies concentrate on the vicinity of the jets) solutions, and so there are subtle new instability types that may emerge. For example, there is no globally regular vacuum counterpart to the solutions we examine \cite{Brennan:2013jla,Teukolsky:1974yv}, so does that mean singularities similar to those seen in the vacuum case would generically develop even in the force-free case when we introduce perturbations? This subtle potential instability, whose underlying source is global in nature, would not have been included in the consideration of typical severe plasma instabilities. 
On the other hand, the solutions we examine do not have extreme features such as a jet boundary, so they are likely less prone to many instabilities that a jet would suffer. In fact, a cursory glance gave us no reason to strongly expect these typical plasma instabilities to severely impact these solutions (we caution however, due to different assumptions regarding the structure of the solution and the nature of the plasma, jet stability results may not be directly applicable to the present study).  
(4) Whereas previous studies mostly concentrate on the stability of a physical feature (i.e. jets), we examine the more restrictive case of the stability of a particular family of exact analytical solutions. Therefore, even though physical processes that slightly alter the exact structure of the jets -- but do not disrupt it -- may not be regarded as serious instabilities, they would in our case be necessarily recognized as a problem, as they would nudge the actual physical configurations away from being precisely the same as the exact solutions.

A consideration of all possible instabilities to an analytical solution is exemplified by the study of Kerr metric stability (see e.g. Ref.~\cite{Dafermos:2010hd} for a summary). The problem can be attacked by first studying the mode stability (see e.g. \cite{2000A&A...355..818A} for an example in the context of jet stability studies) through solving linearized perturbation equations assuming separable solutions. For example, a recent paper \cite{Yang:2014zva} studying perturbations of magnetic monopole and Blandford-Znajek solutions showed no unstable individual FFE modes\footnote{A similar analysis for perturbations over the FFE solutions we are interested in would be much more difficult. These solutions have a null current that couples to the perturbing fields just like the Blandford-Znajek case. However, unlike the current for the Blandford-Znajek solution, this null current is not proportional to a small parameter like the black hole spin, and cannot be treated using the perturbative techniques of Ref.~\cite{Yang:2014zva}.}. 
It is then an arduous task to further prove linear stability, as it is not guaranteed that all linear perturbations can be decomposed into such modes, or that the sum of infinitely many stable individual modes remains stable \cite{Dafermos:2010hd}. A proof for full nonlinear stability is more difficult still. On the other hand, while a rigorous proof is currently out of reach, important evidence of nonlinear stability can often be found using fully nonlinear numerical simulations. 
For example, several studies of the numerical robustness of the Blandford-Znajek process can be found in Refs.~\cite{Palenzuela:2011es,Palenzuela:2010xn,Lehner:2011aa,Komissarov:2007rc}.
Indeed, the ability to examine stability is cited as one of the principal motivations for developing fully time-dependent numerical FFE codes \cite{Spitkovsky2005,Parfrey:2011ta}. 
In this paper, we adopt this more accessible numerical approach to studying nonlinear stability. 

This paper is organized as follows: We begin by introducing the force-free equations in curved spacetimes, as well as the details of the exact analytical solutions that we examine in Sec.~\ref{sec:FFEIntro}. We then introduce in Sec.~\ref{sec:CodeIntro} a new pseudospectral numerical code, used for our stability study. We also present several nontrivial tests demonstrating that the code achieves exponential convergence. 
Then in Sec.~\ref{sec:NumSim}, we evolve a constraint-free perturbed initial data set, and show that the exact analytical solutions are in fact stable, despite their physical specialness. In Sec.~\ref{sec:GPNDs}, we provide some derivations and arguments that provide an intuitive explanation as to why the analytical solutions are not backscattered by spacetime curvature (this section is not necessary for understanding the numerics described in the earlier sections, and utilizes spinors extensively; therefore, readers who are only interested in the numerical aspects of this paper do not need to review Sec.~\ref{sec:GPNDs}). Finally, although not used in the numerical studies in this work, we also provide in Appendix \ref{sec:AugSys} a couple of force-free evolution systems that have improved well posedness properties. 

In this paper, we adopt geometrized units with $G=c=1$, and use $(-+++)$ for the metric signature. The beginning of the lowercase latin alphabet will be used to denote spacetime indices, and the middle of the alphabet denotes spatial indices. 
Capital latin letters will denote spinor indices, while greek letters will index different quantities in different sections, whose meaning will be clear from their context. Bold-faced letters will denote vectors and tensors. The numerical work in this paper is carried out within the pseudospectral code infrastructure of the Spectral Einstein Code (\verb!SpEC!) \cite{SXSWebsite}.

\section{The force-free equations and their exact solutions \label{sec:FFEIntro}}

\subsection{Some useful definitions}
In this paper, we will use the 3+1 form of the metric
\bea
ds^2 = -N^2 dt^2 + g_{ij}(dx^i + \beta^i dt)(dx^j + \beta^j dt)\,,
\eea
where $N$ is the lapse, $\bm{\beta}$ is the shift, and ${\bf g}$ is the (spatial) metric for the spatial hypersurfaces of constant $t$. The extrinsic curvature ${\bf K}$ of these spatial hypersurfaces is given by 
\bea
(\partial_t - \mathcal{L}_{\bm{\beta}}) {\bf g} = -2 N {\bf K}\,, \label{eq:ExCur}
\eea
which is a spatial tensor depending on both the geometry (metric) of the overall four-dimensional spacetime and the way we slice it. For example, the extrinsic curvature of a Minkowski spacetime can be nonvanishing if one picks unusual slicings. The operator on the left-hand side of Eq.~\eqref{eq:ExCur} is the derivative along the normal vector $\mathbb{t}^a$ to the spatial hypersurfaces.

When there is an electromagnetic field represented by the Faraday tensor ${\bf F}$, we can also break it down into a 3+1 form, which will then allow us to write the force-free evolution and constraint equations in terms of spatial tensors in the next section. These equations will then resemble those used in the numerical study of the Einstein equation, and can be handled with the same code infrastructure. We define the electric and magnetic vectors as 
\bea
E^a = F^{ab}\mathbb{t}_b\,, \quad B^d =\frac{1}{2} \epsilon^{abcd}F_{ab}\mathbb{t}_c\,.
\eea
Note that although we have written them as four-vectors in the definitions, they are really only spatial vectors with $E^0$ and $B^0$ vanishing. We will denote them as 3-vectors ${\bf E}$ and ${\bf B}$ below, with the understanding that a projection into the spatial slices has been taken. Within the spatial slices, we will also use traditional vector calculus notations to simplify expressions, with, for example 
\bea
{\bf E} \cdot {\bf B} \equiv E^i B^j g_{ij}, \quad ({\bf E} \times {\bf B})^i \equiv g^{il }E^j B^k \epsilon_{ljk}\,.
\eea

\subsection{The evolution equations \label{sec:EvoEqs}}
In this section, we write down a set of FFE equations with constraint damping capabilities that are numerically robust, although they possess some mathematically undesirable properties that do not appear to hinder their performance in practice. The numerical studies carried out in the main body of this paper use this evolution system. 
Aside from this system of equations, we also provide in Appendix \ref{sec:AugSys} two additional sets of equations with desirable mathematical properties, but which are numerically less forgiving. We go through the derivation of these equations in some details for pedagogical reasons, as existing literature tends to be brief and sometimes leaves out terms that should be included for curved spacetimes.

We begin by writing down the Maxwell equations in curved spacetime, which are 
\bea
(\partial_t-\mathcal{L}_{\bm{\beta}}) {\bf E} &=& N K {\bf E} + \nabla \times (N{\bf B}) - 4 \pi N {\bf J}\,, \\
(\partial_t-\mathcal{L}_{\bm{\beta}}) {\bf B} &=& N K {\bf B} - \nabla \times (N{\bf B})\,, \\
\nabla \cdot {\bf E} &=& 4 \pi \rho\,, \\
\nabla \cdot {\bf B} &=& 0\,.
\eea
To derive the current ${\bf J}$, which is the spatial part of a four-current ${\bf J}^{(4)}$, we 
begin 
with the force-free condition $F^{ab}J^{(4)}_b=0$, which states that the four-force density describing the transfer of energy and momentum between the electromagnetic fields and the charged plasma particles vanishes. This ensures that the stress-energy $T^{ab}_{\text{EM}}$ of the electromagnetic field remains dominant over that of the plasma. Indeed, we can derive the force-free condition starting from the differential conservation of energy and momentum $\nabla_a T^{ab} =0$. Then when $T^{ab}_{\text{EM}}$ is the dominant contribution to $T^{ab}$, we have \cite{Parfrey:2011ta,Palenzuela:2010xn} 
\bea \label{eq:FFCond}
\nabla_a T^{ab}\approx \nabla_a T_{\text{EM}}^{ab} =-F^{ab}J^{(4)}_b=0\,. 
\eea
In a $3+1$ decomposition, this translates into 
\bea
{\bf E}\cdot {\bf J} = 0\,, \quad 
\rho {\bf E} + {\bf J} \times {\bf B} =0\,, \label{eq:LorentzForceEqn}
\eea
where the second equation is the vanishing of the Lorentz force. 
To derive the force-free current ${\bf J}$, we take the cross product between Eq.~\eqref{eq:LorentzForceEqn} and ${\bf B}$ which gives us 
\bea \label{eq:FFECurrent}
4\pi N {\bf J} = 4\pi N ({\bf B} \cdot {\bf J}) \frac{{\bf B}}{B^2} + 4\pi N \rho \frac{{\bf E} \times {\bf B}}{B^2}\,, 
\eea
where the second term above can be seen as the charge density moving at the plasma drift velocity, and we can replace $\rho$ with $\nabla \cdot {\bf E} /4\pi$ using one of the Maxwell constraint equations. To further work out the current ${\bf B} \cdot {\bf J}$ along the ${\bf B}$ field, we note that 
\bea
(B_a/\sqrt{B_bB^b})F^{ab}J^{(4)}_b = 0 \,\, \Rightarrow \,\, {\bf E}\cdot {\bf B} = 0 \,\,\,\, \text{or} \,\,\,\, \rho=0\,,
\eea
and for nonvacuum solutions (vacuum here refers to $\bf{J}^{(4)}=0$, with these solutions satisfying the force-free condition trivially) we would like to enforce the ${\bf E} \cdot {\bf B} =0$ condition, which should be preserved along the timelike normal to the spatial hypersurfaces, and so
\bea
(\partial_t - \mathcal{L}_{\bm{\beta}}) {\bf E} \cdot {\bf B} = 0\,.  
\eea
Using the definition of the extrinsic curvature tensor Eq.~\eqref{eq:ExCur},
and substituting in the Maxwell equations, we obtain an equation for ${\bf B} \cdot {\bf J}$ that reads 
\bea \label{eq:CurrentFromCons}
4\pi N {\bf B} \cdot {\bf J} &=& - {\bf E} \cdot \nabla\times (N {\bf E}) + {\bf B} \cdot \nabla\times (N {\bf B}) \notag \\
&&- 2N K_{ij} E^i B^j + 2N K {\bf E} \cdot {\bf B}\,,  
\eea
(note that the extrinsic curvature terms on the second line appear to be missing in some of the existing literature).
Substituting Eq.~(\ref{eq:CurrentFromCons}) into Eq.~(\ref{eq:FFECurrent}) yields $\bf J$ in terms of $\bf E$ and $\bf B$.  Substituting this expression back into the Maxwell-equations yields the desired minimalist FFE evolution system. 

There is also a set of constraints that needs to be satisfied, which comes from both the Maxwell equations and the force-free condition. 
When deriving the current ${\bf J}$, we have explicitly used $q=\nabla \cdot {\bf E} /4\pi$ as the definition of charge density, so there is no need to enforce this constraint. The nontrivial constraints are $\nabla \cdot {\bf B}=0$ and the force-free constraint ${\bf E} \cdot {\bf B} =0$. These two constraints are preserved automatically by the evolution equations \cite{Gralla:2014yja}. 
Specifically, the $\nabla \cdot {\bf B}=0$ constraint is preserved by the original Maxwell equations and inherited by the force-free specialization. The ${\bf E} \cdot {\bf B} =0$ condition is also preserved, as we have explicitly used the condition $(\partial_t -\mathcal{L}_{\bm{\beta}}){\bf E} \cdot {\bf B} =0$ to derive the current. Physically, this condition fixes the magnitude of the conduction current along the ${\bf B}$ direction \cite{Parfrey:2011ta}, which would short out the ${\bf E}$ field along ${\bf B}$ by redistributing charge to eliminate the potential difference associated with that ${\bf E}$ component, thus enforcing ${\bf E}\cdot {\bf B}=0$. 

Although the $\nabla \cdot {\bf B} =0$ and ${\bf E} \cdot {\bf B} =0$ constraints are preserved by the evolution equations when they were satisfied initially, numerical noise inevitably creates some seed constraint violation that may grow further under the minimal evolution system. It is therefore beneficial to modify the evolution equations so as to be able to clean up the constraint violations as they emerge. 
For the ${\bf E} \cdot {\bf B} =0$ constraint, we adopt a strategy similar in form to Ref.~\cite{Alic:2012df}. Specifically, we add a damping term $-N\delta({\bf E}\cdot{\bf B}) {\bf B}/B^2$ to $\partial_t {\bf E}$, so that the full set of evolution equations becomes 
\bea
(\partial_t - \mathcal{L}_{\bm{\beta}}) {\bf E} &=& N K {\bf E} + \nabla \times (N{\bf B}) 
- \frac{{\bf E} \times {\bf B}}{B^2} N \nabla \cdot {\bf E} \notag \\
&&- N\frac{{\bf B}}{B^2}\left( {\bf B} \cdot \nabla \times {\bf B} - {\bf E} \cdot \nabla \times {\bf E} \right. \notag \\ 
&& \left. 
- 2K_{ij}E^i B^j + 2K {\bf E} \cdot {\bf B} + \delta {\bf E} \cdot {\bf B}\right)\,,   \label{eq:EvoE}\\
(\partial_t - \mathcal{L}_{\bm{\beta}}) {\bf B} &=& N K {\bf B} - \nabla \times (N {\bf E})\,.
\label{eq:EvoB}
\eea
The damping term is proportional to the constraint, and will not affect the physical constraint-satisfying solutions. However, it modifies the constraint evolution equation to a damped form
\bea \label{eq:ConsDampingEvo}
(\partial_t - \mathcal{L}_{\bm{\beta}}) {\bf E}\cdot {\bf B} = -\delta N {\bf E}\cdot {\bf B}\,. 
\eea
We note that our damping strategy differs from that of Ref.~\cite{Alic:2012df}, in that we have kept the original current terms from Eq.~\eqref{eq:CurrentFromCons} and treated the new damping term as an addition instead of a replacement. In contrast, Ref.~\cite{Alic:2012df} removed all of the original current terms from the evolution equations, replacing them with only the damping term. With their strategy, the evolution equations become simpler, but those current terms forcibly removed from the evolution equations will resurface in the constraint evolution equation, specifically on the right-hand side of Eq.~\eqref{eq:ConsDampingEvo}. Therefore Eq.~\eqref{eq:ConsDampingEvo} won't reduce to 
$(\partial_t - \mathcal{L}_{\bm{\beta}}) {\bf E}\cdot {\bf B}=0$ when ${\bf E}\cdot {\bf B}=0$, so that constraint-satisfying FFE solutions at some instance of time can not stay constraint-satisfying as the simulation progresses. 
However, this does not invalidate the results of Ref.~\cite{Alic:2012df}, as the size of the constraint violation would be negatively correlated with $\delta$ (and positively correlated with the magnitudes of the derivatives of ${\bf B}$ and ${\bf E}$), as the damping will be activated when ${\bf E}\cdot {\bf B}$ grows too large. Indeed, Ref.~\cite{Alic:2012df} adopted a $\delta$ greater than the inverse of the time step size and observed a well controlled ${\bf E}\cdot {\bf B}$. Such a strategy however leads to a stiff term in the evolution equations that has to be treated with an implicit-explicit (IMEX) evolution scheme \cite{Alic:2012df}.
As the analytical solutions we examine are exactly constraint-satisfying at all times, and since we do not use an IMEX scheme, we will use the damping term as an addition with a moderate coefficient $\delta = 100$, allowing us to enjoy its constraint cleaning benefits but avoid the aforementioned complications. Finally, note that this damping term does not introduce magnetic monopoles, see Eq.~\eqref{eq:divBEvoEq}. 

We note that the introduction of the additional damping term replaces 
another alternative constraint cleaning strategy of removing the component of ${\bf E}$ along ${\bf B}$ after taking each time step, which has been widely utilized 
(e.g. Refs.~ \cite{Palenzuela:2010xn,Spitkovsky:2006np,Komissarov:2004ms,Parfrey:2011ta}).
Such an alteration of the evolving fields at a discrete set of times (which depend on resolution) will result in a failure of the system to achieve the usual convergence behaviour that would be expected
if the scheme were just applied to a set of differential equations without this alteration \cite{Li:2011zh}. In contrast, a damping term is less intrusive and its properties are more easily understood. We will show that our set of evolution equations displays the expected convergence behaviour in Sec.~\ref{sec:AnaNullTest}. In addition, because this damping term does not contain derivatives, it will not affect the characteristic structure of the evolution equations, which is particularly helpful for our pseudospectral implementation. 

Lastly, we note that aside from the aforementioned ${\bf E}\cdot {\bf B} = 0$ and $\nabla \cdot {\bf B} = 0$ constraints, we have an additional constraint of $E^2 \leq B^2$. When a $E^2 > B^2$ region develops, the plasma particles have to move faster than the speed of light in order to experience a vanishing Lorentz force. 
One can see this from the second term in Eq.~\eqref{eq:FFECurrent}, which can be written as $q {\bf v}_{\rm d}$, where ${\bf v}_{\rm d}$ is the drift velocity for the advection of the charge density \cite{Spitkovsky:2006np,Parfrey:2011ta}. The inequality $E^2 > B^2$ then implies a superluminal ${\bf v}_{\rm d}$. 
For an even simpler demonstration, if we consider the special-relativistic point particle case,
then the requirement for the vanishing of the Lorentz force $q({\bf E} + {\bf v} \times {\bf B})$ would imply $|{\bf v}| >1$ when $E^2>B^2$. Such superluminal motion is unrealistic, but FFE evolution equations cannot prevent it because the current (Eq.~\ref{eq:FFECurrent}) is derived without invoking plasma physics. 
Consequently, the FFE equations have no internal checks that enforce the $B^2 - E^2 \geq 0$ constraint. In other words, the $B^2-E^2 \geq 0$ constraint is not strictly a constraint in the mathematical sense like $\nabla \cdot {\bf B} = 0$ and ${\bf E} \cdot {\bf B} =0$, and solutions satisfying the force-free condition \eqref{eq:FFCond} are not automatically magnetically dominated.  
One symptom of the breakdown of this physical constraint is that, when strong waves interact \cite{Spitkovsky:2006np}, the Alfv\'en mode characteristic speeds become complex, breaking the hyperbolicity of the FFE evolution equations (because these are physical modes, augmentations to the evolution equations like those in Appendix \ref{sec:AugSys} will not be able to change their characteristic speeds and will thus not cure this hyperbolicity breakdown).  
Another example is the formation of current sheets, in which the magnetic fields can vanish and then reverse direction \cite{Spitkovsky:2006np}. 
Physically, the force-free assumption is invalid when $B^2-E^2 < 0$. The actual plasma particles, which have inertia, will experience a nonvanishing Lorentz force and be accelerated. In other words, the system becomes dissipative \cite{Parfrey:2011ta}, averting divergences and possibly restoring magnetic dominance \cite{Spitkovsky:2006np}.
A proper treatment of such regions would require special codes for the plasma \cite{Palenzuela:2010xn} that do not assume the force-free condition being met. 
For the numerical studies carried out in the later parts of this paper, we do not need such a sophisticated treatment as we are guaranteed magnetic dominance by the presence of a magnetic monopole in the solutions we examine. We will discuss this in more details in the next section. In particular, we do not need to adopt the common procedure of scaling down the ${\bf E}$ field after each time step to avoid electric dominance  \cite{Palenzuela:2010xn,Spitkovsky:2006np,Komissarov:2004ms,Parfrey:2011ta}.

We have now a basic set of evolution equations and the associated constraints. However, this does not automatically mean that we can simply plug them into the computer and run simulations, because we need to ensure that the evolution equations are well posed. This is a rather technical discussion, which we have relegated to the appendixes. It turns out that the simple evolution systems given by Eqs.~\eqref{eq:EvoE} and \eqref{eq:EvoB} (which we adopt for our numerical code) is not strictly strongly hyperbolic (see Appendix \ref{sec:Hyperbolicity}), but the violation is insignificant enough that it does not create problems in practice for the simulations carried out in this work. Nevertheless, we have formulated two additional evolution systems that are not only strongly hyperbolic, but also have additional desirable properties in terms of well-posedness. One of them is strongly hyperbolic even when ${\bf E}\cdot {\bf B}\neq 0$, while the other has a particularly simple constraint evolution behaviour. These systems are given in Appendix \ref{sec:AugSys}. 
Unfortunately, they (also including a similar evolution system from Ref.~\cite{Harald}) introduce a 
term containing the second derivative of ${\bf B}$ (see Eq.~\ref{eq:divBEvoEq}) into the evolution equation for $\nabla \cdot {\bf B}$, making it sensitive to high-frequency noises (see Sec.~\ref{sec:ConsEqsHyper}). 
In other words, these mathematically more satisfying systems are numerically less forgiving. 
Therefore, we take a pragmatic approach, and use the simple system as given by Eqs.~\eqref{eq:EvoE}-\eqref{eq:EvoB} for the numerical studies presented in the main text. 
However, the mathematically improved systems may yet prove useful for application in finite-difference and finite-volume FFE codes, which tend to be more forgiving
in the presence of high-frequency noise than pseudospectral schemes. 

\subsection{The exact analytical solutions \label{sec:IntroSol}}
In this section, we introduce the analytical solutions whose stability properties we seek to examine. The solutions are introduced in Ref.~\cite{Brennan:2013jla}, and readers interested in their derivation should consult that reference. 

\begin{figure}[t,b]
\begin{overpic}[width=0.99\columnwidth]{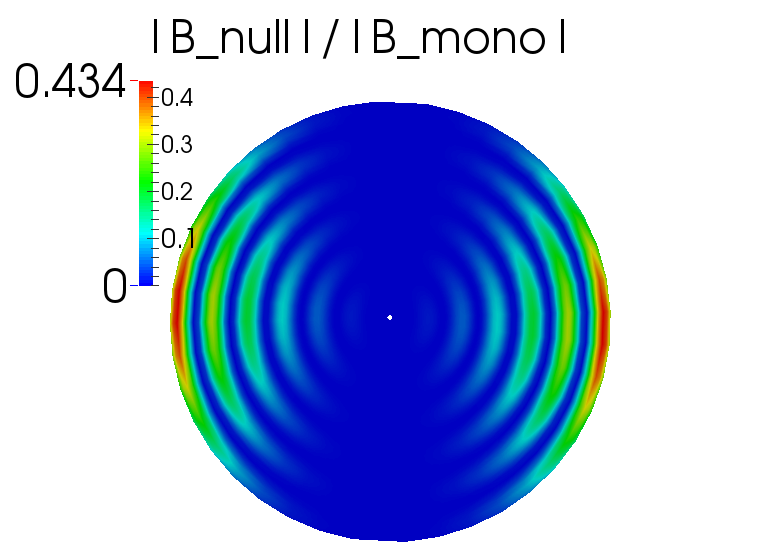}
\end{overpic}
\caption{The ratio $|B_{\text{null}}|/|B_{\text{mono}}|$ for the null$^+$ solution studied, where 
${\bf B}_{\text{null}}$ and ${\bf B}_{\text{mono}}$ are the null and monopole contributions in Eq.~\eqref{eq:FieldTensorWithMono}, while
the norm is defined as $|B_{\text{null}}|=\sqrt{{\bf B}_{\text{null}} \cdot {\bf B}_{\text{null}}}$. Shown is a cross-section in the $xz$-plane, which extends radially from $R_-=1.95$M to $R_+=195$M.  
}
\label{fig:BRatio}
\end{figure}

\begin{figure}[t,b]
\begin{overpic}[width=0.98\columnwidth]{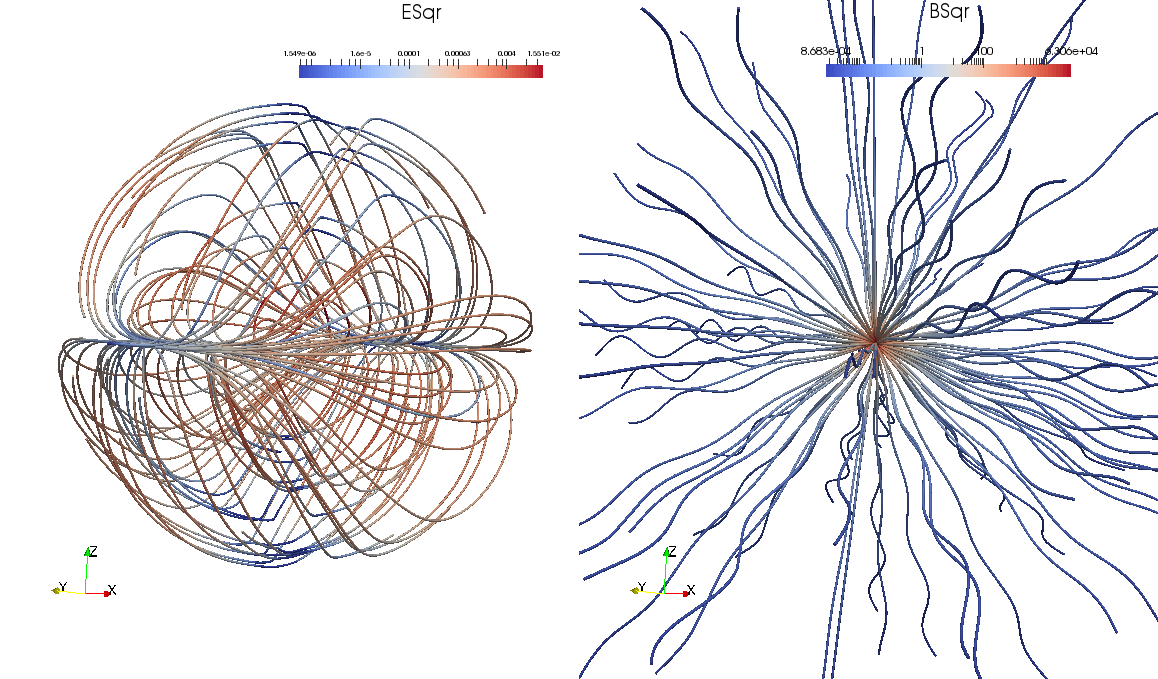}
\end{overpic}
\begin{overpic}[width=0.32\columnwidth]{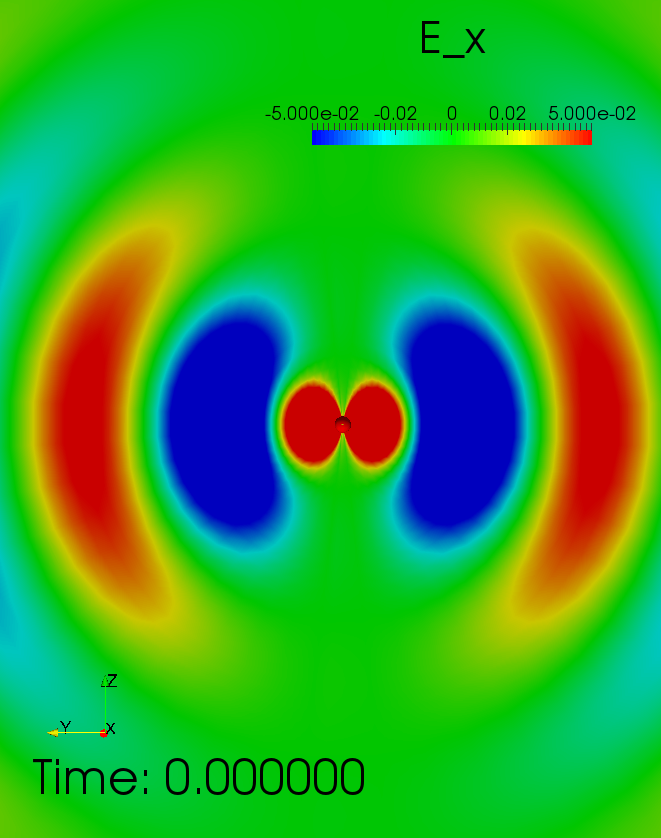}
\end{overpic}
\begin{overpic}[width=0.32\columnwidth]{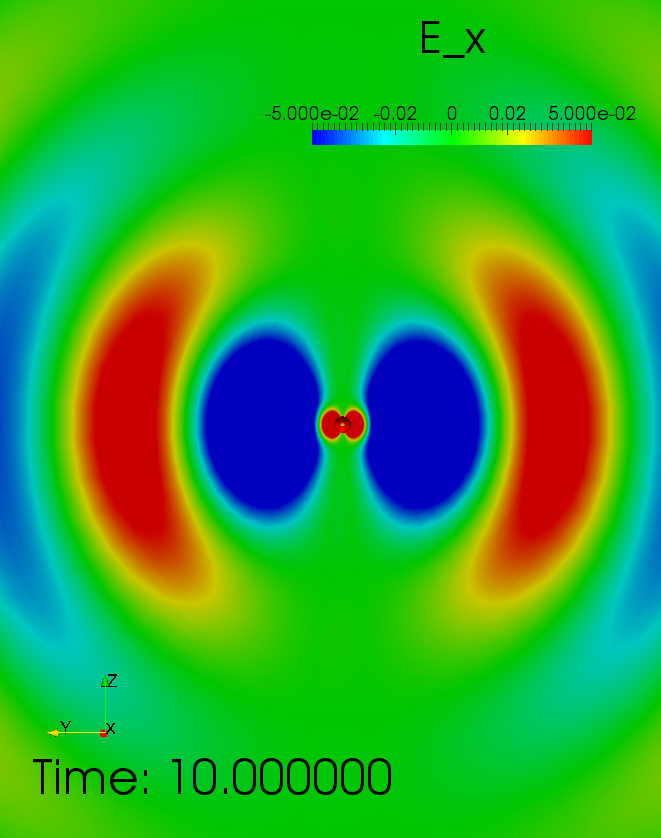}
\end{overpic}
\begin{overpic}[width=0.32\columnwidth]{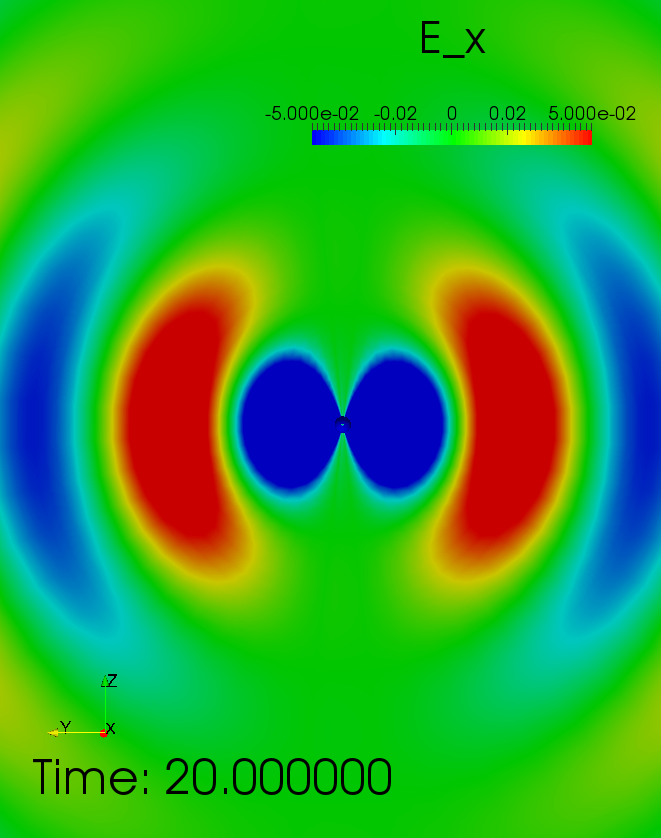}
\end{overpic}
\caption{Top left figure shows the electric field lines, and the top right figure shows the magnetic field lines for the same null$^+$ solution as shown in Fig.~\ref{fig:BRatio}, the lines are coloured by $E^2$ and $B^2$ respectively. The bottom figures plot the time variation of the $E_x$ component, showing the wave propagating inwards.
}
\label{fig:FieldLines}
\end{figure}

We begin by emphasizing that the spacetime background used in this paper is the Schwarzschild spacetime, and the situation in a spinning black hole case is beyond the scope of our current study. This restriction stems from the need to enforce the magnetic dominance condition that we discussed in Sec.~\ref{sec:EvoEqs}. The scatter-avoiding wave-only analytical solutions given in Ref.~\cite{Brennan:2013jla} are the so-called null solutions, which can be seen as generalizations to plane waves (see later in the section), and as such have $E^2=B^2$ just like the plane waves. When we add a small perturbation to it, we can easily end up with $E^2>B^2$ regions which can not be properly accounted for under the force-free approximation. However, this is not a problem because instead of these null solutions, we in fact study the stability of what we call null$^+$ solutions, where a magnetic monopole is superimposed onto the null solutions (this prescription is also provided by Ref.~\cite{Brennan:2013jla}, and \cite{Gralla:2014yja}), so all the scatter-avoiding properties of the null solutions are simply inherited by the wave components of the null$^+$ solution, yet magnetic dominance is maintained even with the presence of perturbations (Figs.~\ref{fig:BRatio} and \ref{fig:FieldLines} depict the basic structure of field quantities for the null$^+$ solutions, and we will discuss them in more details later). Subsequently, we restrict ourselves to the Schwarzschild spacetimes because only when the black hole is not spinning would such a monopole addition not interfere with the null wave component, and the total solution (as a simple superposition of the wave and monopole components) still satisfy the force-free conditions. 
On the other hand, these two components would couple in a nontrivial way in a spinning black hole spacetime, and we do not have a simple null$^+$ solution in that situation to study (the null$^+$ solution in that case has not been worked out yet). 

We note that because in astrophysical applications, we would have a split monopole background present in addition to the energy-carrying waves, we are more interested in the stability of null$^+$ solutions instead of the null ones on physical grounds. So it is the stability of the null$^+$ solutions and not that of the null solutions that is the intended target of study in this paper. The study of the stability of null$^+$ solutions is \emph{not} a surrogate for that of the null solutions. In fact, our stability results would unlikely translate from the null$^+$ to the null solutions, as we will discuss later in the Conclusion section. 
Furthermore, for our numerical study, we will pick the null$^+$ solutions with an ingoing (instead of outgoing
\footnote{The outgoing waves are obtainable through the transformations in Sec.~5 of Ref.~\cite{Brennan:2013jla}.}) wave component, 
as these are regular on the future horizon of the Schwarzschild black hole. The outgoing solutions face the same possible ``fragility due to specialness" issue, but need to be glued onto an astronomically realistic interior solution (not currently available); otherwise, they will represent energy flux emerging from the past horizon \cite{Gralla:2014yja}. 

Before we dive into the details of the null$^+$ solutions, we first introduce some formalism that will be needed. The structure of the solutions is most explicit in the Newman-Penrose (NP) formalism, wherein we express tensors under the NP null tetrad $\{{\bf l, n, m, \bar{m} }\}$ that consists of two real null vectors ${\bf l}$ and ${\bf n}$ usually chosen to be in the outgoing and incoming directions, and two complex null vectors ${\bf m}$ and ${\bf \bar{m}}$. This tetrad can be seen as the null version of an orthonormal tetrad \cite{Zhang:2012ky}, relating to it via a rigid transformation. Under the null tetrad, the metric is a constant matrix 
\bea
\bma 0 & -1 & 0 & 0 \\
-1 & 0 & 0 & 0 \\
0 & 0 & 0 & 1 \\
0 & 0 & 1 & 0 
\ema\,,
\eea
just as the metric is a constant matrix $\text{diag}\{-1,1,1,1\}$  under an orthonormal tetrad. The freedom for choosing the null tetrad is then given by the transformations that preserve the metric, namely the Lorentz transformations. Such a tetrad is just a basis to express components of a tensor in. In particular, the components of the Faraday tensor ${\bf F}$ under the Newman-Penrose null tetrad are the Newman-Penrose scalars $\phi_0$, $\phi_1$ and $\phi_2$, defined by 
\bea
\phi_0 &=& F_{ab} l^a m^b\,, \\
\phi_1 &=& \frac{1}{2} F_{ab} (l^a n^b + \bar{m}^a m^b)\,,\\
\phi_2 &=& F_{ab} \bar{m}^a n^b \,. \label{eq:phi2FromF}
\eea
These three complex numbers are simply the $6$ real components of $F_{ab}$ (three from ${\bf E}$ and three from ${\bf B}$, and these vectors are of course just ${\bf F}$'s components written in the coordinate basis) recombined. 

One reason why the NP tetrad is nicer than a coordinate basis is that we can pick the ${\bf l}$ and ${\bf n}$ in it to be pointing along special directions such as the outgoing and ingoing null directions of a Schwarzschild spacetime (we will be using the so-called Kinnersley tetrad \cite{Kinnersley:1969zza}, which has this property. See Appendix \ref{sec:NullSolKerrSchild} for their detailed expressions). 
This specialness in the basis orientation translates into special meanings for the components of $F_{ab}$ associated with these bases, and subsequently $\phi_0$, $\phi_1$ and $\phi_2$ can be interpreted as the incoming wave, Coulomb background and outgoing wave pieces of ${\bf F}$, respectively. In particular, this identification is physical and gauge (coordinate choice) invariant, as the outgoing and ingoing directions as identified by ${\bf l}$ and ${\bf n}$ in a Kinnersley tetrad are those determined by the spacetime geometry and are unambiguous \cite{Zhang:2012ky}, and not based on the radial direction of arbitrary coordinate systems, which can be subject to random coordinate transformations that change the radial direction. 

As ${\bf E}$ and ${\bf B}$ and the three NP scalars are simply ${\bf F}$ components written in different basis, it is not surprising that we can translate the force-free equations \eqref{eq:EvoE} and \eqref{eq:EvoB} into equations for $\phi_0$, $\phi_1$ and $\phi_2$, and that the equations' structure clarifies tremendously because all the quantities in it now have clear physical meaning. This is the approach taken in \cite{Brennan:2013jla}. 

For the ingoing null$^+$ solutions that we are interested in for this work, we need outgoing $\phi_2 =0$, while the ingoing wave $\phi_0$ and background $\phi_1$ are nonvanishing. It turns out however that in a Schwarzschild background, the equations for $\phi_0$ and $\phi_1$ decouple, so we can solve for $\phi_0$ first, assuming $\phi_1 =0$, and then add a monopolar $\phi_1$ back later. The ansatz Ref.~\cite{Brennan:2013jla} uses when solving for $\phi_0$ is then $\phi_1 = 0= \phi_2$, as well as an additional condition (guess) that the current flows in the ingoing null direction ${\bf n}$. Under these assumptions, we get an equation for $\phi_0$ we can actually solve. 

After obtaining an expression for $\phi_0$ (like the one we will write down later in the section), we can reconstruct the Faraday tensor as
\bea  \label{eq:FieldTensor1}
F_{ab} = 4\Re\left(\phi_0 \bar{m}_{[a}n_{b]} \right). 
\eea
which can be shown to satisfy 
\bea
\frac{1}{2} F^{ab}F_{ab} &=& -\frac{1}{2} {}^*F^{ab}{}^*F_{ab} = B^2 - E^2 =0\,, \label{eq:NullCond1}\\ 
\frac{1}{4} F_{ab}{}^*F^{ab} &=& {\bf E} \cdot {\bf B} = 0\,. \label{eq:NullCond2}
\eea
These conditions form the definition for a null wave. We have the following observations for our ingoing null wave solutions (which are also inherited by the wave component of the null$^+$ solutions)
\begin{enumerate}
\item Locally, the two null wave conditions imply that the {\bf E} and {\bf B} fields are orthogonal to each other, and share the same amplitude just like a plane wave. These solutions are of course not really plane waves, see the ${\bf E}$ field lines in Fig.~\ref{fig:FieldLines} (also note the ${\bf B}$ field lines in that figure are for null$^+$ and not null solutions, and are not applicable for our present discussion). 

\item The amplitude of ${\bf E}$ and ${\bf B}$, and thus the wave, is given by $\phi_0$. 

\item This ``local plane wave'' travels along the ingoing ${\bf n}$ direction (radially inwards), which is also the direction of the current. 

\item The presence of the current is an extra freedom not available to vacuum Maxwell equations, so while FFE null solutions can be globally regular, singularities would develop in a vacuum null solution \cite{Brennan:2013jla}. 

\item $\phi_1$ and $\phi_2$ vanish everywhere in this solution, meaning that the null wave is not backscattered (in which case an outgoing component would be generated) by either the current or the spacetime curvature. And this property is shared by outgoing solutions, which makes them highly efficient channels for transferring energy. 

\item The magnitude of $\phi_0$ (and thus the amplitudes of ${\bf E}$ and ${\bf B}$) is not limited. The solution represents fully nonlinear waves satisfying the nonlinear FFE equations. This allows the outgoing variant of these solutions to carry an arbitrarily large energy flux from the interior region through the magnetosphere. If energy flux travels via scattered solution on the other hand, we would likely see much of the energy turn back and be swallowed by the black hole. This is not what simulation of binary merger in magnetospheres appears to show.  

\item When we add a monopole background, and take the small amplitude/linearized limit, these waves reduce to the travelling waves discussed in Ref.~\cite{2014PhRvD..90j4022Y}. And if we further take the eikonal limit such that the wavelength is much smaller than the radius of curvature, these waves become Alfv\'en waves travelling along the background magnetic field \cite{Yang:2015ata}. 
\end{enumerate}

Now that we have a broad picture of what these solutions look like, we turn to the details of a particular representative example that we will study in this work. There are infinitely many solutions to the $\phi_0$ equation, and we cannot numerically examine the stability of each and every one of them. So instead, we pick an arbitrary representative solution which does not have any special properties (it shares the same basic current/field structure and other physical properties listed above with all of its siblings, and possesses no special symmetries) that would make it more stable than any other solution. Therefore its stability should be seen as strong evidence that most, if not all, of the other solutions obtained in the same manner should also be stable. 

One such representative solution is give in Sec 4.2.4 of Ref.~\cite{Brennan:2013jla}, whose $\phi_0$ is given by 
\bea \label{eq:phi0}
\phi_0 = \frac{f^R+if^I}{\Delta \rho}\,,  
\eea
where 
\bea
\Delta = r^2 - 2Mr\quad {\rm and} \quad \rho = -\frac{1}{r}
\eea
in the ingoing Kerr (Eddington-Finklestein) coordinates $(\nu,r,\theta,\psi)$, 
with $M$ being the mass of the background Schwarzschild black hole. 
The quantity $f^R$ is a real function of the form 
\bea \label{eq:fR}
f^R = 15 F(\nu) \sin^2 \theta \cos \theta \cos \psi \,,  
\eea
where $F(\nu)$ is an arbitrary function specifying essentially the time dependence of the solution ($\nu$ is the null coordinate in ingoing Kerr). 
The quantity $f^R$ also determines a companion function 
\bea \label{eq:fI}
f^I &=& \frac{1}{\sin\theta} \int (\partial_{\psi}f^R)d \theta.  \notag  \\
&=& - 5 F(\nu) \sin \psi \sin^2 \theta\,. 
\eea

Given these expression we have now 
\bea \label{eq:phi0explicit}
\phi_0 &=& \frac{\sin^2 \theta}{\Delta \rho} \left[5F(\nu)(3\cos\theta\cos\psi-i\sin\psi) \right]\,,
\eea
while the current is
\bea
J&=& \frac{1}{2\pi \sqrt{2} \Delta}\left[ 20F(\nu)\cos\psi\sin\theta(2\cos^2\theta-\sin^2\theta)\right]
\,,
\eea
flowing along the ingoing congruence tangential to the ingoing null base ${\bf n}$. 
 
When we numerically implement this solution, we will use the Kerr-Schild coordinate system $(\tilde t, x,y,z)$, instead of the ingoing-Kerr coordinates originally utilized in Ref.~\cite{Brennan:2013jla}. 
In Kerr-Schild, we have 
\bea
\phi_0 &=& \frac{3 (x^2+y^2) F(r+t)}{r^3} \left[3 z \sin \left(\tan ^{-1}\left(\frac{y}{x}\right)-\frac{1}{2} \pi  \text{sgn}(x)\right)
\right. \notag \\ && \left.
+i r \cos \left(\tan ^{-1}\left(\frac{y}{x}\right)-\frac{1}{2} \pi  \text{sgn}(x)\right)\right]
\eea
where $r=\sqrt{x^2+y^2+z^2}$. 
We can also now add a magnetic monopole piece
\bea \label{eq:MagMonopole}
\phi_1= -\frac{i}{2} \frac{q}{r^2}\,, 
\eea
(it has the same expression in ingoing Kerr and Kerr-Schild coordinates) to obtain our final magnetically dominated null$^+$ solution. 
We can do this (despite the equations being nonlinear) because the null and monopole solutions decouple in a Schwarzschild spacetime in the sense that the monopole has no currents, and its field tensor ${\bf F}$ does not exert a force on the current of the null solution \cite{Gralla:2014yja}. Therefore, the superposition of the null and monopole solutions still satisfy the force-free condition, and yield another FFE solution. 
(We note that although a magnetic monopole is not physically realistic, a more realistic split-monopole solution can be constructed by gluing two copies of the null$^+$ solutions together, with a current sheet on the interface \cite{Brennan:2013ppa,Gralla:2014yja}.)
The monopole only contributes to the Coulomb part, 
and the field tensor is now given by 
\bea  \label{eq:FieldTensorWithMono}
F_{ab} = 4\Re\left(\phi_0 \bar{m}_{[a}n_{b]} + \phi_1 m_{[a}\bar{m}_{b]} \right)\,. 
\eea
When combined with the expressions for the tetrad basis summarized in Appendix \ref{sec:NullSolKerrSchild}, one can generate the Faraday tensor and subsequently the ${\bf E}$ and ${\bf B}$ fields via
\bea \label{eq:EBFromF}
E^a = F^{ab}T_b\,, \quad B^a = (1/2)\epsilon^{abcd}F_{cd}T_d 
\eea
where ${\bf T}$ is the timelike normal to the Kerr-Schild spatial slices. We do not reproduce the full expressions for ${\bf E}$ and ${\bf B}$ in Kerr-Schild coordinates, as they are long and tedious. We can however, plot figures that further illustrate the properties of the null$^+$ solutions.

For concreteness, we first need to pick the parameters in our $\phi_0$ and $\phi_1$. Because 
$\phi_1$ drops off as $\sim 1/r^2$ (see Eq.~\ref{eq:MagMonopole}) while $\phi_0$ drops at a slower rate of $\sim 1/r$ (see Eq.~\ref{eq:phi0explicit}), we pick $q=1000$ and some arbitrary values (so the solution has no special stability properties as compared to the rest in the family) $F(\nu) = A \cos(\Omega \nu)$ with $A=1$ and $\Omega=0.1$ (i.e. the solution is time-dependent), 
so that the monopole is large enough to ensure magnetic dominance at the outer edge of our computational domain (see the next section). The ratio between the null and monopole contributions to ${\bf B}$ is shown in Fig.~\ref{fig:BRatio}, where we see that the two contributions are comparable in magnitude in the outer regions of the computational domain.

In Fig.~\ref{fig:FieldLines}, we plot the field lines for ${\bf E}$ and ${\bf B}$ for the null$^+$ solution with the parameters set in the last paragraph. The top left panel of the figure shows the ${\bf E}$ field lines, which clearly demonstrate that the solution is fully three dimensional, without any axisymmetry. The top right figure shows the magnetic field lines, which is clearly affected by the background monopolar contribution. The bottom panels show the time variation of the $E_x$ component of the electric field, demonstrating the time dependence of our solution and the fact that the wave is propagating inwards. 
Looking back at Fig.~\ref{fig:BRatio}, one also sees that the wave component of this particular null$^+$ solution is rather isotropic and not concentrated near the poles so there are no jet-like features. 

\section{A pseudospectral numerical code \label{sec:CodeIntro}}
In this section, we briefly introduce the pseudospectral code used for evolving the force-free equations. As is noted in Refs.~\cite{Parfrey:2011ta}, a pseudospectral code is especially suited to the task of examining stabilities, as it avoids erroneous instabilities that may be triggered by the larger
numerical noise present in finite-difference or finite-volume schemes. An example is pointed out by Ref.~\cite{Parfrey:2011ta}, that during a study of the Sweet-Parker reconnection in Ref.~\cite{Ng:2011uia}, a spectral code was observed to not suffer from secondary island formations resulting from a tearing-mode instability, in contrast to results obtained using finite-difference and finite-volume schemes. 

In addition to our main evolution code, we also implement an initial data solver to ensure that the constraints are properly satisfied. We note that this is an additional improvement relative to existing codes, as many previous studies using finite-difference or finite-volume codes have evolved initial data that violated the constraints 
(%e.g. \cite{Palenzuela:2009hx,Alic:2012df}, 
they however employ constraint cleaning schemes to remove the violation later during evolution). 

\subsection{The code infrastructure \label{sec:CodeInfra}}
In the interest of completeness, we briefly introduce the basic infrastructure employed in this work. Our force-free code is a module of the \verb!SpEC! code, which was developed primarily to study binary black holes in full general relativity. Readers interested in the details are encouraged to consult \cite{SXSWebsite} and the list of articles shown there. 

\begin{figure}[t,b]
\begin{overpic}[width=0.98\columnwidth]{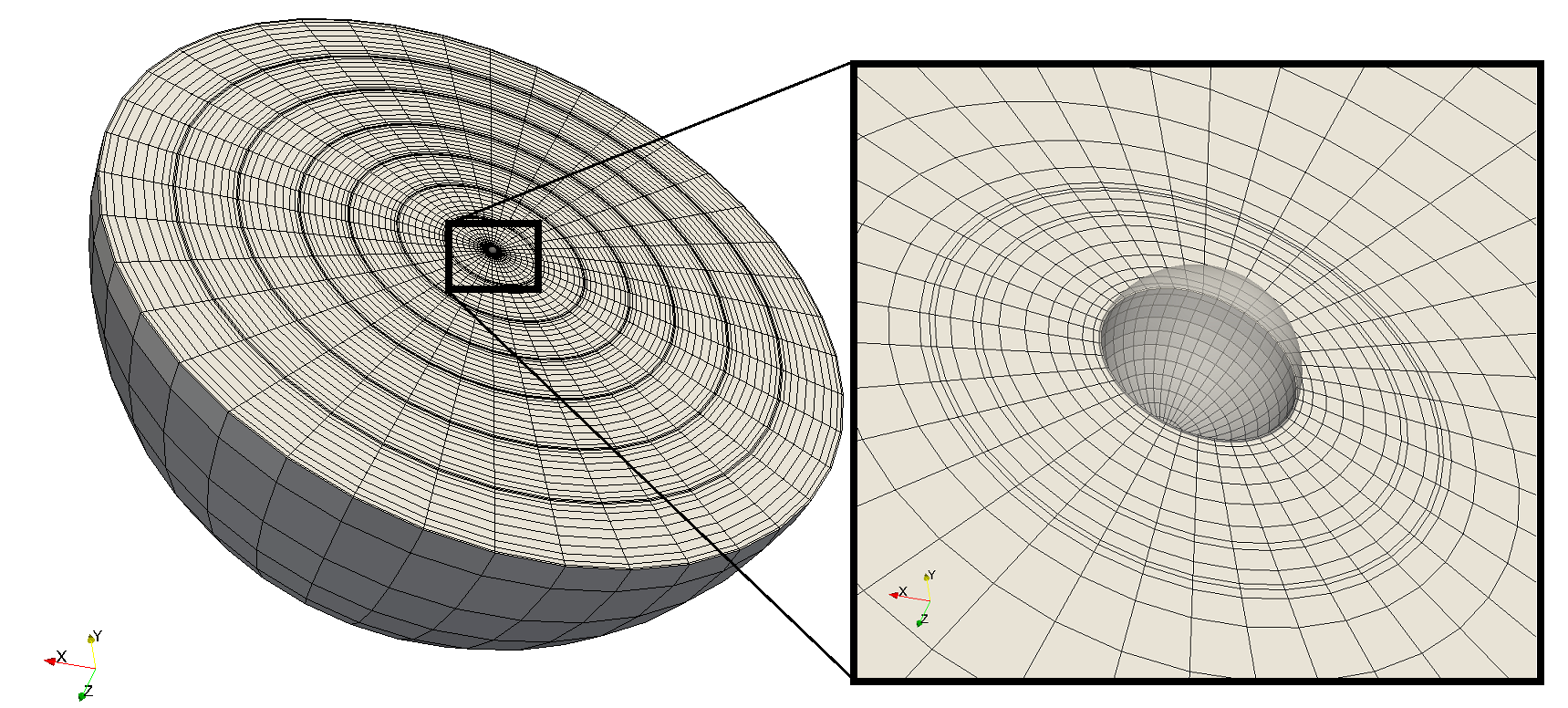}
\end{overpic}
\caption{Half of the computational domain and the spectral grid (spectral collocation points are at the intersection of the black lines), the semitransparent sphere represents the event horizon of the Schwarzschild black hole.
}
\label{fig:Grid}
\end{figure}

The basic setup of the computational domain used in this work is a spherical shell (see Fig.~\ref{fig:Grid}), whose inner edge is at $R_-=1.95$M, and whose outer edge extends to $R_+=195$M, where M is the mass of the Schwarzschild black hole, 
which sets the length and time scales for the simulations, and will be used as a unit in the plots. The inner edge terminates just inside the event horizon at $2$M (the semitransparent surface in Fig.~\ref{fig:Grid}), effectively excising the singularity from the computational domain, so we won't run into related numerical issues. The inner edge being inside the event horizon means there will be no information coming through that boundary into the computational domain, and so we do not need to impose boundary conditions there. 

For the outer boundary, the evolution variables ${\bf E}$ and ${\bf B}$ are first translated into the characteristic modes (see Appendix \ref{sec:Hyperbolicity} for their detailed expressions), which can be seen as waves propagating normal to the boundary, carrying information into and out of the computational domain. To ensure there are no additional incoming perturbations during the simulation, which would create the false impression of instabilities, we use the analytical expressions for ${\bf E}$ and ${\bf B}$ from the exact null$^+$ solutions to compute the clean incoming characteristic modes and impose them as boundary conditions (so there is no additional perturbation being carried by these waves into the computational domain). On the other hand, there are no conditions imposed on the outgoing characteristic modes. This way, there is no new perturbations coming into the computational domain, but the perturbations already inside are allowed to exit. 

In order to carry out parallel computation, we break the entire computational domain into eight concentric spheres, whose boundaries are seen as dense concentrations of black circles in Fig.~\ref{fig:Grid}. Each subdomain is handed to a separate processor for computations. The communication between subdomains is also done via characteristic modes, where the outgoing characteristic modes of one subdomain are matched onto the ingoing characteristic modes of its neighbouring subdomains via a penalty method 
\cite{Hesthaven1997,Hesthaven1999,Hesthaven2000,
Gottlieb2001}
so that any discontinuity across the subdomain boundaries is forced to vanish over time. 
Within each subdomain, the data is represented pseudospectrally through an expansion into Chebyshev polynomials in the radial direction and spherical harmonics in the angular directions. When we take spatial derivatives, we simply use the analytical expressions for the derivatives of the individual basis functions and sum the results up using the expansion coefficients. The code is pseudospectral and not fully spectral in that we do not evolve the series expansion coefficients, but instead keep the values of ${\bf E}$ and ${\bf B}$ on a set of collocation points (at the intersections of the black lines in Fig.~\ref{fig:Grid}) optimized for translating back and forth into expansion coefficients (so that we can go quickly into the series expansion representation and take spatial derivatives there, before jumping back). This way, we can implement the evolution equations in their natural spacetime form, yet still take advantage of the high accuracy of spectral derivatives. The number of collocation points represents the resolution of the simulation and corresponds to the highest order of the basis functions used (e.g. the largest $l$ in $Y^{lm}$). We label the different levels of resolution using an integer $k$ which changes linearly with the total number of collocation points in each spatial dimension. In the radial direction, the number of collocation points is given by $k+6$. In the spherical directions, the highest $l$ in the harmonics used is $2k+7$, which translates into $2k+8$ collocation points in the $\theta$ direction, and $4k+16$ collocation points in the $\phi$ direction. To give readers an intuitive feel of the density of collocation points, Fig.~\ref{fig:Grid} plots the grid for $k=6$.

\subsection{Initial data solver \label{sec:IDSolver}}
It is frequently the case with FFE evolutions that the initial ${\bf B}$ field is not divergence free, and subsequently $\nabla \cdot {\bf B}$ is cleaned using some additional cleaning field (see Appendix \ref{sec:HyperCleaning}) during evolution. We do not implement such a cleaning field, and instead properly solve the constraints for our initial data. This is only necessary for the evolution of the perturbed solutions carried out in Sec.~\ref{sec:NumSim}, and is not used for the numerical tests of Sec.~\ref{sec:NumTests}, where we simply use exact constrain-satisfying analytical solutions as initial data. 

The divergence $\nabla \cdot {\bf B}$ can be removed by solving the Poisson equation 
\begin{equation}
\label{eq:Poisson}
\nabla^2 \Phi = - \nabla\cdot {\bf B} 
\end{equation}
on the initial spatial hypersurface, and then ${\bf B}+\nabla \Phi$ will be a divergence-free field. We solve Eq.~\eqref{eq:Poisson} with the multidomain spectral method described in Ref.~\cite{Pfeiffer2003}, and set the Dirichlet boundary condition $\Phi =0$, so as to preserve the original ${\bf B}$ as much as possible by avoiding any unnecessary $\nabla \Phi$ pointing between different segments of the boundaries. 
 
We also tune the ${\bf E}$ field so that the FFE constraint ${\bf E} \cdot {\bf B} = 0$ is also satisfied by the initial data. This is achieved easily through an algebraic operation 
\begin{eqnarray} 
{\bf E} \rightarrow {\bf E} - \frac{{\bf E} \cdot {\bf B}}{B^2}{\bf B}\,. \label{eq:PostSteppingCleaning1} 
\end{eqnarray}
As this only modifies the ${\bf E}$ field, it will not interfere with the earlier divergence cleaning step. We emphasize that this is the same operation as is given by Eq.~(15) of Ref.~\cite{Palenzuela:2010xn}, except we are strictly applying the operation on the initial data, whereas in \cite{Palenzuela:2010xn} this operation is applied at every time step of the evolution. 

\subsection{Numerical tests \label{sec:NumTests}}
\subsubsection{Constant $B$ field in flat spacetime}
\begin{figure}[t,b]
\begin{overpic}[width=0.99\columnwidth]{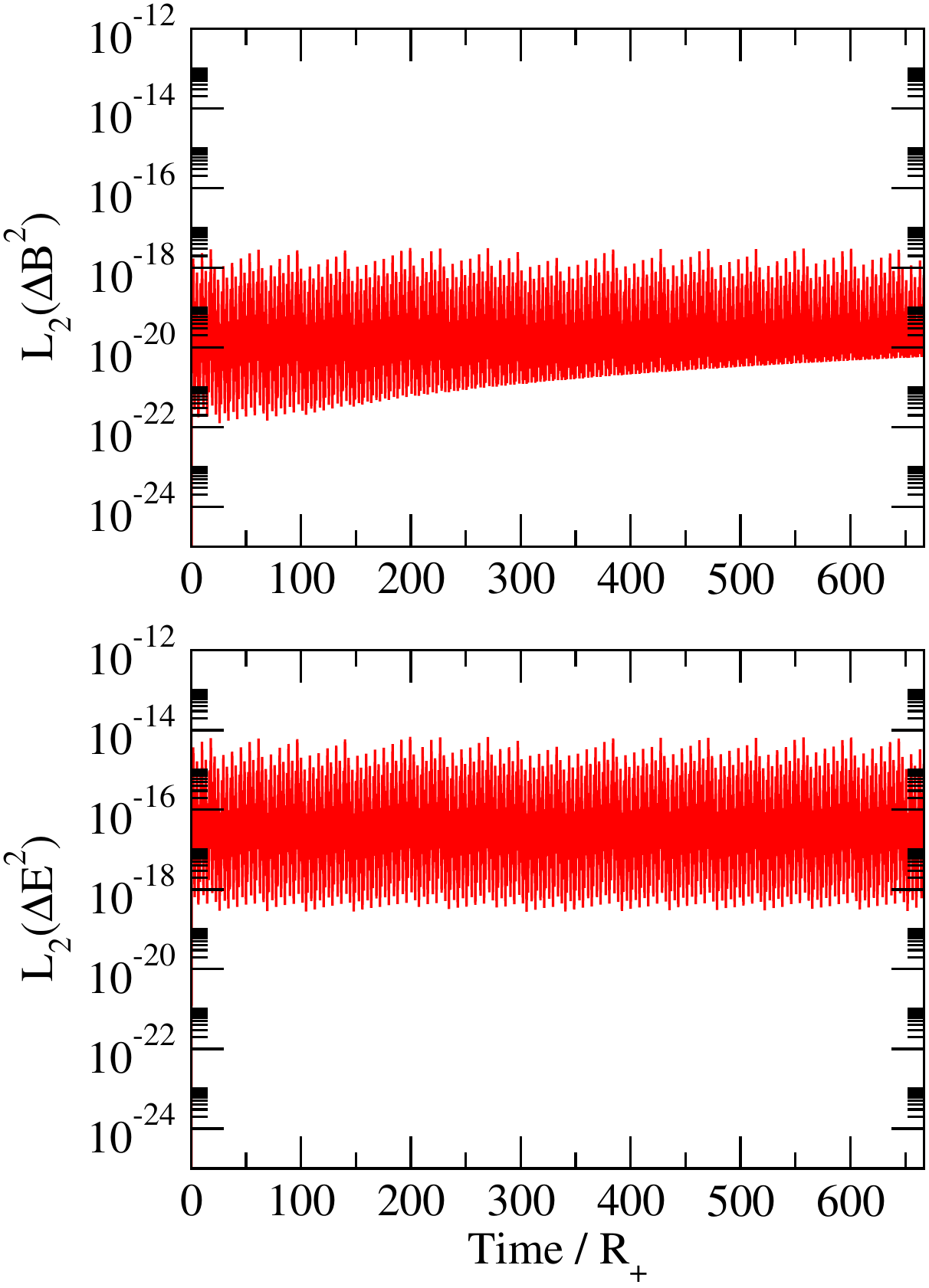}
\end{overpic}
\caption{Top: The $L_2$ norm of the error measure $\Delta {B}^2$ for the constant ${\bf B}$ field simulation. 
Bottom: The $L_2$ norm of the error measure $\Delta {E}^2$. 
The simulation time is shown in units of the light-crossing time $R_+$ from the origin to the outer boundary $R_+$. There is only one curve being plotted in each panel, which oscillates quickly giving the appearance of a band. 
}
\label{fig:ErrorPlotConstB}
\end{figure}
A particularly simple solution to the FFE equations is a constant ${\bf B}$ field with vanishing ${\bf E}$ field in a Minkowski background spacetime. We choose the spatial slicings such that ${\bf K}=0$, and adopt a Cartesian coordinate system in which $N=1$, ${\bm \beta} =0$, and ${\bm g}$ is the unit matrix. This is a physically trivial yet numerically interesting solution. 
Here and for the rest of the paper, the computational domain is a spherical shell, which is broken down into ``subdomains'' of concentric spherical shells (four shells extending from a radius of $R_-=1.9$ to $R_+=15$ code units for this test; the speed of light is unity in the code units). Therefore, the constant ${\bf B}$ field subtends all possible angles with the normal ${\bf \hat{n}}$ of the subdomain boundaries, including the special case of $({\bf \hat{n}} \cdot {\bf B})^2 = ({\bf \hat{n}}\times {\bf E})^2 = 0$, when the evolution system prescribed by Eqs.~\eqref{eq:EvoE} and \eqref{eq:EvoB} becomes ill posed. 

For initial conditions, we use an analytical constant ${\bf B}$ field of unit strength without any added noise or perturbation, which is also imposed as a Dirichlet boundary condition on the incoming (into the computational domain) characteristic modes during the evolution, on the external boundaries of the computational domain (at $R_-=1.9$ and at $R_+=15$). 
The interior of the computational domain is evolved with Eqs.~\eqref{eq:EvoE} and \eqref{eq:EvoB}, and the differences between the numerical and analytical values of ${\bf B}$ and ${\bf E}$ provide a precise measurement of the simulation error. Letting $\Delta {\bf B}$ and $\Delta {\bf E}$ be the differences between the numerical and analytical values of ${\bf B}$ and ${\bf E}$, we compute the $L_2$ norms of $\Delta {\bf B} \cdot \Delta {\bf B}$ and $\Delta {\bf E} \cdot \Delta {\bf E}$:
\bea
L_2(s) = \sqrt{\frac{\int_{\Sigma} |s|^2 dV}{\int_{\Sigma}  dV} }\,, \quad dV = \sqrt{|{\rm det}({\bf h})|} d x^3\,,
\eea
where ${\bf h}$ is the spatial metric and ${\Sigma}$ is the computational domain. The values of $L_2(\Delta {B}^2)$ and $L_2(\Delta {E}^2)$
over the entire computational domain are plotted in Fig.~\ref{fig:ErrorPlotConstB}. From this figure, we see that despite not being strictly strongly hyperbolic, the evolution system can be evolved stably for a long period of time. 

The constant ${\bf B}$ field test is, however, not suitable for examining the convergence behavior of our pseudospectral code with increasing resolution, because the spatial derivatives of the evolution variables vanish identically in approximations to the spatial derivatives at any order.   
For this task, we turn to the nontrivial analytical solutions given in Ref.~\cite{Brennan:2013jla}. Once again, we utilize analytical solutions because they provide precise references for comparison, allowing for more rigorous convergence tests. 

\subsubsection{Analytical null$^+$ solutions \label{sec:AnaNullTest}}
For a nontrivial analytical solution that's more suitable for testing convergence, we turn to the time-dependent fully three-dimensional wave given at the end of Sec.~\ref{sec:IntroSol}, whose structure is plotted in Fig.~\ref{fig:FieldLines}. This null$^+$ wave solution has a large amplitude and is fully nonlinear. 

\begin{figure}[t,b]
\begin{overpic}[width=0.99\columnwidth]{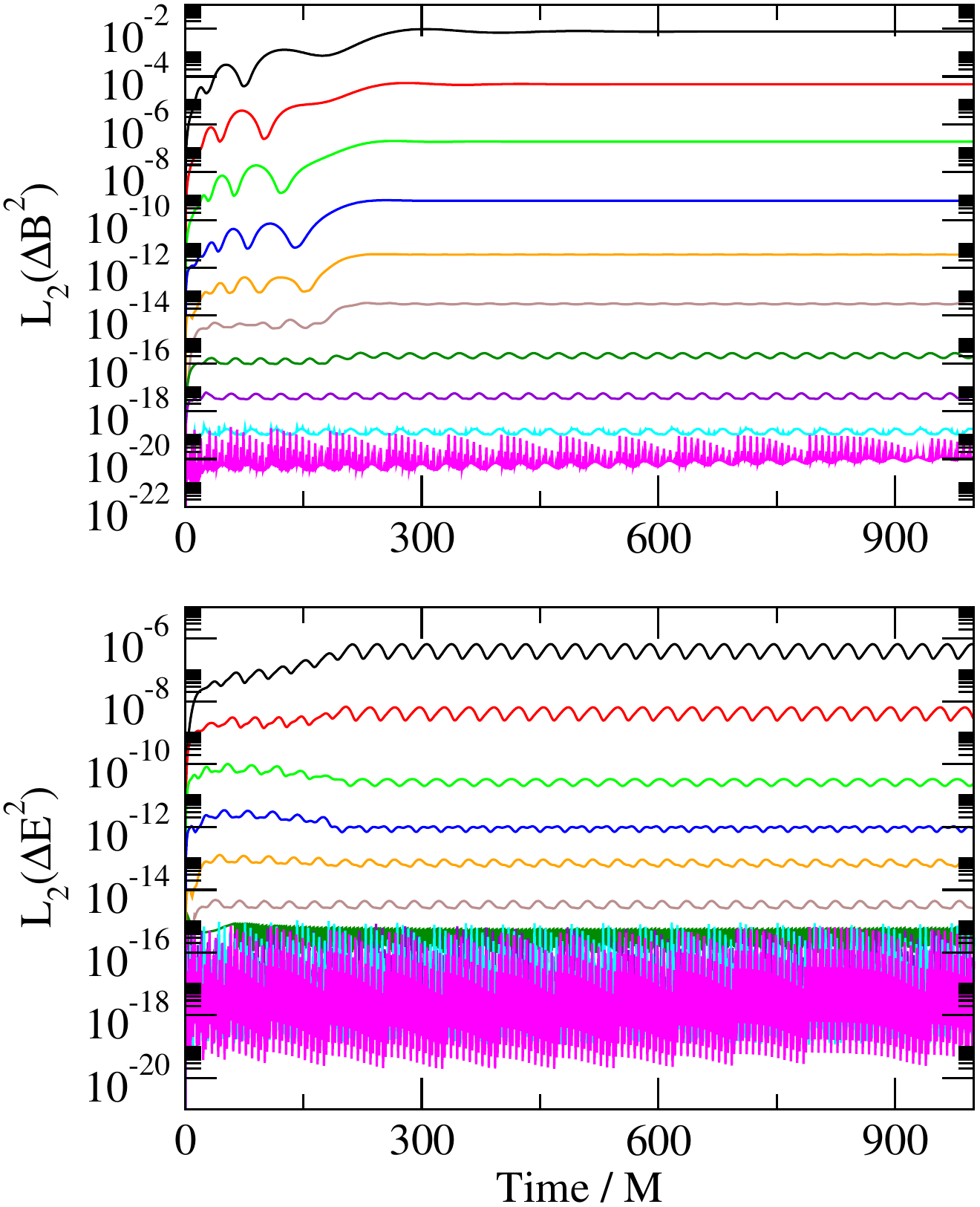}
\end{overpic}
\caption{Top: The $L_2$ norm of the error measure $\Delta {B}^2$ for the null$^+$ simulations. 
Bottom: The $L_2$ norm of the error measure $\Delta {E}^2$. 
There are ten resolutions being plotted, with the number of radial collocation points given by $k+6$, and the maximum $l$ for the angular $Y^{lm}$ decomposition given by $2k+7$. 
In both panels, the error decays with increasing $k$, so that $k=0$ corresponds to the topmost (black) line, and the lower lines are, in turn, $k=1,2,3,\cdots$.
The errors decline approximately linearly in log scale, indicating that our pseudospectral code achieves exponential convergence as expected over a wide range of resolutions, with the highest resolution being limited in accuracy by machine precision.
}
\label{fig:Convergence}
\end{figure}

In Fig.~\ref{fig:Convergence}, we plot the $L_2$ norms of the errors in ${\bf B}$ and ${\bf E}$ for our simulations of the null$^+$ solution. Due to constraints on computational resources, we evolve the simulations to $1000$M, which is around $16$ cycles for the time-dependent $F(\nu)$. We carry out evolutions at ten different resolutions where the number of radial collocation points is given by $k+6$, $k\in\{0,\cdots,9\}$. Note our maximum $l_{\text{max}}$ for the higher resolutions becomes excessive and the improvements in accuracy come mainly from the increase in radial resolution, but we keep the same $l_{\text{max}}$ vs. $k$ relationship to ensure consistency across all resolutions. 
Recall from Sec.~\ref{sec:CodeInfra}, the number of collocation points in each spatial dimension scales linearly with $k$. On the other hand, Fig.~\ref{fig:Convergence} shows that the errors decline approximately exponentially with $k$, so our FFE implementation achieves exponential convergence for this time-dependent, nontrivial, but smooth (without current sheets) test case, as expected from its pseudospectral nature.  
 
\subsubsection{Quasinormal modes of the black hole magnetosphere}
Next, we present a test case with richer dynamics. The test is designed to demonstrate the code's ability to correctly treat various types of waves allowed by the force-free equations, and in particular account for their interactions with the spacetime curvature to a high accuracy. As mentioned in the Introduction section, the null$^+$ solutions we examine are envisaged to play the role of carrying energy across magnetospheres of black holes or neutron stars, therefore a more direct demonstration of the relativistic effects is beneficial. 

Specifically, we launch a train of ingoing FFE waves towards the black hole, and examine the magnetosphere quasinormal modes (QNMs) being excited. Similar modes are intensely studied for gravitational waves, and constitute the main postmerger signal after a coalescence of two black holes. More recently, they have been shown to also be present in a black hole magnetosphere, where the so-called trapped FFE modes (in the short-wavelength/eikonal limit, they become the fast magnetosonic waves, so they will be referred to as generalized magnetosonic waves below) can be excited and trapped by the gravitational potential well of the black hole, and leak out over time with an exponentially decaying amplitude. The frequency and decay rates of these modes 
are completely determined by the spacetime geometry and the nature of the wave, and is independent of how they are excited. In particular, this means whatever the frequency of the wave train we send in, we should observe the same frequency and decay rate for the QNM it excites. 

When we launch a wave train consisting of in general the long wavelength generalizations of both Alfv\'en and magnetosonic waves towards the black hole, part of the wave is swallowed by the black hole, while the rest scatters around it and travels back out. During the process, some generalized magnetosonic waves are excited as QNMs, which do not have a counterpart consisting of generalized Alfv\'en waves. Our code needs to be able to correctly distinguish the propagation behaviour of both types of waves, especially their different interactions with spacetime curvature, in order to reproduce the frequency and decay rates predicted by analytical calculations. 

\begin{figure}[t,b]
\begin{overpic}[width=0.49\columnwidth]{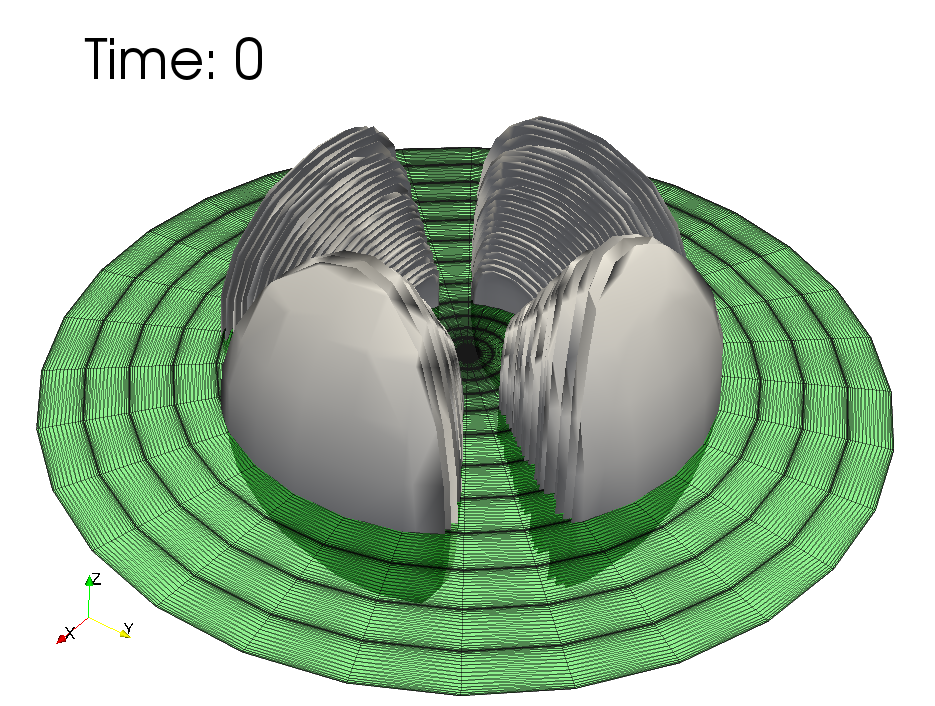}
\end{overpic}
\begin{overpic}[width=0.49\columnwidth]{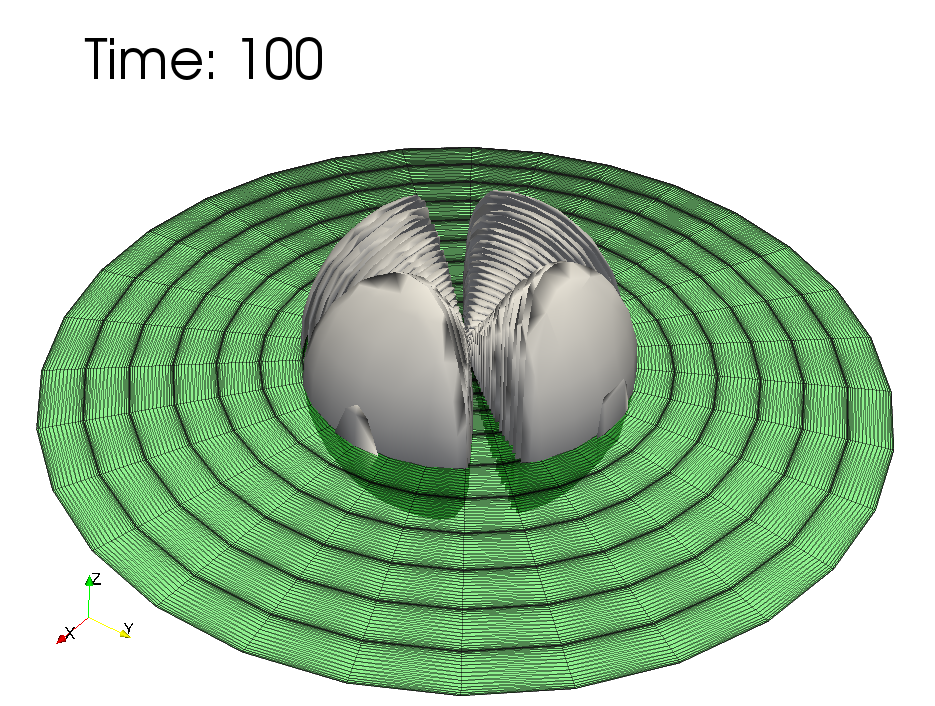}
\end{overpic}
\begin{overpic}[width=0.75\columnwidth]{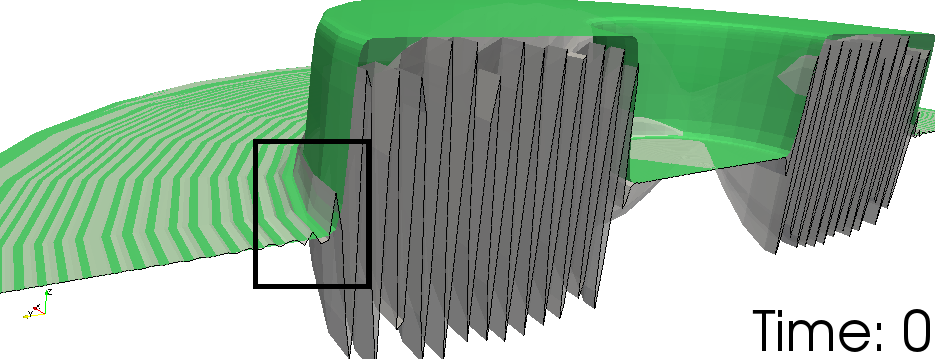}
\end{overpic}
\caption{Top row: Contours of $\Re\phi_0$ that show the structure of the wave train. It has a $Y^{2,2}$ angular profile, and consists of fast radial oscillations with frequency $\Omega$. It is confined into a radial top-hat-style envelope of width $200$M. The shape of the wave train shows that our test is fully three dimensional, without any axisymmetry. The figures on the left and right depict the wave train at earlier and later times, showing that it is initially moving inwards toward the black hole. The green semitransparent surface is the equatorial slice of the computational domain, included to indicate its extent. The grid structure on this slice is also plotted to provide a visual description of the density of collocation points at $K=7$. 
Bottom figure: The $\Re\phi_0$ is depicted by warping half of the equatorial slice with a vertical displacement proportional to the $\Re\phi_0$ values at collocation points, and then connecting them via ``straight lines'' into the gray surface (i.e. the gray surface is a trivial interpolation between the collocation points and is \emph{not} the actual $\Re\phi_0$ as represented by the spectral functions). The green surface is the tophat-like radial envelope function. This figure also corresponds to $K=7$.
}
\label{fig:WaveTrain}
\end{figure}

The wave train is constructed by first building a $\phi_0$ (the NP scalar representing ingoing waves) that has a $Y^{22}$ spherical harmonic angular profile and a sinusoidal radial profile $A\sin(\Omega (r-r_0))$ with $r_0=200$M, enclosed in a top-hat-like radial envelope extending from $100$M to $300$M.  
We also add a $q=1000$ magnetic monopole which enters through the Coulomb background $\phi_1$ as in Eq.~\eqref{eq:MagMonopole}. Together, these scalars allow us the build the Faraday tensor via Eq.~\eqref{eq:FieldTensorWithMono} and then ${\bf E}$ and ${\bf B}$ via Eq.~\eqref{eq:EBFromF} (Fig.~\ref{fig:WaveTrain} provides a visual depiction of the wave train and its time evolution). In order to accommodate the wave train, we use a slightly different domain structure from that of Sec.~\ref{sec:IntroSol} and the rest of the paper (including the stability studies later). Namely we have 16 concentric spherical shells extending from $1.05$M to $495$M, with the number of radial collocation points given by $5+2K$ and those for the $\theta$ and $\phi$ directions by $7+K$ and $14+2K$ (we have used capital $K$ to distinguish it from our usual resolution label of $k$ used in Sec.~\ref{sec:AnaNullTest} and later stability test sections). 

The bottom panel in Fig.~\ref{fig:WaveTrain} shows both the power and the limitation of the spectral methods. The gray surface in that figure is constructed by linearly interpolating between the collocation points (done by the visualization software, and is \emph{not} the actual spectral representation of $\Re\phi_0$), which is not smooth at all because there are very few collocation points inside each radial oscillation period. The actual spectral ``interpolation'' using our Chebyshev polynomials on the other hand, gives very smooth functions (see Figs.~\ref{fig:QNMDiffFreq}, \ref{fig:QNMDiffFreqFit} and \ref{fig:QNMConvergence} below). This shows that the spectral methods are capable of providing accurate representations of functions with very few collocation grid points, in other word their effective resolution, given a fixed number of grid points, is very high. On the other hand, the black square in that figure highlights the region close to the rather sharp corner of the top-hat-like envelope of the form 
\bea
\frac{1}{\left(1+e^{-100 \frac{r-r_-}{r_+-r_-}}\right)\left(1+e^{-100 \frac{r_+-r}{r_+-r_-}}\right)}\,,
\eea 
with $r_+$ and $r_-$ being the outer and inner boundaries of the envelope respectively. The spectral methods behave rather differently from finite difference methods, where the sharp kinks are simply under-resolved and rounded due to numerical dissipation. With the present low radial resolution, the steep envelope cannot be resolved, and leads to Gibbs oscillations bleeding into the surrounding region. While the shape of the waveform is thus not perfectly represented, we note that nevertheless the shape is propagated without diffusive broadening \cite{Boyle2006} (also see this reference for an additional suite of tests on the numerical behaviour of the pseudospectral infrastructure underlying our force-free code), as typically caused by numerical viscosity of finite-difference codes. 

\begin{figure}[t,b]
\begin{overpic}[width=0.99\columnwidth]{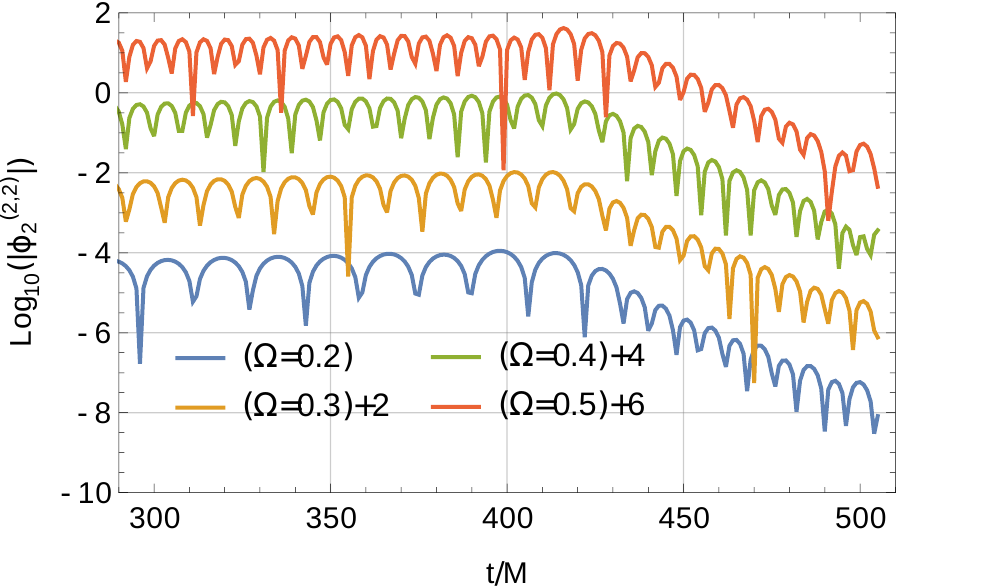}
\end{overpic}
\caption{The log of $\phi_2^{(2,2)}$ plotted as a time series. The curves corresponding to different wave train frequencies $\Omega$ are shifted slightly relative to each other vertically, so that the curves won't lie right on top of each other, and we can see more clearly. The flat region for the curves correspond to the wave train itself, moving back outwards after having scattered around the black hole. The sloped region corresponds to the exponentially decaying QNM. We can see that the wave train has very different frequencies, but the QNMs excited in all cases share the same frequency and decay rate. 
}
\label{fig:QNMDiffFreq}
\end{figure}

To examine the QNMs, we need to extract the outgoing wave component $\phi_2$ of the Faraday tensor (QNMs escaping from the trapping gravitational potential will show up as an exponential tail to the wave train in $\phi_2$). We do so by extracting the $\phi_2$ values via Eq.~\eqref{eq:phi2FromF} on a sphere located at $100$M from the coordinate origin, and integrate this $\phi_2(\theta, \phi)$ distribution against the $Y^{22}$ spherical harmonic to get a single scalar value representing the $(2,2)$ harmonic component of $\phi_2$. We do so at each time step and obtain a time series for this harmonic coefficient of the outgoing wave, which we denote $\phi_2^{(2,2)}(t)$. We note that in order to make comparison with perturbation theory that predicts the QNM frequencies, we use a small wave amplitude of $A=0.0001$. There is however still nonlinear interactions between the wave and the background magnetic monopole.  

\begin{table}[t,b]
\begin{tabular}{c|c|c|c|c|c}
& Analytical Ref. & $\Omega =0.2$ & $\Omega =0.3$ & $\Omega =0.4$ & $\Omega =0.5$ \\ \hline\hline
$\omega$ & 0.457596 & 0.444071 & 0.453783 & 0.453742 & 0.44466 \\ \hline
$\gamma$ & 0.0950044 & 0.0922016 & 0.0885479 & 0.0893736 & 0.0901566 \\ \hline
\end{tabular}
\caption{The fitted values of the frequency $\omega$ and decay rate $\gamma$ of the QNMs excited by wave trains of different $\Omega$, as well as the reference analytically computed values.}
\label{tb:Freqs}
\end{table} 

\begin{figure}[t,b]
\begin{overpic}[width=0.99\columnwidth]{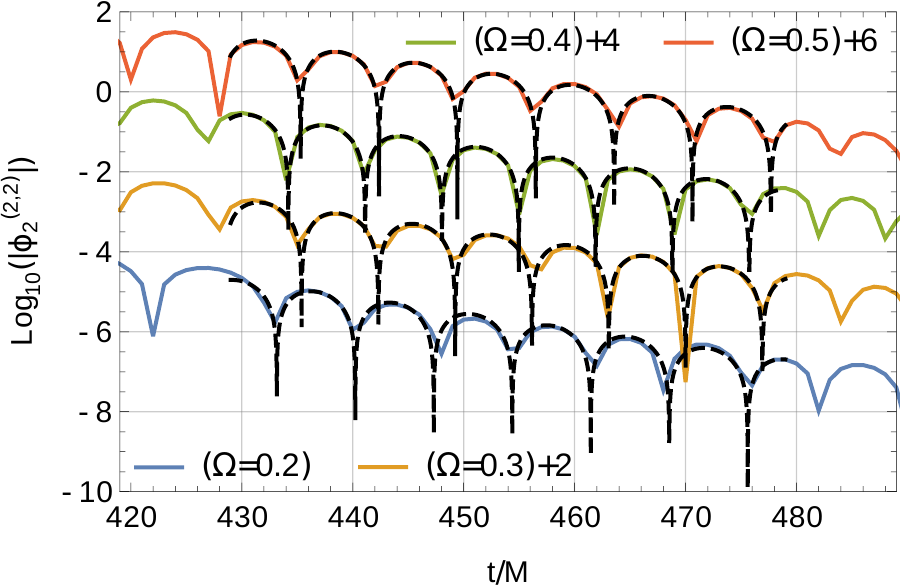}
\end{overpic}
\caption{The dashed black lines indicate the fitting results (they extend only over the fitting interval) corresponding to the fitted parameters in Table \ref{tb:Freqs}, while the curves being fitted are the negatively sloped segments of the curves in Fig.~\ref{fig:QNMDiffFreq}. 
}
\label{fig:QNMDiffFreqFit}
\end{figure}

We first test whether the correct QNM wave is produced in our code, specifically, whether it has the correct frequency and decay rate as predicted by analytical models. To this end, recall that these values are independent of the specifics of the initial wave train, so a stringent test consists of launching initial wave trains with different frequencies $\Omega$, and see if we get the same set of QNM parameters despite this major difference. If we do obtain the same parameters as expected, we can then be confident that the code has managed to cleanly excite and extract the QNMs, and the decaying tail is not some other leftover feature from the wave train (such as the oscillations seen in the black square of Fig.~\ref{fig:WaveTrain}, which is in reality much smaller in amplitude and drops off faster than the QNMs). We simulate four cases with $\Omega =0.2\,,0.3\,,0.4$, and $0.5$ at $K=7$, and plot their $\phi^{(2,2)}_2$ in a log plot (so exponential decay shows up as a straight line with a negative slope equaling the decay rate) in Fig.~\ref{fig:QNMDiffFreq}. By visual inspection, we can already see that despite the rather different frequencies of the wave train, the frequencies $\omega$ and decay rates $\gamma$ of the exponential tails in the four cases are indeed the same. We can also carry out a more quantitative fitting of these quantities using \verb!Mathematica!'s FindFit function. The results and the analytical reference values are shown in Table.~\ref{tb:Freqs}. 
(We also plot in Fig.~\ref{fig:QNMDiffFreqFit} the fitting results (dashed black lines) inside the fitting interval to provide a visual depiction of the fitting quality. )
We indeed find a good match between the measured and predicted values. Aside from truncation error, the excitation of QNMs of other overtone numbers $n>1$ (we have only considered the fundamental overtone $n=1$ that decays the slowest) is likely a main source of any residual errors. 

\begin{figure}[t,b]
\begin{overpic}[width=0.99\columnwidth]{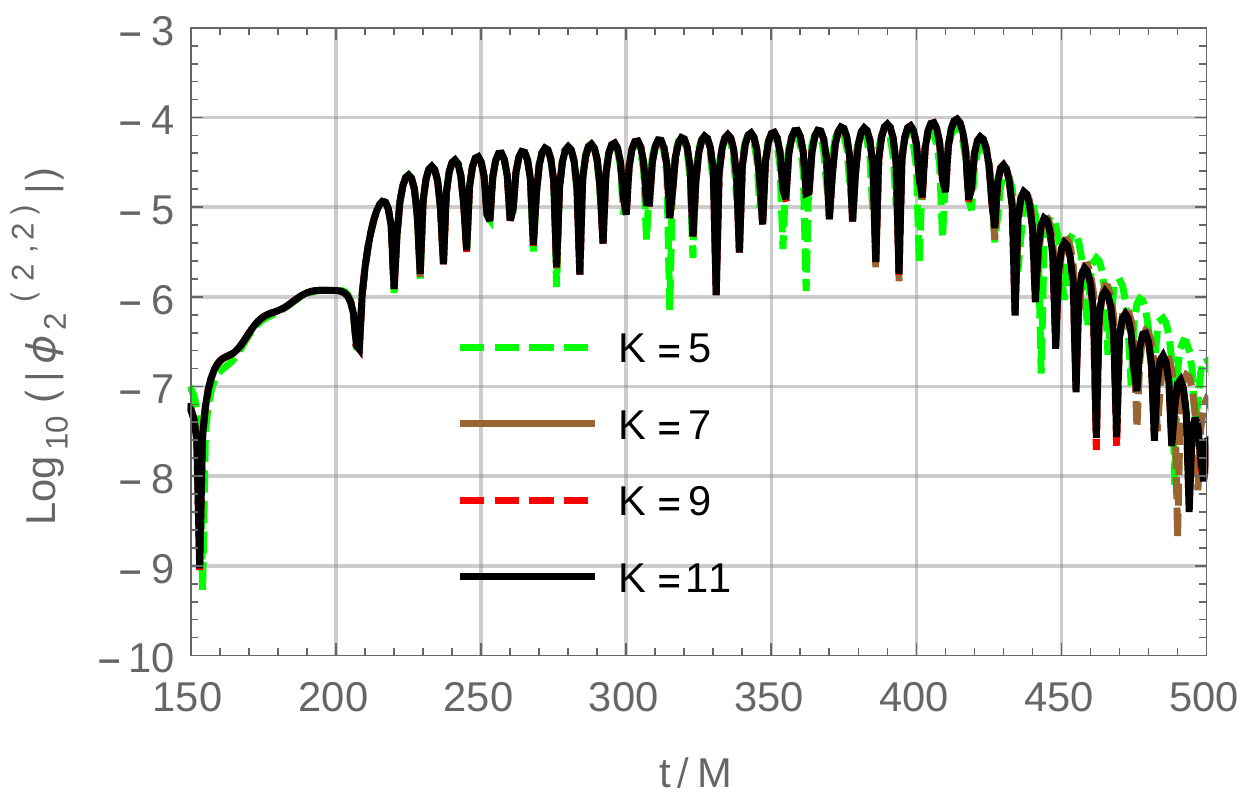}
\end{overpic}
\begin{overpic}[width=0.99\columnwidth]{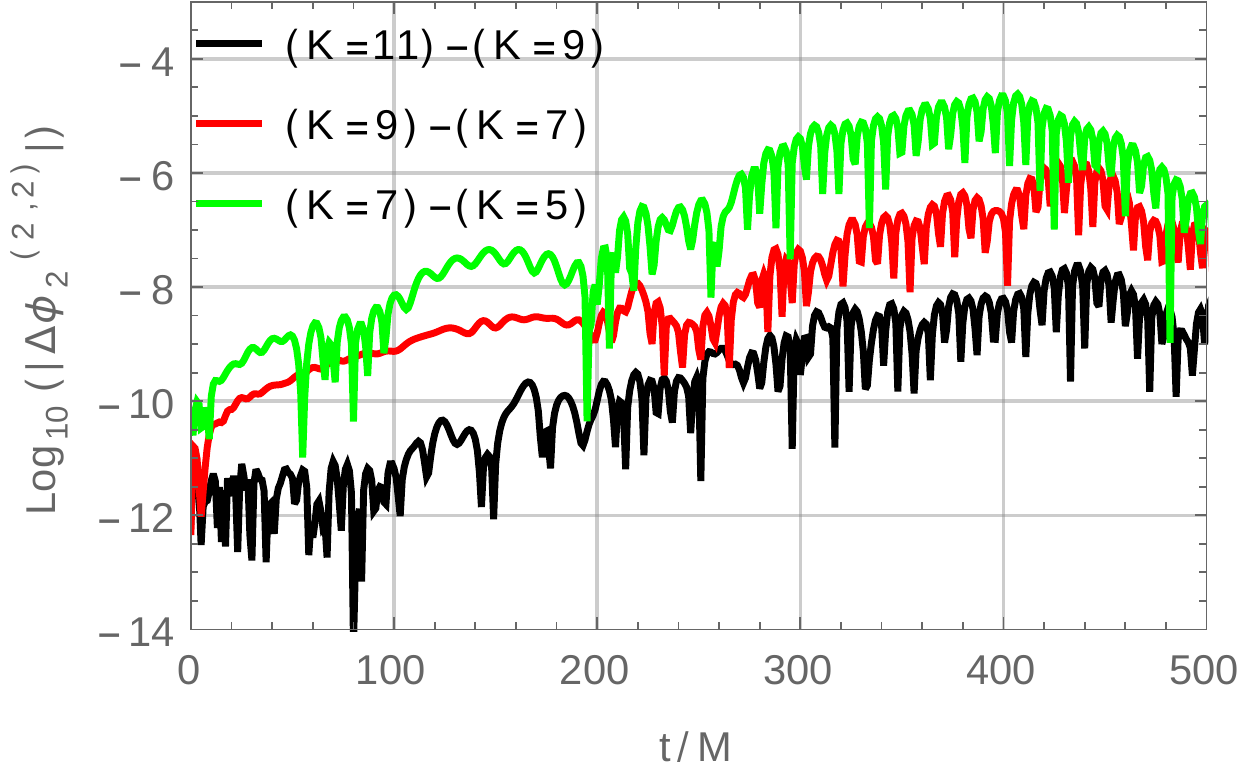}
\end{overpic}
\caption{The top figure shows the wave train and the QNM tail for the four resolutions of the $\Omega=0.4$ simulation. The bottom figure shows the differences between resolutions. Because the differences are approximately equally displaced in the vertical direction, which is plotted in log scale, the simulation indeed achieves exponential convergence. 
}
\label{fig:QNMConvergence}
\end{figure}

The second test we carry out is a convergence test, to verify that our code still achieves exponential convergence with this dynamically richer setup. To this end, we fix $\Omega=0.4$ and perform four simulations at $K=5$, $K=7$, $K=9$ and $K=11$. The log plots of $\phi_2^{(2,2)}$ for these runs and their differences are shown in Fig.~\ref{fig:QNMConvergence}, which shows that exponential convergence is indeed achieved.

\section{Stability of the null$^+$ solutions \label{sec:NumSim}}
In this section, we utilize our new FFE code and carry out numerical simulations in order to determine whether the null$^+$ solutions are stable, in the sense of whether a perturbed solution will asymptote to an exact null$^+$ solution over time, or diverge from it. 
As an underlying unperturbed null$^+$ solution, we use the same configuration shown in Sec.~\ref{sec:IntroSol} and studied in Sec.~\ref{sec:AnaNullTest}, with $q=1000$, $F(\nu) = A \cos(\Omega \nu)$ and $A=1$. However, here, we will vary $\Omega$. Note that we have chosen a large absolute magnitude for the Faraday tensor in order to carry out a more stringent test, as small magnitudes will diminish the significance of the nonlinear terms. 
We also use the same domain structure as in Secs.~\ref{sec:IntroSol} and \ref{sec:AnaNullTest}, namely eight spherical shells extending from $1.95$M to $195$M. 

We denote the Faraday tensor for the unperturbed solution as ${\bf F}$ and impose it as a Dirichlet boundary conditions on the incoming characteristic modes as discussed in Sec.~\ref{sec:CodeInfra}. In other words, we enforce the condition that there are no incoming perturbative modes. On the other hand, there is no restriction on the outgoing characteristic modes, so overall, we have a purely outgoing boundary condition for the perturbations, just as when one studies the mode stability of the Kerr metric by calculating its QNMs, for example. 

\subsection{Perturbation in $\phi_0$ \label{sec:Pertphi}}
\begin{figure}[!b]
  \centering
  \begin{overpic}[width=0.99\columnwidth]{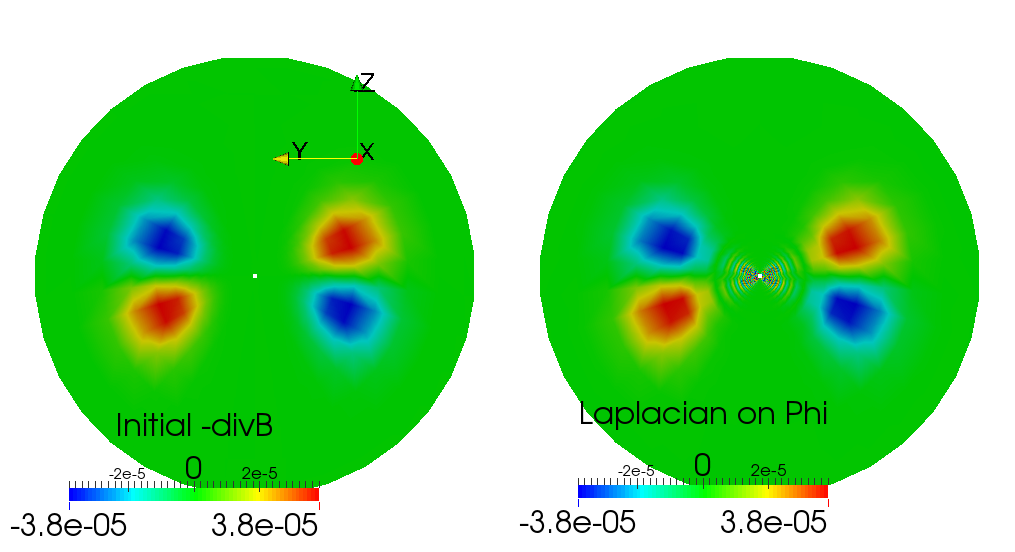}
  \end{overpic}
  \caption{Left: The right-hand side of Eq.~\eqref{eq:Poisson}, or explicitly, the initial $-\nabla \cdot {\bf B}$ value on a vertical slice of the computational domain before we apply our initial data solver (the $z$-axis corresponds to $\theta=0$ and $\theta=\pi$). Right: The left-hand side of Eq.~\eqref{eq:Poisson}, which is the value of $\nabla^2 \Phi$ that emerges after solving for the initial data. 
}
	\label{fig:NoScatteringZeroBCIDSolve}
\end{figure}

The simplest perturbation to the null$^+$ solution is a variation in the initial $\phi_0$ profile, so that it no longer satisfies Eq.~\eqref{eq:phi0explicit}. Specifically, we generate a Faraday field ${\bf F}^{A}$ from an altered $\phi^A_0$ (in addition to an unperturbed monopole component) that is perturbed away from the exact solution (Eq.~\ref{eq:phi0explicit}) by 
\begin{equation} 
\delta\phi_0 = 0.25 \frac{\sin^2\theta}{\Delta\rho} F(\nu) i \sin\psi\,,
\end{equation}
where we choose the same $F(\nu)$ function as the unperturbed solution. 
For simplicity, we also choose $\Omega = 0$ so that the unperturbed background solution is time independent. Such a solution does not have 
the usual character that one would associate with a travelling wave, although it is still technically a wave in the same sense that a constant would satisfy a simple 1-D wave equation.  Nonetheless, this solution is still physically interesting, as the energy flux does not vanish, so that the solution describes an ``electromagnetic wind'' or ``Poynting wind'' \cite{Gralla:2014yja}. We will examine a time-dependent case later in Sec.~\ref{sec:TimeDepPert}. 

\begin{figure}[t,b]
\begin{overpic}[width=0.99\columnwidth]{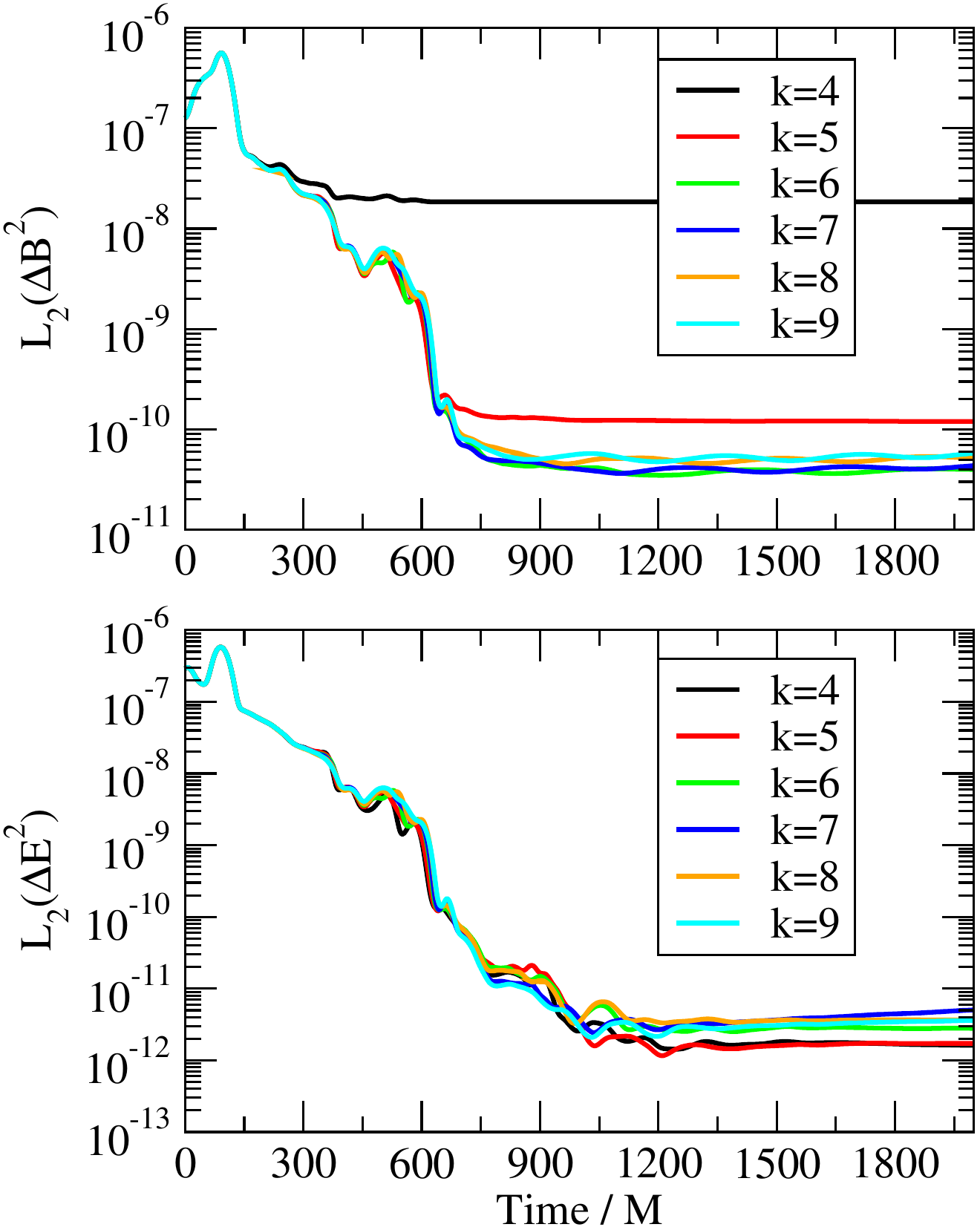}
\end{overpic}
\caption{
Top: The $L_2$ norm of the difference measure $\Delta {B}^2$ for the null$^+$ simulations initially perturbed in the $\phi_0$ values. 
Bottom: The $L_2$ norm of the difference measure $\Delta {E}^2$. The arrangement of the collocation points is the same as for Fig.~\ref{fig:Convergence}. 
}
\label{fig:PertErrNoScattering}
\end{figure}

Given the altered field $\phi^A_0$, we now define our perturbed
  initial data.  Because we employ boundary conditions based on the unperturbed background solution, we require the perturbed initial data ${\bf F}^P$ to approach
the unperturbed solution at the boundaries (strictly speaking we don't need a boundary condition at the inner boundary as it is inside the event horizon, but we use the Dirichlet condition there for simplicity).  We achieve this by blending between ${\bf F}^{A}$ and ${\bf F}$  via
\bea
{\bf F}^P = f\; {\bf F}^{A} + (1-f)\, {\bf F}\,, 
\eea
where the weighting function is
\begin{equation} \label{eq:PertWeightinga}
f = P^8(2-P)^8Q^8(2-Q)^8\,,
\end{equation}
with
\begin{equation} \label{eq:PertWeightingb} 
P=2\frac{r-R_-}{R_+-R_-} \quad {\rm and} \quad Q=2 \frac{\theta}{\pi}\,.  
\end{equation}
Equation~(\ref{eq:PertWeightinga}) ensures $0\le f\le 1$ such that the unperturbed solution dominates on the domain boundaries, ensuring a smooth transition to the Dirichlet boundary conditions imposed there. The angular dependence in $A$ ensures that the unperturbed solution also dominates on the vertical axis ($\theta=0$ and $\theta=\pi$). This feature is not constructed to satisfy any requirements of the present form of perturbation, but instead is included for later convenience in Sec.~\ref{sec:PertPropDir}.

The perturbed solution does not automatically satisfy the constraints; since the monopole resides inside the event horizon and outside of our computational domain, the magnetic field should be divergence-free within the computational domain. We are therefore free to solve the Poisson equation in Sec.~\ref{sec:IDSolver} to remove any divergence. 
In Fig.~\ref{fig:NoScatteringZeroBCIDSolve}, we plot the right and left-hand sides of Eq.~\eqref{eq:Poisson} and show that the initial data solver performs as designed by removing $\nabla \cdot {\bf B}$. Note that there is some high-frequency noise in the output of this elliptic solver (see the center of Fig.~\ref{fig:NoScatteringZeroBCIDSolve} (b)), which we partially remove by passing the initial data through a spectral filter that reduces the high-frequency spectral coefficients in the radial direction. The removal is only partial as the filtering strength is chosen to be conservative so it does not introduce new divergence into the magnetic field. As a second step, we also use Eq.~\eqref{eq:PostSteppingCleaning1} to impose the FFE constraint ${\bf E}\cdot {\bf B} =0$. 

\begin{figure}[!b]
\begin{overpic}[width=0.49\columnwidth]{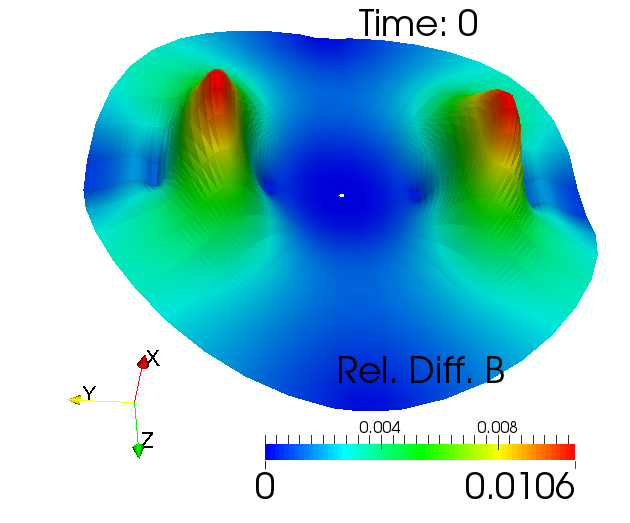}
\end{overpic}
\begin{overpic}[width=0.49\columnwidth]{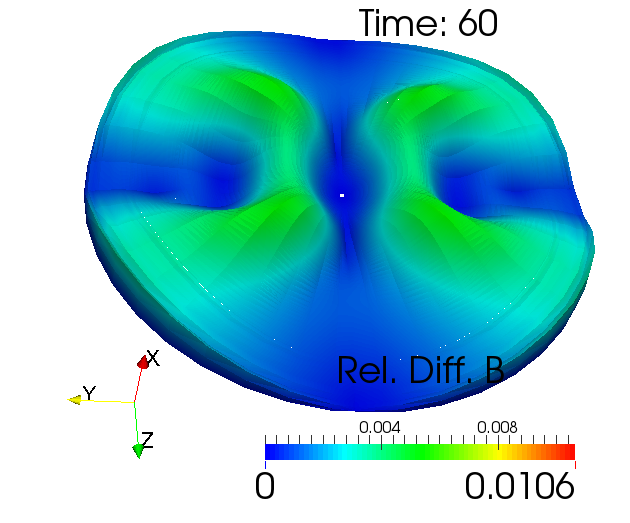}
\end{overpic}
\begin{overpic}[width=0.49\columnwidth]{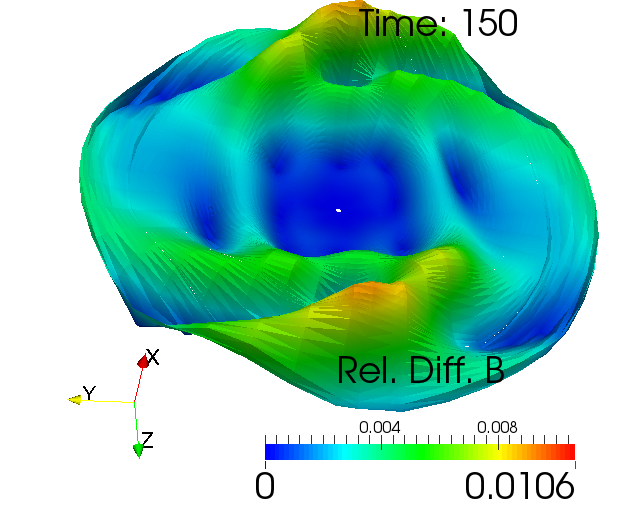}
\end{overpic}
\begin{overpic}[width=0.49\columnwidth]{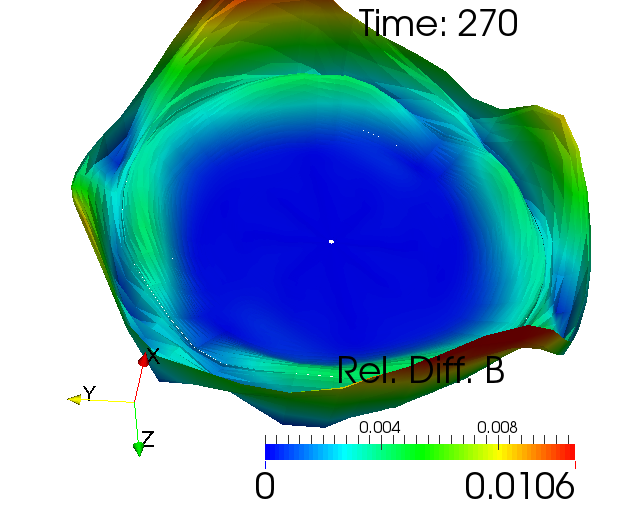}
\end{overpic}
\begin{overpic}[width=0.49\columnwidth]{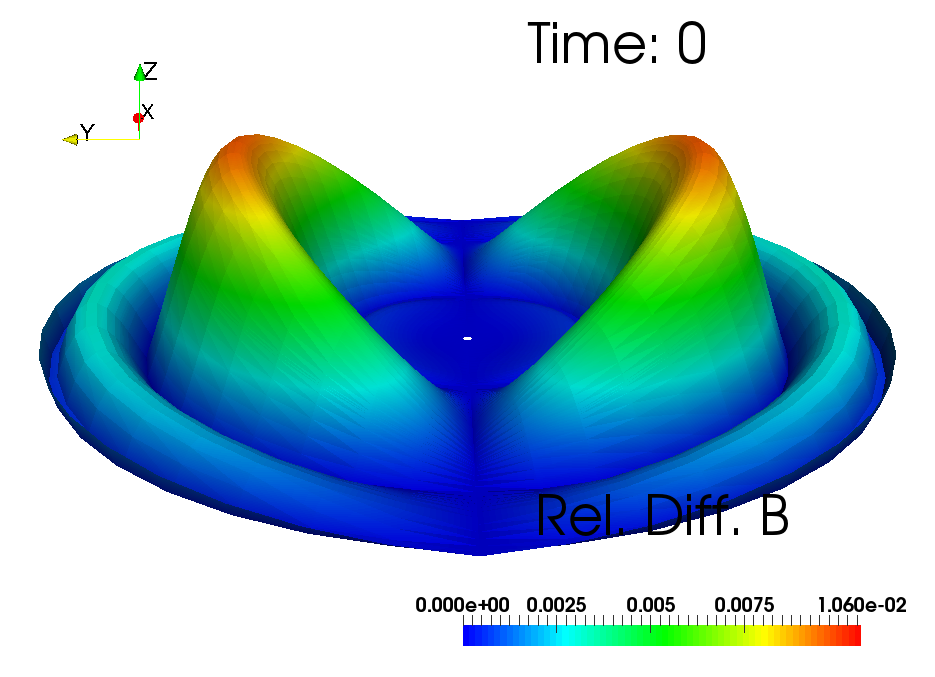}
\end{overpic}
\begin{overpic}[width=0.49\columnwidth]{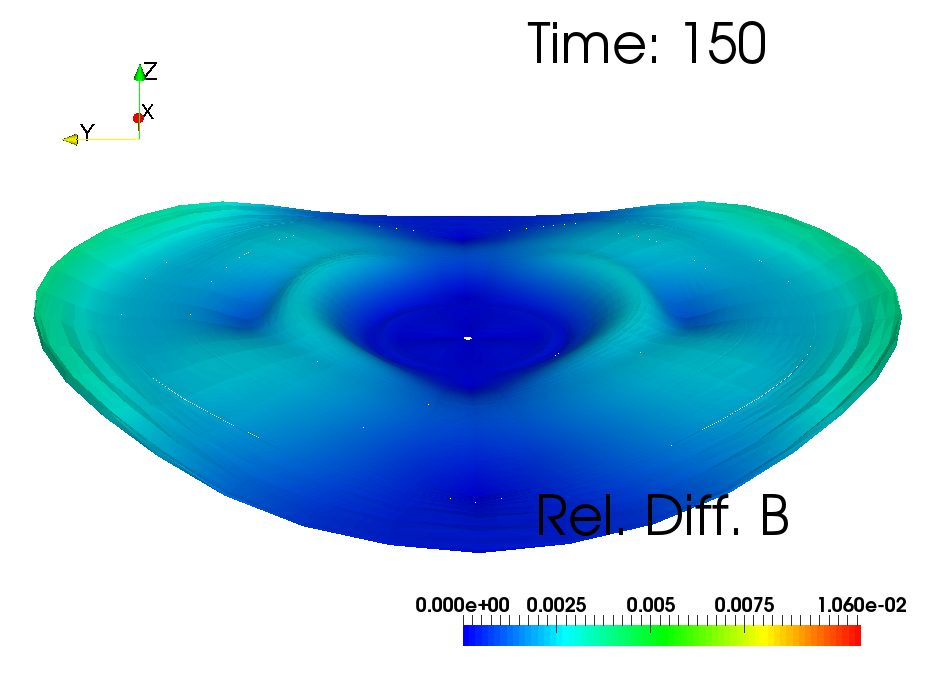}
\end{overpic}
\caption{Top four panels: The relative difference $|\Delta \bf{B}|/|\bf{B}|$ on a vertical slice of the computational domain. The magnitude of the relative difference is indicated by the height of the surface as well as the colouring. Different panels correspond to different times. The initial perturbation seen at $t=0$ propagates inwards creating the pattern shown at $t=60$. After scattering around the black hole, the perturbations propagate outwards, forming the pattern seen at $t=150$, and eventually begin to exit the computational domain as seen at $t=270$. 
The bottom two panels show the initial and later perturbations on the horizontal equatorial slice of the computational domain, they show that the perturbations are three dimensional in nature, without axisymmetry. 
}
\label{fig:PertEvoNoScattering}
\end{figure}

In Fig.~\ref{fig:PertErrNoScattering}, we plot the evolution of $L_2(\Delta B^2)$ and $L_2(\Delta E^2)$, where $\Delta \bf{B} = \bf{B}^P-\bf{B}$ is the difference between the evolved $\bf{B}^P$ and its unperturbed counterpart $\bf{B}$ (computed analytically). $\Delta \bf{E}$ is defined similarly. 
We observe that the differences drop by several orders of magnitudes as time progresses. A more detailed distribution of $\Delta \bf{B}$ is shown in Fig.~\ref{fig:PertEvoNoScattering}, which plots the relative difference $|\Delta \bf{B}|/|\bf{B}|$ on a vertical slice of the computational domain. The magnitude of the relative difference is indicated by the  height of the surface as well as the colouring. The panels correspond to different times and suggest that the perturbation does not diverge, but instead propagates inwards to begin with, and then outwards after scattering around the black hole, before exiting through the outer boundary. 

\begin{figure}[t,b]
\begin{overpic}[width=0.75\columnwidth]{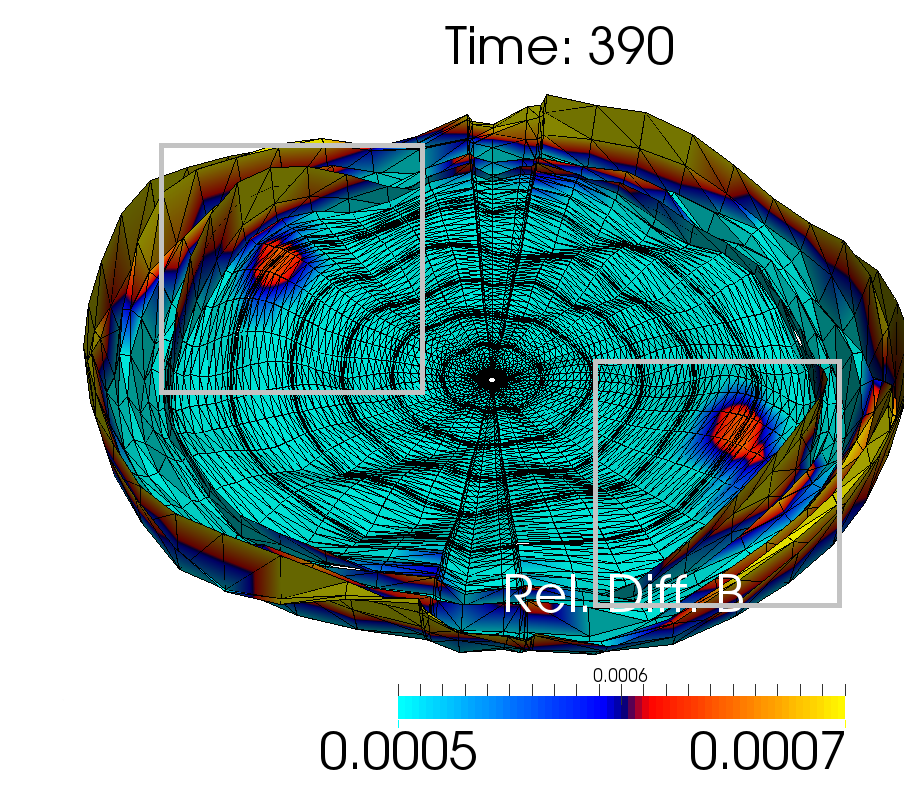}
\end{overpic}
\caption{The numerical error in $\bf{B}^P$ (as seen in $|\Delta \bf{B}|/|\bf{B}|$) being generated at subdomain boundaries (signified by dense concentrations of black lines). The gray frames highlight the locations of the error on such boundaries. Note that the warping scale and the colour map are different from Fig.~\ref{fig:PertEvoNoScattering}, but the plane shown is the same as those appearing in Fig.~\ref{fig:PertEvoNoScattering}. 
}
\label{fig:BError}
\end{figure}

Consequently, our observation indicates that there are no diverging modes excited by our initial perturbation at the fully nonlinear level, and so the null$^+$ solution is stable against our perturbation. Furthermore, the eventual exit of the perturbation is consistent with the null$^+$ solution being asymptotically stable (attracting). We note however that we cannot rigorously prove the latter stronger stability, as $L_2 (\Delta B^2)$ and 
$L_2 (\Delta E^2)$ values at late times do not reach their small sizes in the unperturbed convergence test case shown in Fig.~\ref{fig:Convergence}. The main culprit appears to be the high-frequency noise in the initial data we see in Fig.~\ref{fig:NoScatteringZeroBCIDSolve} (b), which is absent from the analytical initial data used for the convergence tests. The fact that the convergence behaviour improves after we apply spectral filtering to the initial data before starting the evolutions provides evidence for this conclusion. In particular, the \verb!SpEC! code utilizes a penalty method \cite{Hesthaven1997,Hesthaven1999,Hesthaven2000,
Gottlieb2001} to enforce consistency across internal boundaries, which allows for discontinuities to exist temporarily. It has been noted in several previous studies  \cite{Zhang:2012ky,Zhang:2013gda} that high-frequency noise tends to induce large discontinuities at the internal boundaries, thereby creating errors in $\bf{B}$. The situation is the same for our FFE evolution system, as can be seen in Fig.~\ref{fig:BError}. The high-frequency noise also destroys the convergence at the highest resolutions (see Fig.~\ref{fig:PertErrNoScattering}), because lower resolution (smaller $k$) acts as an effective spectral filter, shielding the lower resolution simulations from high-frequency noise. We expect this complication to disappear as we ascertain the source of the high-frequency noise (one possibility is the boundary condition of $\Phi =0$ being too simplistic) and further improve our procedure of solving Eq.~(\ref{eq:Poisson}).

\subsection{Perturbation in the propagation direction \label{sec:PertPropDir}}
With numerical experiments, we can only state that there are no diverging modes being excited by the particular perturbations that we introduce into the initial data. Therefore, in order to provide as strong evidence as possible for stability (meaning the nonexistence of diverging modes in general), it is important that we consider a set of initial perturbations that is as general as possible. In the last section, although we started with specific modifications to $\phi_0$, our initial data construction procedure nevertheless has to go through the blending and constraint solving stages, which means the resulting perturbation is in fact rather general (containing many modes).

\begin{figure}[t,b]
  \centering
  	\includegraphics[width=0.75\columnwidth]{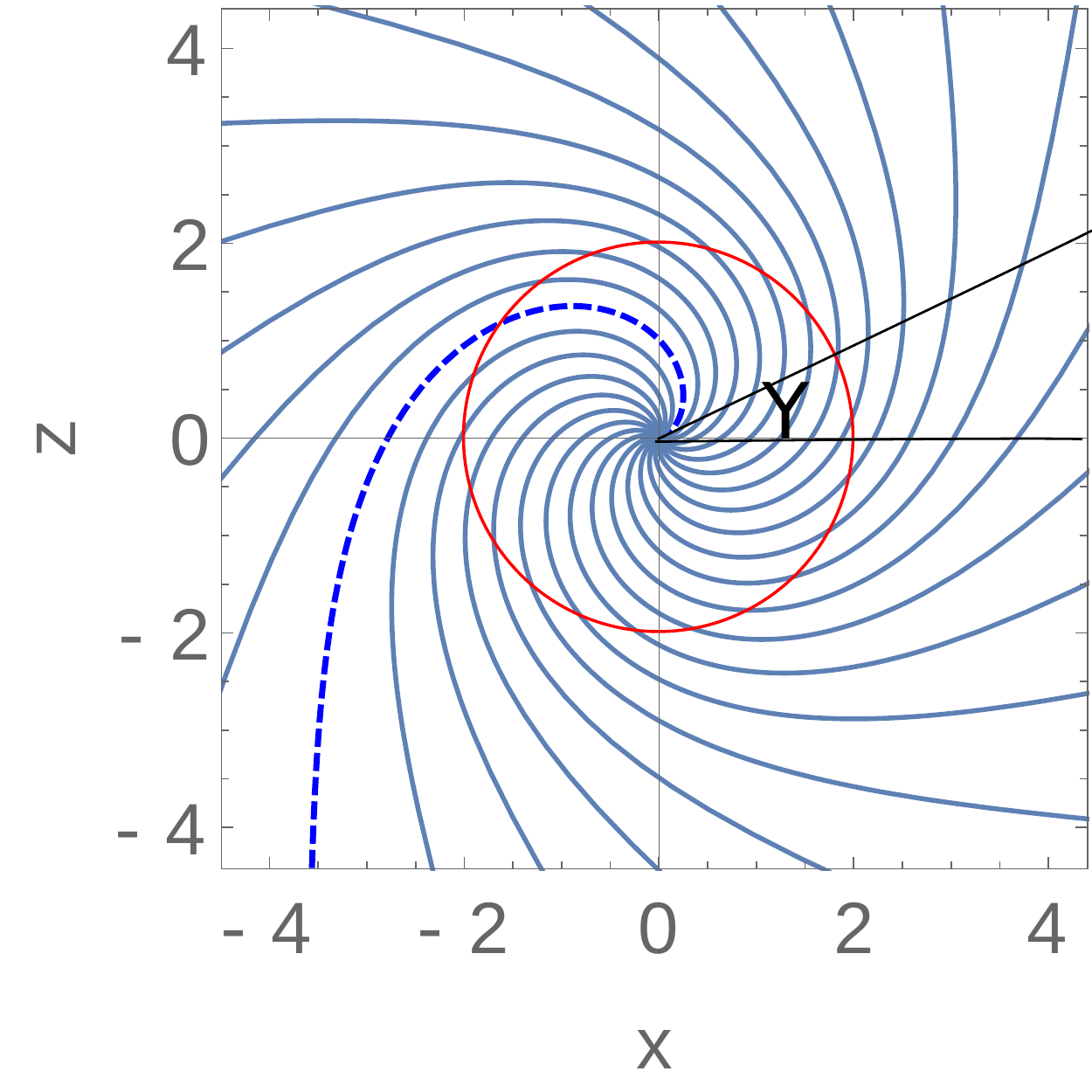}
  \caption{Example congruence of geodesics in the equatorial plane of the Schwarzschild black hole, with a nonvanishing impact parameter $b$ translating into $\mathcal{P}=2/27M^2$ (not the value we use for the simulations, but instead chosen to accentuate the perturbation). The red circle is the location of the event horizon. Angle $Y$ is the angle at which the dashed geodesic strikes the origin.}
	\label{fig:GeodesicInSchwarzschild}
\end{figure}

\begin{figure}[!b]
  \centering
  \begin{overpic}[width=0.85\columnwidth]{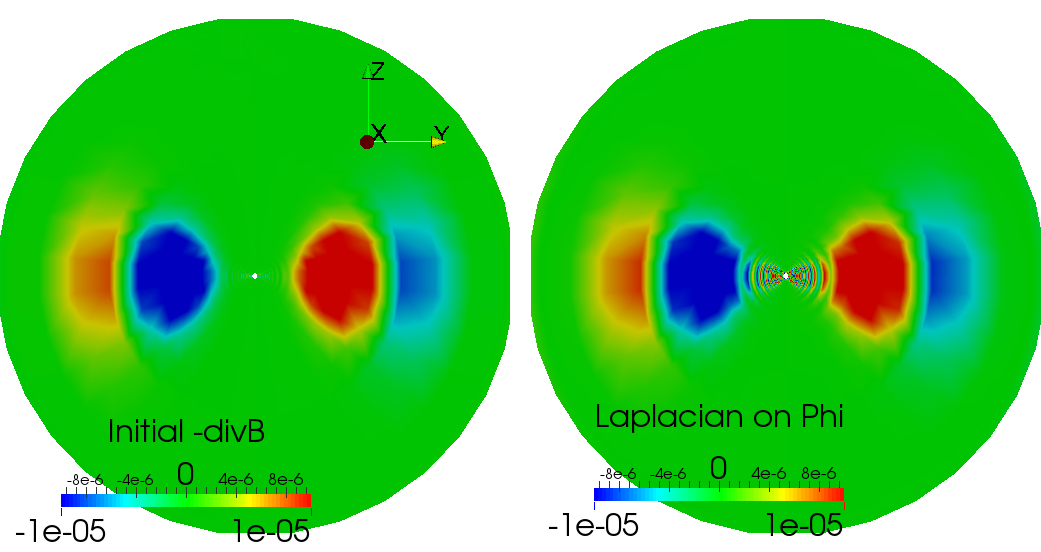}
  \end{overpic}
  \caption{
Similar to Fig.~\ref{fig:NoScatteringZeroBCIDSolve}, but for initial data with a perturbed propagation direction. 
 Left: The right-hand side of Eq.~\eqref{eq:Poisson}. Right: The left-hand side of Eq.~\eqref{eq:Poisson}.}
	\label{fig:IDSolve}
\end{figure}

However, the wave component (the $\phi_0$ piece of the Faraday tensor) of the simple perturbation studied in Sec.~\ref{sec:Pertphi} still follows the GPNDs initially as we have retained the use of the Kinnersley tetrad when constructing the perturbed Faraday tensor ($\phi_0$ is by definition the wave propagating along the tetrad's ingoing null direction, which for the Kinnersley tetrad is in a doubly degenerate GPND direction) 
 and is therefore (at least initially) not necessarily backscattered by the spacetime curvature (see Sec.~\ref{sec:GPNDs}). In other words, the perturbation may be restricted to a rather specific submanifold in the solution space. In principle, it is possible that even though null$^+$ solutions are stable on this submanifold, they may still be unstable under perturbations off of it. Although we would expect numerical truncation errors to induce perturbations away from this submanifold regardless of whether the initial perturbation places us on it or not, the time scale for this occurring may simply be too large so we fail to observe the possible instabilities in the previous test, and a more appropriate off-the-submanifold initial perturbation is desirable. In this subsection, we construct perturbations that are nontrivial in the sense that the wave component of the initial data does not follow degenerate GPNDs everywhere and is therefore generically backscattered by spacetime curvature. In other words, the initial data do not possess one of the core features of the null$^+$ solutions, and the perturbation is not restricted to some specific submanifold in the solution space. 

\begin{figure}[t,b]
\begin{overpic}[width=0.99\columnwidth]{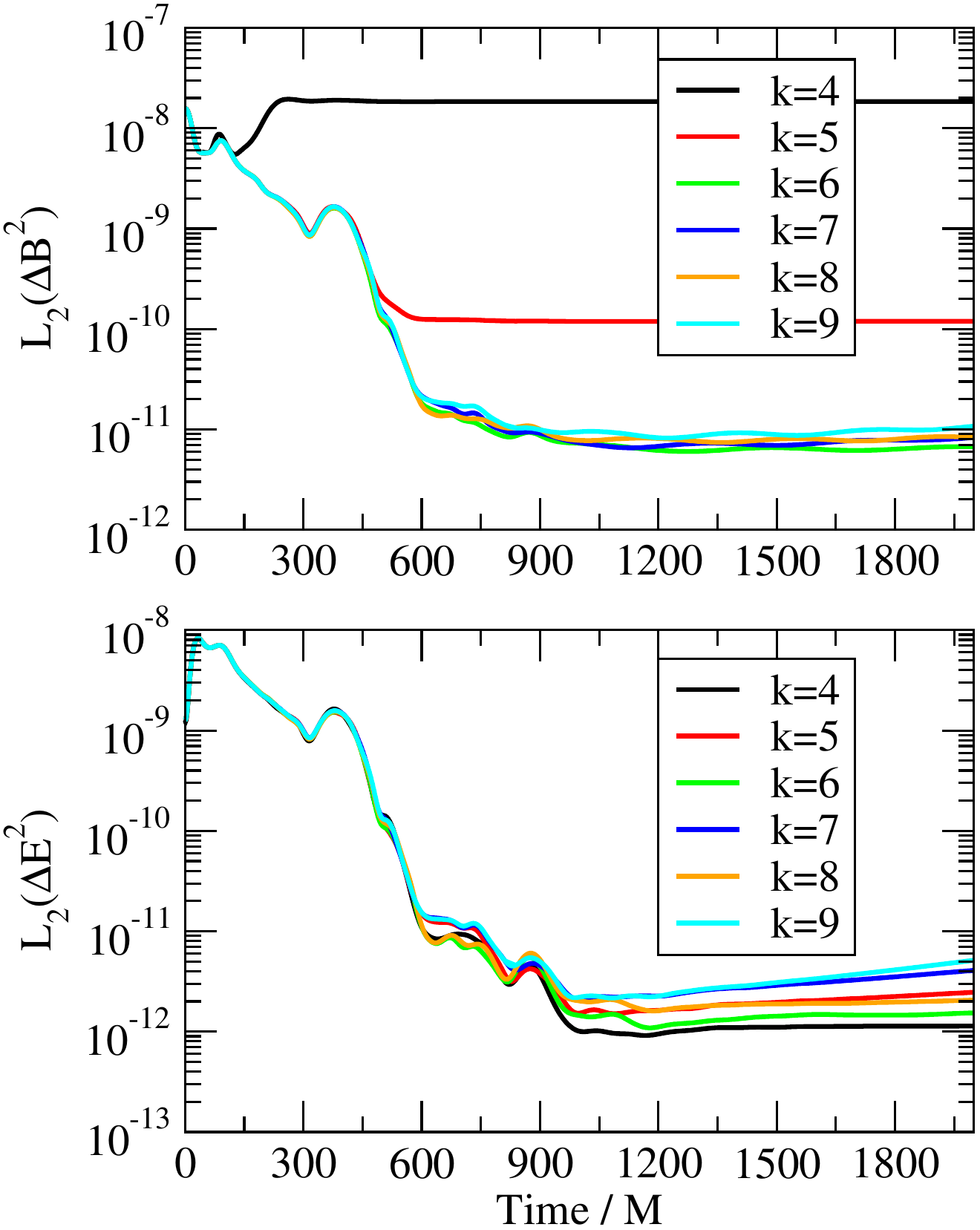}
\end{overpic}
\caption{Top: The $L_2$ norm of the difference measure $\Delta {B}^2$ for the null$^+$ simulations initially perturbed in the propagation direction. Bottom: The $L_2$ norm of the difference measure $\Delta {E}^2$. 
}
\label{fig:PertErrScattering}
\end{figure}

For the initial wave propagation direction, we use a congruence of null geodesics (that is generally not tangential to degenerate GPNDs) in the Schwarzschild spacetime, which admits analytical descriptions. 
We will begin by introducing the procedure for generating a single geodesic within the congruence, building an adapted Newman-Penrose null tetrad along it whose ingoing null basis vector is tangential to the geodesic, and constructing the ${\bf F}^P$ field whose wave component $\phi_0$ (as defined with respect to that newly built tetrad) follows the geodesic direction. Later, we will describe how to obtain the entire congruence, thus filling in the ${\bf F}^P$ field everywhere. 
We begin by choosing Boyer-Lindquist/Schwarzschild coordinates $(r,\theta,\phi)$ so that the geodesic lies on the equatorial plane.  The null geodesic then satisfies the equation \cite{Whittaker,0264-9381-29-6-065016,Hagihara}
\bea \label{eq:EOMGeodesic}
\left(\frac{dy}{d\phi} \right)^2 = 4y^3 - g_2 y-g_3 \,,
\eea 
where 
\bea
y = \frac{M}{2r}-\frac{1}{12}, \quad g_2 = \frac{1}{12}, \quad g_3 = \frac{1}{216}- \left(\frac{M}{2} \right)^2 \mathcal{P}\,,
\eea
with $\mathcal{P}=1/b^2$ for impact parameter $b$. With small impact parameter ($\mathcal{P} > 1/27M^2$), the null geodesic will be absorbed by the black hole, and the shear-free principal congruence tangential to the degenerate pair of incoming GPNDs corresponds to $b =0$. 
The solution to Eq.~\eqref{eq:EOMGeodesic} is then given by $y(\phi)=\wp(\phi+Y | g_2,g_3)$,  where $\wp$ is the Weierstrass elliptic function and $(g_2,g_3)$ are its invariants.  
The angle $Y$ is the angle at which the geodesic strikes the origin $r=0$.

\begin{figure}[!b]
\begin{overpic}[width=0.49\columnwidth]{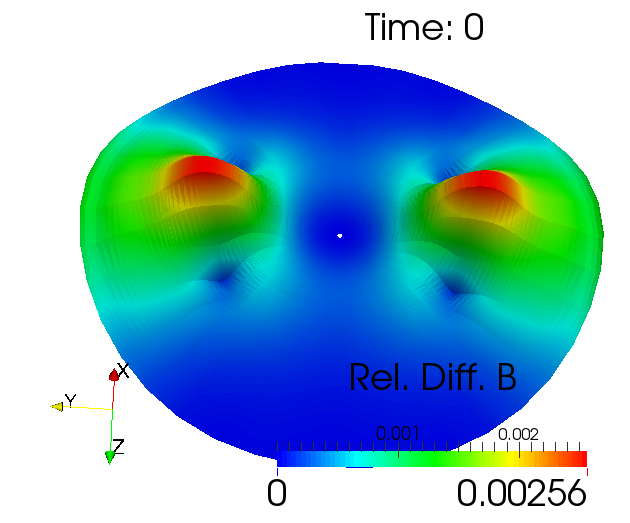}
\end{overpic}
\begin{overpic}[width=0.49\columnwidth]{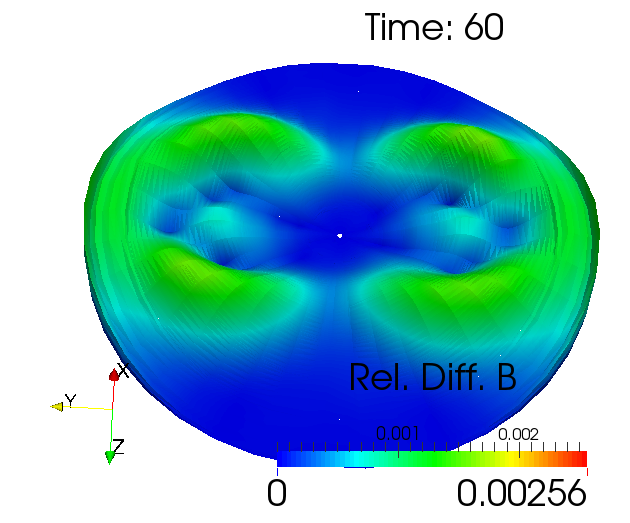}
\end{overpic}
\begin{overpic}[width=0.49\columnwidth]{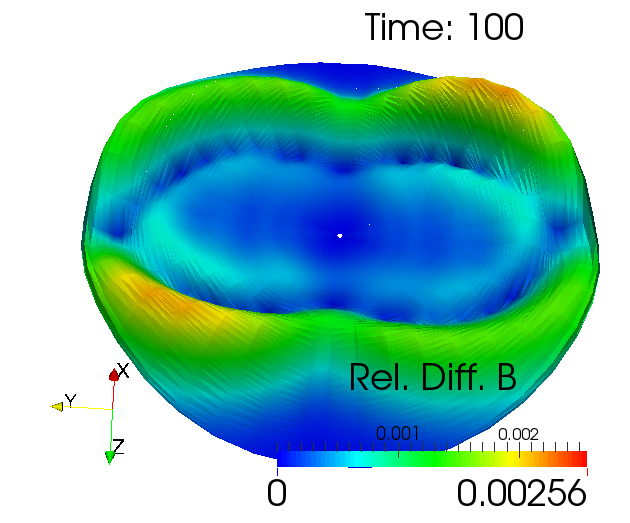}
\end{overpic}
\begin{overpic}[width=0.49\columnwidth]{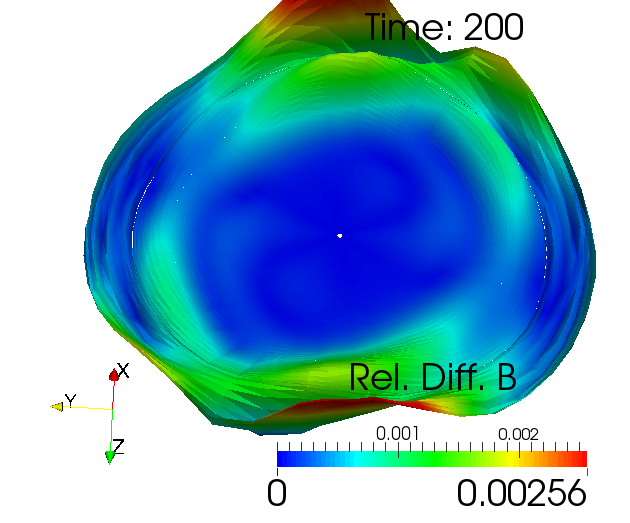}
\end{overpic}
\caption{The relative difference $|\Delta\bf{B}|/|\bf{B}|$ on a vertical slice of the computational domain. The initial perturbation seen at $t=0$ propagates mostly outwards, creating the patterns seen at $t=60$ and $t=100$, and exits the computational domain as seen at $t=200$. 
}
\label{fig:PertEvoScattering}
\end{figure}

Armed with the geodesic, we can now build a Newman-Penrose null tetrad adapted to it. First, we calculate the spatial tangent to the geodesic, and then we convert it into a null $4$-vector and apply the Jacobian from Boyer-Lindquist to Kerr-Schild coordinates (see Appendix \ref{sec:NullSolKerrSchild} for details). 
We will let the ${\bf n}$ basis vector be in the direction of this four-dimensional null tangent, while keeping ${\bf l}$ the same as that of the Kinnersley tetrad. The scaling of ${\bf n}$ is then fixed by $l^an_a = -1$. The remaining ${\bf m}$ and ${\bf \bar{m}}$ bases can be fixed using a Gram-Schmidt procedure \cite{Zhang:2012ky}. We first define two spatial vectors ${\bf C}$ and ${\bf D}$, so that ${\bf m}= 1/\sqrt{2}({\bf C}+i{\bf D})$. Let ${\bf G}$ and ${\bf H}$ be the Kinnersley tetrad versions of ${\bf C}$ and ${\bf D}$, respectively. 
We can then achieve proper orthonormality for the new tetrad by setting 
\bea
\hat{G}^a = G^a + G^b l_b n^a + G^b n_b l^a, \quad 
C^a = \frac{\hat{G}^a}{\sqrt{\hat{G}^b \hat{G}_b}}\,,
\eea
and then 
\bea
\hat{H}^a &=& H^a + H^b l_b n^a + H^b n_b l^a - H^b C_b C^a\,, \\ 
D^a &=& \frac{\hat{H}^a}{\sqrt{\hat{H}^b \hat{H}_b}}\,.
\eea
Under this tetrad, we choose a preliminary Faraday field ${\bf F}^A$ according to Eq.~\eqref{eq:FieldTensor1}, with the $\phi_0$ distribution prescribed by Eq.~\eqref{eq:phi0explicit}, which now describes an incoming wave (there is no $\phi_2$) travelling in the new ${\bf n}$ direction that is different from the GPNDs provided $b\neq 0$.  

We now turn to filling the entire computational domain with geodesics and the Faraday field tensor.
To this end, we begin by specifying an impact parameter $b$ and populate the equatorial plane with null geodesics by varying $Y$ (see Fig.~\ref{fig:GeodesicInSchwarzschild}). 
We then take the $x\ge 0$ portion of Fig.~\ref{fig:GeodesicInSchwarzschild} and rotate it around the $z$-axis, thus filling the entire 3D space. 
However, when the impact parameter does not vanish (for our perturbative study, we choose $b=M/\sqrt{10}$), the resulting congruence will be singular on the vertical axis. We eliminate this problem by constructing an unperturbed null solution with the same $\phi_0$ (but with $b=0$), and blend its Faraday tensor ${\bf F}$ with the ${\bf F}^{A}$ associated with the $b \neq 0$ congruence, so that only the unperturbed null solution $\bf{F}$ that is regular on the vertical axis is present there. 
Explicitly, we set 
\bea
{\bf F}^P = f\; {\bf F}^{A} + (1-f)\, {\bf F}\,,
\eea
with the weighting function $f$ given by Eqs.~\eqref{eq:PertWeightinga} and \eqref{eq:PertWeightingb} (the choice of $Q$ in those equations was made in anticipation of this regularization procedure on the poles). 
Furthermore, we will add an unperturbed magnetic monopole with $q=1000$ to obtain a perturbed null$^+$ solution just as in Sec.~\ref{sec:Pertphi}. 
We also carry out the same constraint enforcement procedure as in Sec.~\ref{sec:Pertphi}. For this section we will keep the background solution time independent with $\Omega=0$, and leave the time-dependent case to the next section. 

We note that just as before, the blending and constraint solving stages ensure that the perturbation we start the simulation with is in fact rather general, with many radial and spherical modes excited. Our procedure differs from an explicit sum of modes with random coefficients in that the specific modifications to the wave propagation directions essentially introduce correlations into the mode coefficients, so that the modes don't accidentally cancel out, leaving us with a perturbation without propagation direction change. It also helps to avoid the initial data solver simply removing the uncorrelated constraint-violating modes and bringing us back to the unperturbed exact solution. 

The output of the initial data solver is displayed in Fig.~\ref{fig:IDSolve}, while the evolution of $\Delta\bf{B}$ and $\Delta\bf{E}$ is shown in Figs.~\ref{fig:PertErrScattering} and 
\ref{fig:PertEvoScattering}. Despite having a different type of initial perturbation, we observe a similar behaviour as seen in Sec.~\ref{sec:Pertphi}, with no diverging modes occurring, and with the perturbation eventually exiting through the outer boundary.

\subsection{time-dependent background solution \label{sec:TimeDepPert}}
\begin{figure}[t,b]
  \centering
  \begin{overpic}[width=0.85\columnwidth]{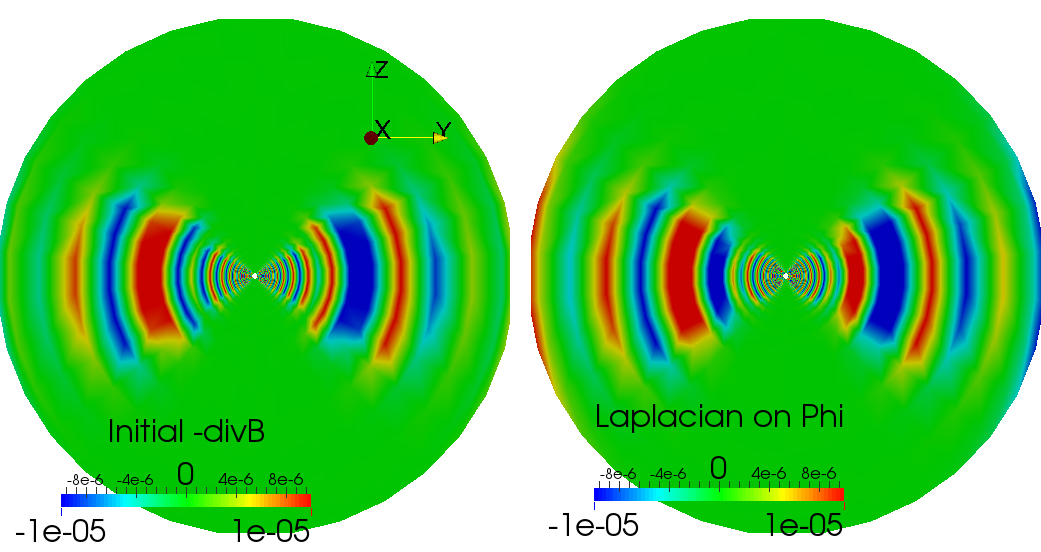}
  \end{overpic}
  \caption{
Similar to Fig.~\ref{fig:IDSolve}, but for initial data with a perturbed propagation direction based on a time-dependent unperturbed solution. 
 Left: The right-hand side of Eq.~\eqref{eq:Poisson}. Right: The left-hand side of Eq.~\eqref{eq:Poisson}.}
	\label{fig:IDSolve3}
\end{figure}

\begin{figure}[!b]
\begin{overpic}[width=0.99\columnwidth]{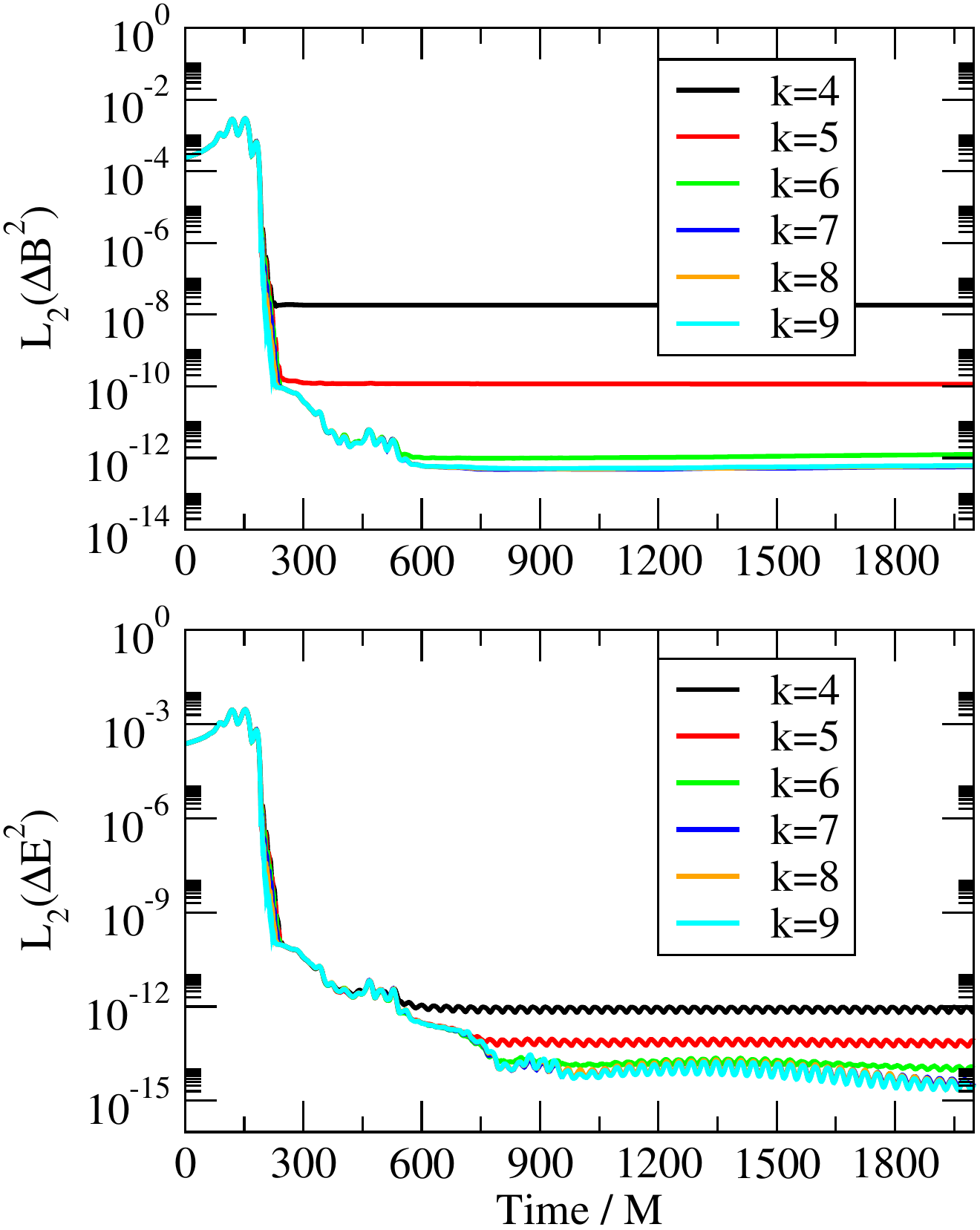}
\end{overpic}
\caption{Top: The $L_2$ norm of the difference measure $\Delta {B}^2$ for the initially perturbed time-dependent null$^+$ simulation.
Bottom: The $L_2$ norm of the difference measure $\Delta {E}^2$. 
}
\label{fig:PertErrScatteringWithTime}
\end{figure}

\begin{figure}[t,b]
\begin{overpic}[width=0.49\columnwidth]{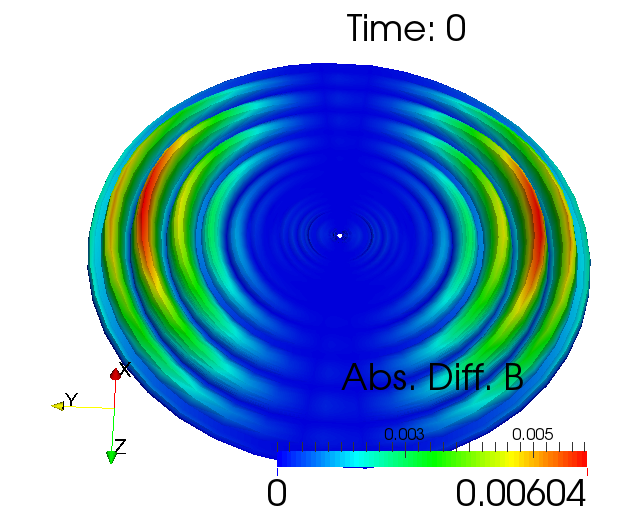}
\end{overpic}
\begin{overpic}[width=0.49\columnwidth]{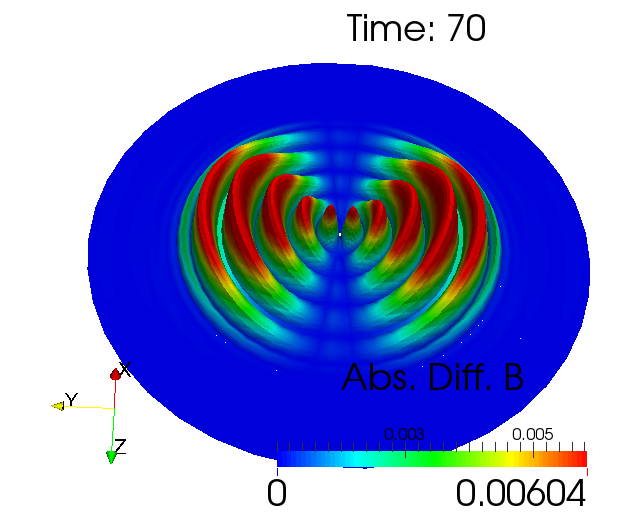}
\end{overpic}
\begin{overpic}[width=0.49\columnwidth]{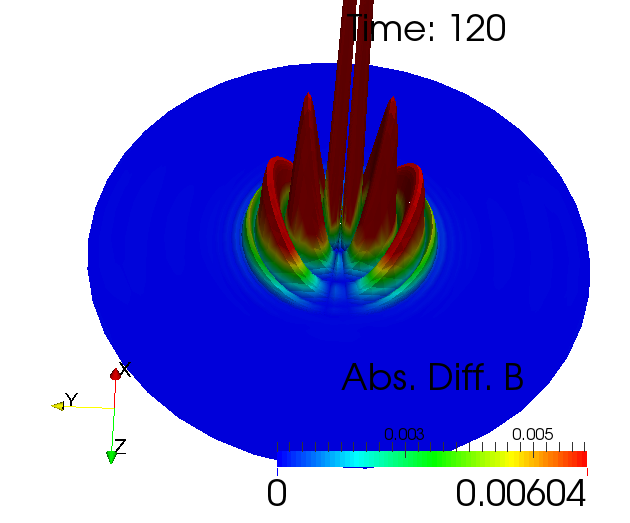}
\end{overpic}
\begin{overpic}[width=0.49\columnwidth]{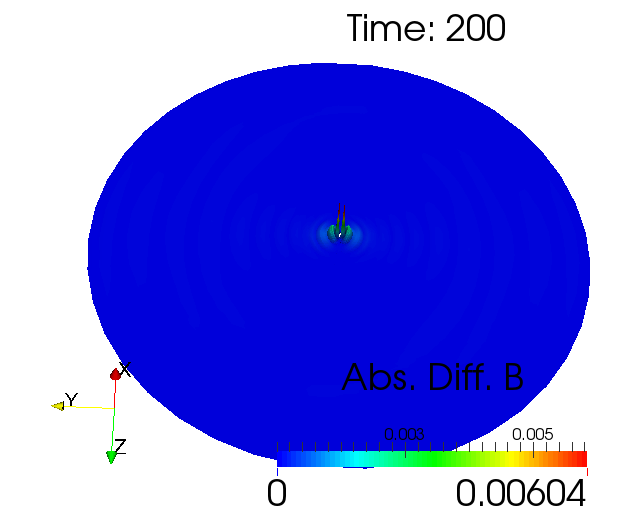}
\end{overpic}
\caption{The absolute difference $|\Delta \bf{B}|$ on a vertical slice of the computational domain. The initial perturbation seen at $t=0$ propagates inwards creating the patterns seen at $t=70$ and $t=120$. By $t=200$, the perturbation has been almost entirely absorbed by the black hole. We have shown the absolute, rather than the relative difference, because $|\bf{B}|$ increases quickly when we approach the black hole, so it is more difficult to see what is going on in the inner regions with a relative difference plot. 
}
\label{fig:PertEvoScatteringWithTime}
\end{figure}

For our final numerical setup, we introduce a time dependence with $\Omega=0.1$, so that the background solution has the familiar character of a travelling wave. The procedure for introducing perturbations is otherwise identical to Sec.~\ref{sec:PertPropDir}, so that the waves are initially travelling in directions different from the degenerate GPNDs.  

The output of the initial data solver and the evolution codes are displayed in Figs.~\ref{fig:IDSolve3}, \ref{fig:PertErrScatteringWithTime} and \ref{fig:PertEvoScatteringWithTime}.
Despite the 
change of energy flux character from electromagnetic winds to waves, our simulation suggests that the now time-dependent null$^+$ solutions are also stable. The most noticeable difference with the two earlier cases is that the perturbation propagates almost entirely inwards initially, and is absorbed by the black hole almost completely after one light-crossing time.

\section{Scattering by spacetime curvature and the role of GPNDs \label{sec:GPNDs}}

In this section, we seek to shed some light on the question of what feature of the analytical solutions examined in this work allows them to avoid being backscattered by spacetime curvature. 
Because we will only carry out analytical studies on the unperturbed exact solutions in this section, the null and null$^+$ solutions are exactly the same in terms of their wave propagation properties. So we will consider pure null solutions for brevity, with the understanding that the conclusions translate to the wave component of the null${}^+$ solutions trivially.

The answer to the scatter-avoidance question is interesting in that it provides guidance to the search for similar FFE solutions in other spacetime backgrounds, or solutions to non-FFE equations. 
For example, when several analytical solutions \cite{Menon:2011zu,Michel1973} to the FFE equations were first found, it was not immediately clear why such simple solutions exist \cite{Brennan:2013jla,Michel1973}, given that the FFE equations are nonlinear. 
Furthermore, such scatterless null solutions are closely related to important advances in mathematical physics, such as the 
discovery of new solutions to the Einstein equations \cite{RobinsonTrautman1,RobinsonTrautman2}, and the definition of twistors \cite{Penrose1986V2}. Therefore it is informative to try to understand the core features of these solutions at an intuitive level. 

A hint on the answer to this question is provided by the Goldberg-Sachs theorem \cite{2009GReGr..41..433G}, which states that scatter-avoiding waves must propagate along a repeated principal null direction (GPND\footnote{We denote these principal null directions as GPNDs, rather than PNDs, to emphasize that they are gravitational PNDs, and distinguish them from the electromagnetic PNDs that we will encounter later.}) of the Weyl curvature tensor. However, as far as the authors are aware, there is no explicit analysis in the literature of the reverse question, i.e. whether all waves propagating along GPNDs are to some extent scatter avoiding. Aside from shedding some light on the scatter avoidance puzzle, this analysis also fills a gap in the literature by providing a simple physical intuition on the  concept of GPNDs, which underlies such important constructs and results as the Petrov classification of spacetimes 
and the peeling theorem \cite{Sachs1961,Sachs1962,Penrose1963,Penrose1965,Penrose1986V2}. 

Before diving into the technical details, we first summarize the results of this section, and provide some intuitions as to why one should expect these results. Readers only interested in the conclusions and their uses can skip the derivations presented later in this section. The conclusions are 
\begin{enumerate}
\item Null solutions that propagate along repeated GPNDs of multiplicity 3 or above will not experience scattering by spacetime curvature at all. This requires the background spacetime to be of Petrov type III or type N \cite{Petrov2000,PetrovBook} (and of course type O which is flat) \footnote{Petrov type I: four different GPNDs; type II: two degenerate GPNDs and two nondegenerate GPNDs; type D: two sets of doubly degenerate GPNDs; type III: a triply degenerate set of GPNDs and a nondegenerate one; type N: four-fold degeneracy in GPNDs; type O: Weyl curvature tensor vanishes.}, and for the null waves to be propagating along the special direction in those spacetimes identified by the repeated GPNDs. 

Such a situation arises when analysing the coincident gravitational and electromagnetic wave signals from distant sources in the context of multimessenger astronomy. The two types of waves travel in the same direction for the majority of their journey, and the gravitational wave gives a metric perturbation of type N with the four-fold degenerate GPND pointing in the propagation direction. This means one does not need to worry about the gravitational wave changing the electromagnetic wave during their long journey to Earth. 

\item When the spacetime is less special in the sense that the repeated GPNDs have less multiplicity, the null electromagnetic solutions travelling along the repeated GPNDs will in general experience some scattering by the spacetime curvature. However, the more severe backscattering (where a wave propagating in the opposite direction and/or a Coulomb background piece is created by the scattering) can be avoided, and the scattering only manifests itself as an influence on the wavefront profile. 

As Schwarzschild and Kerr spacetimes are of Petrov type D (the multiplicity of the degenerate GPNDs is two), this is the situation relevant for the exact analytical solutions (more precisely the null part of the null$^+$ solutions) presently under examination. The avoidance of backscattering allows for, e.g., setting $\phi_2$ (the outgoing wave component of the Faraday tensor) and $\phi_1$ (the Coulomb background piece) to zero, and only solving for $\phi_0$ (the ingoing wave piece) in the force-free equations, which significantly reduces the complexity of the solution finding process. We expect similar features to also be available when solving the equations of other field theories in these important spacetimes. 

\end{enumerate}

\begin{figure}[t,b]
\begin{overpic}[width=0.55\columnwidth]{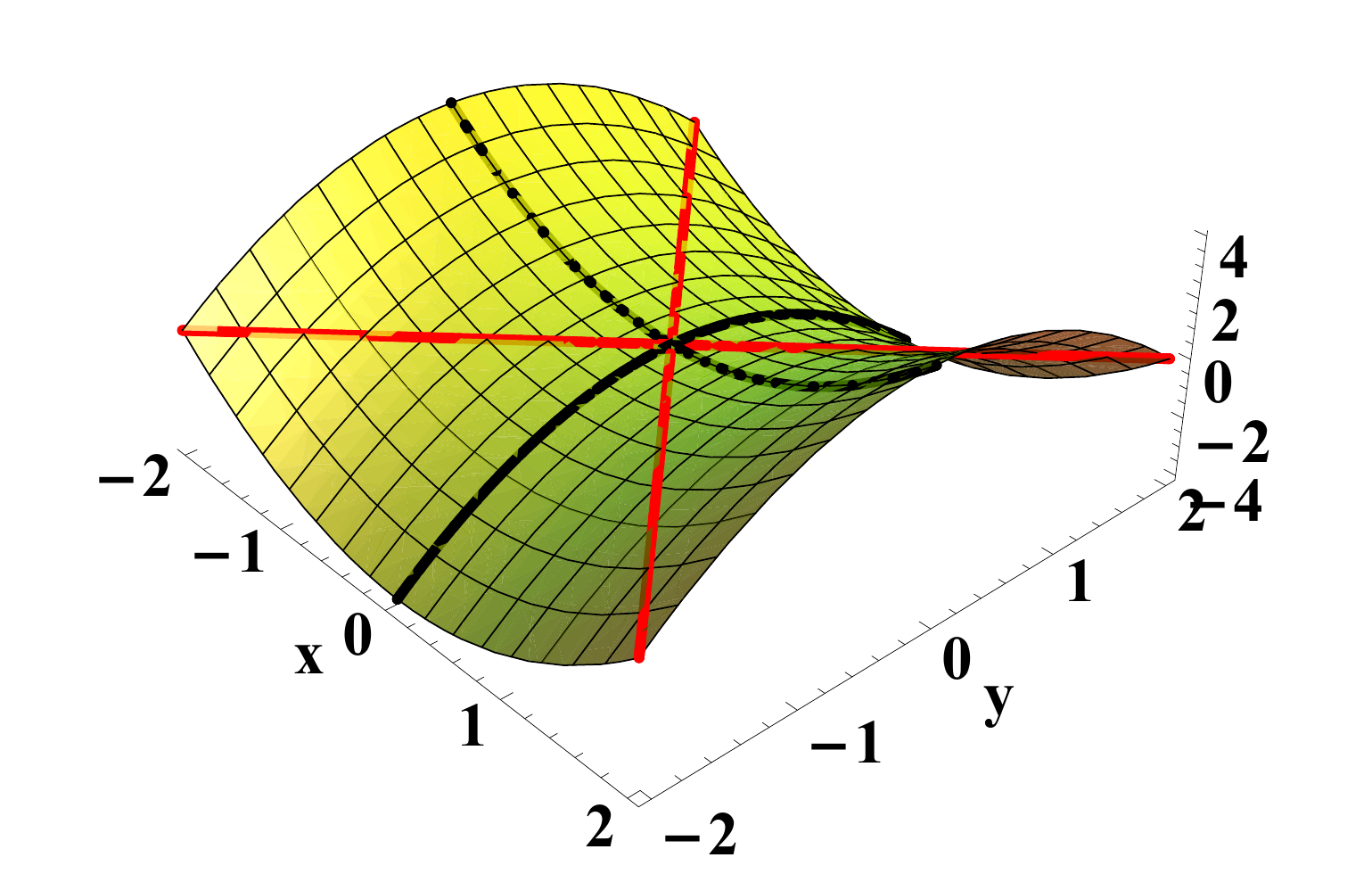}
\end{overpic}
\begin{overpic}[width=0.4\columnwidth]{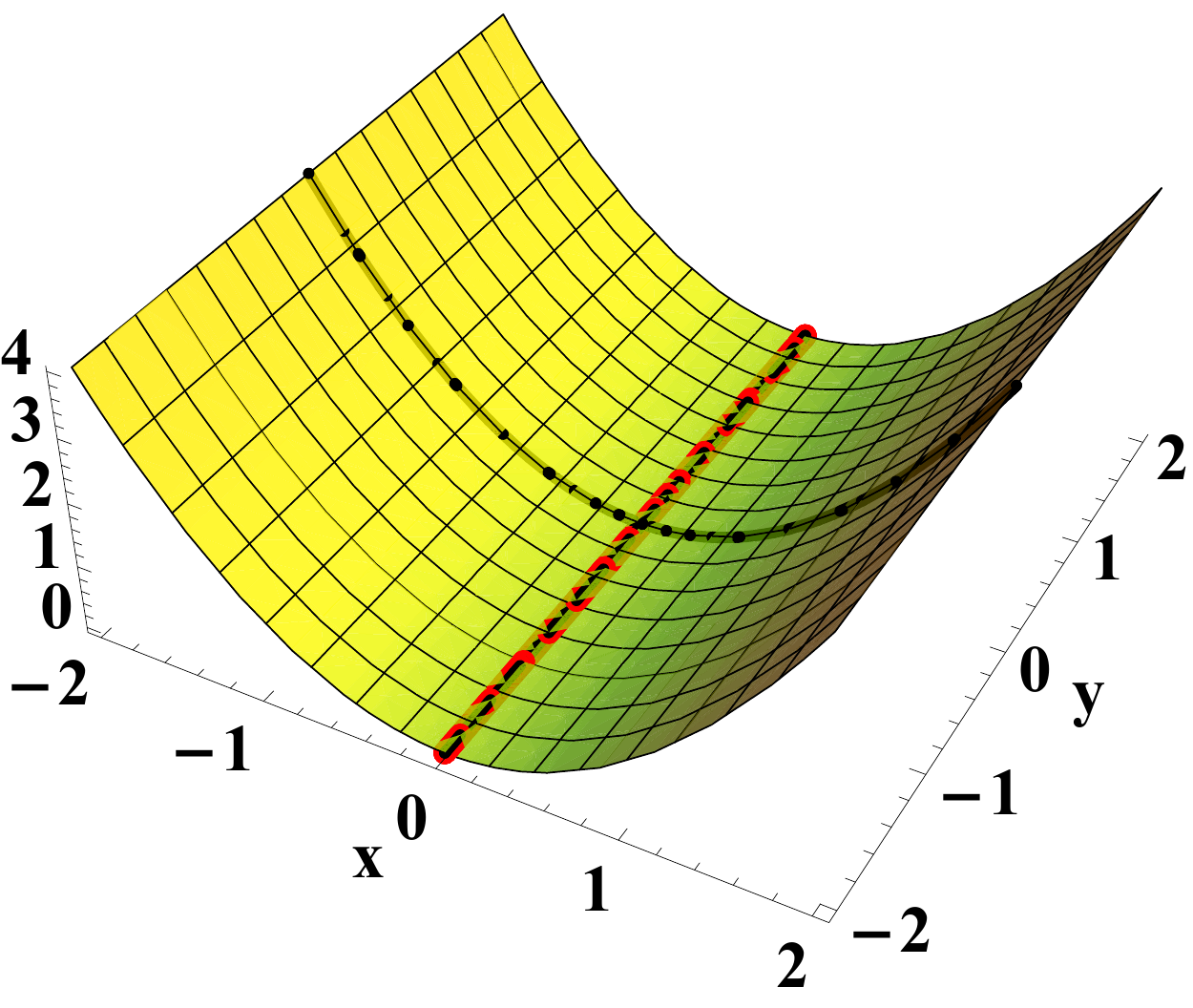}
\end{overpic}
\caption{The principal directions (black lines) and the asymptotes (red lines) on curved surfaces. Figure on the left depicts the situation with a more generic surface, while the figure on the right depicts a more degenerate surface where two asymptotes become coincident and the principal direction in between them becomes flat. 
}
\label{fig:Surface}
\end{figure}

These conclusions create the obvious impression that the GPNDs are somehow particularly flat directions of spacetime, such that waves propagating along them see less of the spacetime curvature. This intuition can be made more visual by invoking an analogy with a curved surface embedded in a three dimensional (flat) ambient space. In fact, the name ``principal null directions'' already invoke this analogy, but we will argue that instead of the principal directions, it is really the asymptotes on the curved surface that these GPNDs are more akin to. Given our quick scan of the literature, our discussion appears to be the first to state this way of intuiting the GPNDs, and we hope it would be helpful to researchers newly acquainted with these important quantities. 

The curving shape of a surface embedded in the ambient (see Fig.~\ref{fig:Surface}) is given by the extrinsic curvature tensor ${\bf K}$, which is a rank two tensor. The twice contraction of a vector tangential to the surface at a point (the center/origin in Fig.~\ref{fig:Surface}) with ${\bf K}$ gives the curvature of a geodesic on the surface developed along that vector direction. The principal directions are defined to be those directions whose associated geodesics have the maximum or minimum (most negative/bending the other way) curvature among the different direction choices, and are in fact the eigenvectors of ${\bf K}$. Their associated geodesics are the thick black lines on the surfaces in Fig.~\ref{fig:Surface}, which are clearly the most curved (either in the up or down direction) directions on the surface. The asymptotes on the other hand are defined to be those directions whose twice contraction with ${\bf K}$ vanishes. These are the red curves in Fig.~\ref{fig:Surface}, and are clearly locally flat/straight. Later in the section, we will show that the contraction of quantities representing GPNDs with those representing the spacetime curvature tensor gives us zeros instead of maxima and minima, so the GPNDs are really more akin to the asymptotes. 

The principal directions are always half way in between the asymptotes, so these two sets of quantities are trivially related. When two asymptotes coincide, the principal direction in between them is then forced to become coincident with both asymptotes and thus take up vanishing curvature (the right panel of Fig.~\ref{fig:Surface} shows this situation, where there is also a thick black line aligned with the coinciding red lines at the bottom of the ``valley''), so the spacetime becomes flatter in a sense as one of its extremally curved directions become flat (in the case of the curved surface, the surface on the right of Fig.~\ref{fig:Surface} that has coinciding asymptotes becomes developable -- its intrinsic curvature vanishes and the surface can be unfolded into a simple flat plane). The analog with this on the spacetime side is that more coinciding GPNDs signal that the spacetime is ``flatter'', especially along the direction of those coinciding GPNDs (this is essentially the intuitive meaning of the Petrov classification). It is then not entirely surprising that FFE waves travelling along the ``extra flat'' degenerate GPNDs will experience less curvature-induced scattering. 

This analogy can be developed much further
\footnote{We brief mention why the name ``principal null direction'' is historically given to the GPNDs, even though they are more like asymptotes, and not the principal directions on a surface. That name assignment comes from their being related to some eigenvalue problem of the spacetime curvature tensor when written in the tensor language. However, when we migrate to the spinor language, the eigenvalue problem switches into one over some other directions that sit in between the GPNDs (they are the null basis vectors of the so-called canonical transverse tetrads \cite{Penrose1986V2}, and they are really the things that are akin to the principal directions on a surface), just like the principal directions on a surface sit in between the asymptotes. So in the spinor language, the ``true'' nature of the GPNDs become more apparent, and using something called sectional curvatures \cite{Ehlers,Hall1987}, we can 
show that it is in the spinor and not the tensor language, that the eigenvalue problem is more closely related to the curvatures of geodetic submanifolds of the spacetime (analogs to the curves on a surface), and so the spinor language is the one that's more appropriate for building analogies. }
, but a detailed discussion will lead us too far on a digression away from the main content of this paper, so we stop here and turn to some derivations directly relevant for curvature-related scattering that back up the conclusions listed above, which also provide some more concrete examples of the type of vanishing contractions underlying our analogy-based intuitive picture. 

The discussion below will rely heavily on the spinor formalism, which reveals the characteristic structure of the Weyl curvature tensor in a significantly more transparent manner than the tensor formalism. We include a brief summary of spinors in Appendix \ref{sec:SpinorIntro}. The most important feature for us is that the spinors can roughly be seen as ``square-roots'' of null vectors with the tensor product $\o^{A}\bar{\o}^{A'}$ of a pair of complex conjugate spinors $\o^{A}$ and $\bar{\o}^{A'}$ corresponding to a null vector.  In addition, the self-contractions of the spinors vanish ($\o_{A}\o^{A}=0=\bar{\o}_{A'}\bar{\o}^{A'}$) just as they do with null vectors. 
 
The null solutions of Ref.~\cite{Brennan:2013jla} are the FFE counterparts to the vacuum null electromagnetic solutions described in Ref.~\cite{Robinson61}. A null solution is defined by the property that the two principal null directions (EPNDs\footnote{We will use EPND to refer to principal null directions of electromagnetic solutions.}) of the Faraday tensor are coincident, just like simple plane waves in flat spacetime. 
These solutions can thus be seen locally as generalized plane waves (see the discussion following Eq.~(22) of Ref.~\cite{Brennan:2013jla} for more explicit local similarities with plane waves), whose propagation directions follow ingoing or outgoing shear-free null congruences \cite{Robinson61}. In a curved spacetime, this implies that they must evade being backscattered by the spacetime curvature, or else they cannot remain purely ingoing or outgoing. As a concrete example, the solution given by Eqs.~\eqref{eq:FieldTensor1}-\eqref{eq:fI} has only the ingoing wave component $\phi_0$, while the outgoing wave component $\phi_2$ and the Coulomb background $\phi_1$ vanish identically throughout space and time. To understand how backscattering is avoided, we recall that the ingoing solution as specified by $\phi_0$ follows the Kinnersley tetrad ${\bf n}$ basis direction, which is tangential to a geodetic shear-free null congruence \cite{ChandrasekharBook}. By the Goldberg-Sachs theorem \cite{2009GReGr..41..433G}:

\noindent 
\emph{A strictly nonflat vacuum metric has a multiple principal null direction $\ell^a$ iff $\ell^a$ is geodetic and shear free,}

\noindent
this ${\bf n}$ direction must also be the direction of two or more degenerate GPNDs. We now show explicitly that following degenerate GPNDs is responsible for the simplifications to scattering by spacetime curvature.  

We begin by recalling how electromagnetic waves (with allowance for current and charge, so FFE waves are included as a special case) scatter off of spacetime curvature. 
The wave equation satisfied by the Faraday tensor is given by \cite{Tsagas:2004kv}
\bea \label{eq:WaveEq}
\nabla^c \nabla_c F_{ab} &=& -2R_{acbd} F^{cd} + R_a{}^cF_{cb} + F_a{}^{c} R_{cb} \notag \\
&&+ \nabla_b J_a - \nabla_a J_b\,,
\eea
where $\nabla^c \nabla_c$ is the generalized covariant Laplacian operator. 
The scattering by spacetime curvature is described by the first three terms on the right-hand side of Eq.~\eqref{eq:WaveEq}, and is a consequence of the tensorial nature of ${\bf F}$ that allows it to couple to the spacetime curvature through the Ricci identities, which when applied to our case gives  
\bea
2\nabla_{[a}\nabla_{b]} F_{cd} = R_{abce}F^e{}_d + R_{abde} F_c{}^e\,,
\eea
and subsequently yields the aforementioned terms. 
These scattering terms imply that, generically, the electromagnetic waves can propagate inside as well as on the future null cone of a light source, as secondary ingoing waves can be created by scattering, and so they do not satisfy Huygens's principle \cite{DeWitt:1960fc,Friedlander,GuntherBook,Tsagas:2004kv}.
The remaining terms on the right-hand side of Eq.~\eqref{eq:WaveEq} describe scattering by charge and current, and are not the scattering we are interested in here. In other words, our consideration in this section concentrates on the scattering shared by vacuum (without current), FFE (with current), as well as other electromagnetic solutions with more generic currents, so that our conclusions are not confined to the FFE case.  
 
Since we do not include the stress-energy tensor of the electromagnetic field or the plasma in the gravitational sector, as per the simplifying convention in FFE computations \cite{Brennan:2013jla}, we have $R_{ab} =0$, and $R_{abcd}$ reduces to the Weyl tensor $C_{abcd}$. 
Both $C_{abcd}$ and $F_{ab}$ can be written in the spinor formalism as \cite{Penrose1992} 
\bea
F_{ab} &=& \phi_{AB}\epsilon_{A'B'} + \epsilon_{AB}\bar{\phi}_{A'B'}\,, \label{eq:EMSpinor}\\
C_{abcd} &=& \Psi_{ABCD}\epsilon_{A'B'}\epsilon_{C'D'} 
+ \bar{\Psi}_{A'B'C'D'}\epsilon_{AB}\epsilon_{CD}\,.
\label{eq:SpinorWeylForm}
\eea 
Note that as per convention, we have left out the soldering forms like $\sigma_a^{AA'}$ for brevity, with the understanding that spinor index pairs like $AA'$ correspond to tensor index $a$ (same letter but lowercase). 
The multi-indexed spinors can be further factorized into products of their respective principal spinors (relating to the principal null directions of the original tensors via Eq.~\ref{eq:NullVecVsSpinor}):
\bea 
\phi_{AB} &=& \o_{(A}\iota_{B)}\,,  \label{eq:GrGPNDFactor1} \\
\Psi_{ABCD} &=& \alpha^{(1)}_{(A}\alpha^{(2)}_{B}\alpha^{(3)}_{C}
\alpha^{(4)}_{D)} \,, \label{eq:WeylPrincipal}
\eea
where more specialized Petrov classes of spacetimes have more of the $\alpha^{(\cdot)}$s (corresponding to the GPNDs) being coincident. 

Now a simple calculation shows that the scattering term translates into the spinor language as 
\bea
C_{acbd} F^{cd} = 
\left(\Psi_{ABCD}\phi^{CD}\right)\epsilon_{A'B'} + c.c.\,,
\eea
where $c.c.$ stands for complex conjugation. Substituting in Eqs.~\eqref{eq:GrGPNDFactor1} and  \eqref{eq:WeylPrincipal}, we find that the spinor counterpart to the scattering term is
\bea \label{eq:PairWise}
\Psi_{ABCD}\phi^{CD} = \frac{1}{12}\sum_{i,j} \left( 
\alpha^{(i)}_{C} \o^C \right)
\left( \alpha^{(j)}_{D} \iota^D \right) \alpha^{(k)}_{(A}\alpha^{(l)}_{B)}
\eea
where $(i,j)$ are an unordered and unequal pair of numbers from $\{1,2,3,4\}$, while $(k,l)$ are an ordered pair consisting of the remaining two numbers, with $k > l$. 
From Eq.~\eqref{eq:PairWise}, and recalling that the contraction of coincident spinors vanish, it is clear that the more pairs of EPNDs and GPNDs that are coincident, the more terms in the sum will vanish, leaving us with a scattering term that is simpler in its composition, meaning it has fewer independent components. 

Such an effect is particularly strong with null electromagnetic waves, defined by the property that the two EPNDs are coincident or $\o \propto \iota$ (Sec.~5.1 of \cite{Penrose1992}). 
This implies that $\phi_{AB} \phi^{AB} =0$ due to Eq.~\eqref{eq:SpinorSelfContraction}. Through Eq.~\eqref{eq:EMSpinor}, this further implies that the two real invariants of the Faraday tensor must vanish \cite{Penrose1986V2}, i.e. Eqs.~\eqref{eq:NullCond1} and \eqref{eq:NullCond2} must be satisfied. 
In particular, these conditions are consistent with the force-free constraints, so there can be FFE null solutions. More specifically, using Eqs.~\eqref{eq:EMSpinor}, \eqref{eq:GrGPNDFactor1}, \eqref{eq:TetradVSpinor} and \eqref{eq:SympDyad}, it is easy to verify that for a purely ingoing null solution, 
we can write 
\bea \label{eq:FField}
F_{ab}= 
\phi_0 {\o}_A {\o}_B \epsilon_{A'B'} 
+c.c.\,,
\eea
where $\phi_0$ is extracted under any dyad (corresponding to a Newman-Penrose null tetrad via Eq.~\ref{eq:TetradVSpinor}) that has $\o$ as a member of its basis. We will denote such a dyad as $(\o^A,\zeta^A)$.

We now consider a null electromagnetic wave travelling in a purely radiative (Petrov type N) 
\cite{Petrov2000,PetrovBook} spacetimes\footnote{Petrov type I: four different GPNDs; type II: two degenerate GPNDs and two nondegenerate GPNDs; type D: two sets of doubly degenerate GPNDs; type III: a triply degenerate set of GPNDs and a nondegenerate one; type N: fourfold degeneracy in GPNDs; type O: Weyl curvature tensor vanishes.}
with all four GPNDs coinciding. If furthermore, the electromagnetic wave travels in the same direction as the gravitational wave, so that the doubly degenerate EPNDs coincide with the four-fold degenerate GPNDs, then all of the contractions in Eq.~\eqref{eq:PairWise} will vanish, and there will not be any curvature scattering term left. Such a complete disappearance of scattering can also happen when the spacetime is of type III. 

For the Petrov type D Kerr spacetime, the scattering term in  Eq.~\eqref{eq:PairWise} does not vanish completely,  so 
there is still some residual scattering, 
but of a simplified structure 
\bea \label{eq:Coupling2}
C_{acbd} F^{cd} = \phi_0 \Psi_2 \o_A \o_B \epsilon_{A'B'} + c.c.\,,    
\eea 
that does not necessarily lead to \emph{back}scattering. 
In particular, the scattering term only contains $\o_A$ and no $\zeta_A$ in its spinor form, which helps prevent a contamination of $F_{ab}$ by $\phi_1$ and $\phi_2$, as the spinor counterpart of $F_{ab}$ is given by 
\bea
\phi_{AB} = 
\phi_0 {\o}_A {\o}_B 
- 2\phi_1 {\o}_{(A} {\zeta}_{B)} 
+ \phi_2 {\zeta}_A {\zeta}_B\,,
\eea
so that $\phi_1$ and $\phi_2$ need to multiply with  $\zeta_A$ in order to pick up spinor and subsequently tensor indices. A rigorous proof for the existence of backscatterless solutions in type-D spacetimes is provided by the the general theorem 
(see Refs.~\cite{Robinson61,Sommers1976} and (7.3.14) of Ref.~\cite{Penrose1986V2}):

\noindent \emph{If $\ell^a$ is a geodetic and shear-free null congruence and analytic, then there is a nonzero solution of the vacuum Maxwell equations which is null along $\ell^a$.}

\noindent This result applies in the vacuum case and, to a more limited extent (restricted to Kerr spacetimes \cite{Brennan:2013jla}), in the force-free case. The proof of the theorem above and the discovery of the FFE solutions are rather technical in nature, but we hope that our discussion regarding the contraction/annihilation between EPNDs and GPNDs and the subsequent simplification/elimination of the scattering term will serve to help build intuition for these important results.

\section{Conclusion}
In this paper, we have introduced a new pseudospectral fully 3D curved spacetime FFE code, with an initial data solver and an improved constraint damping mechanism. Using this code, we have shown through numerical experiments that the backscatterless analytical FFE solutions found by Ref.~\cite{Brennan:2013jla} are stable against a variety of perturbing scenarios, which we selected to avoid restricting ourselves to special subspaces of the FFE solution space. However, with any simulation, one can only constrain the growth rate of unstable modes, as the simulations can not be performed for an infinite amount of time. We have carried out our simulations for around ten light crossing times, and observed the perturbations to exit the computational domain after two or three. If there exist unstable modes, and they are excited to an appreciable initial amplitude, we would expect them to remain in the computational domain and then gradually grow. This does not appear to be the case. Nevertheless, we should be cautious and only place an upper bound on the growth rate of any unstable modes assuming the worst scenario. Assuming that the unstable modes exist but are not excited by our initial perturbation at all, and instead started at the floor level of numerical noise with $\Delta E^2$ and $\Delta B^2 \sim 10^{-20}$ (see Fig.~\ref{fig:Convergence}), our final values for these quantities are around $10^{-11}$ (see Figs.~\ref{fig:PertErrNoScattering}, \ref{fig:PertErrScattering} and \ref{fig:PertErrScatteringWithTime}) after $2000$M of simulation, which gives an upper bound on the growth rate for $\sqrt{\Delta E^2}$ and $\sqrt{\Delta B^2}$ at around $0.005$. For comparison, the fundamental $l=1$ quasinormal mode decay rate for the magnetosphere in our study is $0.09$.

In order to carry out concrete numerical studies, we had to select particular solutions from the family of infinitely many null$^+$ solutions that can be obtained. We have chosen arbitrary representative solutions that share all the important physical features listed in Sec.~\ref{sec:IntroSol} with the rest of their siblings that are not explicitly simulated, and these solutions do not have any special features or symmetries that would make them more (or less) stable than the rest. We have also examined both classes -- the time-dependent ``waves" and the time-independent ``winds" -- of the solution family. Therefore, we expect the stability conclusion to be generalizable to the majority, if not all, of the solutions in the parent family. 
In addition, we have chosen the magnitudes of the unperturbed solution in such a manner as to ensure the tests are carried out within the nonlinear regime of the FFE equations, so our results suggest full nonlinear stability. 
Furthermore, even though we cannot make mathematically rigorous statements as to the asymptotic stability of the null$^+$ solutions due to the presence of numerical noise, we note that with all three different types of perturbations that we have examined, the perturbations are seen to exit the computational domain after two light-crossing times, and the final $\Delta {B}^2$ and $\Delta {E}^2$ are small. From a practical point of view, even if the final ``steady-state'' solutions are not exactly the same as the null$^+$ solutions, they would be well approximated by them, so that the null$^+$ solutions can be considered effectively asymptotically stable. Therefore, despite their physical specialness, these solutions are not fragile, and can in fact describe physically realistic scenarios like the outer magnetospheres of pulsars. We note that one feature of the FFE null$^+$ solutions that made our numerical study possible is the fact that they are globally regular, and are therefore amenable to numerical simulations. In contrast, the vacuum null solutions in curved spacetimes would contain singularities if extended globally \cite{Brennan:2013jla,Teukolsky:1974yv}. 

Unlike the jet stability (despite many plasma instabilities that can be present in that case, the observed jets from active galactic nuclei can extend to several hundred kpc), the stability of the null$^+$ solutions is perhaps less surprising. The null$^+$ solutions are more or less isotropic, so there is no need to maintain a narrow collimated energy flux by e.g. ambient gas pressure, and no sharp boundaries in the form of a jet surface exist. Therefore, we do not have problems like dangerous surface Kelvin-Helmholtz modes associated with the vortex sheet on the jet surface \cite{2000A&A...355..818A, 2000A&A...355.1201L}. Nevertheless, there are many interesting physical properties of the null$^+$ solutions, such as having a null current and being scatter avoiding, which we did not know would be stablizing or destablizing. Given the stability result from our numerical experiment, one can now ask interesting questions such as whether the current being null will actually help prevent the onset of current-driven instabilities? Would an electromagnetic wave propagating close to a degenerate GPND direction in fact tend to end up moving exactly along it during its journey (in which case the null$^+$ solutions will really have a preferential status)? These will become interesting topics for future studies. 

We caution however, that the stability of the null$^+$ solutions does not necessarily translate into that of the null solutions. This is because first of all, the null exact solutions will only satisfy the magnetic dominance condition marginally, so any perturbation would likely introduce non-FFE influences such as current sheets. In the study of jet stability, the presence of current sheets at the jet surfaces is long known to encourage instabilities \cite{2015MNRAS.450..982K}, and in a more recent study, it has been seen that dynamical current sheets have the tendency to drastically rearrange the underlying force-free solutions even when the initial data provided are already a solution to the FFE equations \cite{2015arXiv150306788Y}. 
Such current-sheet-induced instabilities may well also beset the null solution stability. 
Secondly, taking the limit where the null waves become Alfv\'en waves, the background monopole ${\bf B}$ field in a null$^+$ solution provides a preferred direction for the waves to travel in. It is not clear whether the null solutions would be more susceptible to changing propagation directions when this guidance is taken away. Lastly, it has been observed in the study of jet stabilities that increased magnetization has a stabling effect \cite{2015MNRAS.450..982K}, which can also be gone in the null case. As in astrophysical situations, we expect a split monopole/dipole-like background field to be present, the stability of the null$^+$ solutions observed in this work would hopefully mean that they can serve as efficient channels for carrying energy across magnetospheres unhindered. However, because the background field drops off faster than the null wave component in a null$^+$ solution, eventually, this stability may be lost (if the edge of the magnetosphere is not reached first), and the ensuing instability may play a role in how the transported energy finally turns into observational signals. 

Finally, on the analytical front, we have carried out an explicit analysis of the scattering of an electromagnetic wave by the spacetime curvature, with emphasis on the role played by the GPNDs of the Weyl curvature tensor. We  showed that waves propagating along the degenerate GPNDs experience simpler forms of curvature scattering, thereby providing some intuition into the perplexing existence of backscatterless null solutions (although general theorems on this subject already exist, their proofs are highly technical). One interesting new conclusion is that the scattering can vanish completely in a Petrov type III or type N spacetime. This would remove a potential complication (though may or may not be significant in the first place) associated with analysing the coinciding electromagnetic counterpart to a gravitational wave, as the gravitational wave would not in fact try to scatter its electromagnetic companion. 
The understanding of curvature scattering gained here should also prove useful when constructing new analytical solutions to the FFE equations and other field theories where a lack of scattering is desirable, perhaps for the sake of reducing computational complexity.

\acknowledgements
We thank Zachariah Etienne and Matthew Duez for useful discussions, and the anonymous referee for many helpful comments.
The numerical simulations presented in this work were performed on the WVU cluster Spruce Knob and the Caltech clusters SHC and Zwicky. Spruce Knob is supported in part by National Science Foundation EPSCoR Research Infrastructure Improvement Cooperative Agreement No. 1003907, the state of West Virginia (WVEPSCoR via the Higher Education Policy Commission) and WVU. Zwicky is funded by NSF MRI-R2 Award No.~PHY-PHY-0960291.
F. Z. is supported in part by National Natural Science Foundation of China Grant No.~11443008, Fundamental Research Funds for the Central Universities Grant No.~2015KJJCB06, and a Returned Overseas Chinese Scholars Foundation grant. 

\appendix
\section{Hyperbolicity of the evolution system \label{sec:Hyperbolicity}}
The study of the characteristic structure of a set of evolution equations arises from the need for the initial value problem to be well posed. For our pseudospectral implementation, there is an added urgency because the boundary conditions for the overall computational domain \cite{Rinne2006,Rinne2007,Lindblom:2007} as well as between the adjacent subdomains \cite{Hesthaven1997,Hesthaven1999,Hesthaven2000,Gottlieb2001} are imposed on the characteristic modes. 
Recall that the evolution system can be written as a collection of coupled first order differential equations in the form of  
\bea \label{eq:EvoEquGen}
\partial_t U_{\alpha} + A^{i \beta}{}_{\alpha} \partial_i U_{\beta} = R_{\alpha}\,,
\eea
where $i$ is the spatial index, and $\alpha$ is the internal variable index. In our case, we can see $U_{\alpha}$ as an abstract six-dimensional state vector that is an alternative formulation of $F_{ab}$.  For convenience, we will frequently express such vectors as a pair of three dimensional vectors. 
The Eq.~\eqref{eq:EvoEquGen} is \emph{strongly hyperbolic} iff for \emph{all} unit-vectors $\bf{\hat{n}}$, the matrix $\bf{\hat{n}} \cdot \bf{A}=\hat{n}_i A^{i}$ has only real eigenvalues and a complete set of eigenvectors. It is furthermore \emph{symmetric hyperbolic}, if there exists a definite positive symmetric matrix $\bf{S}$ (a symmetrizer), such that the product of $\bf{S}$ and $\bf{A}$ is symmetric. Symmetric hyperbolicity implies strong hyperbolicity, and a strongly hyperbolic system is well posed \cite{lrr-1998-3}. 
We note that as the spacetime metric is treated as a background quantity, its derivatives do not contribute to the principal part of the FFE evolution equations. Comparing the curved spacetime evolution equations with their flat spacetime limit, we see that 
\bea \label{eq:FlatVsCurved}
\left(\hat{n}_j  \tilde{A}^j\right)^{\alpha}{}_{\beta} =  N \left(\hat{n}_j A^j\right)^{\alpha}{}_{\beta} -\left(\beta^j \hat{n}_j\right) \delta^{\alpha}{}_{\beta}\,,
\eea
where tilde denotes curved spacetime expressions and $\bf{A}$ the flat spacetime counterpart. Therefore the eigenvalues (characteristic speeds) in curved spacetime are simply given by
\bea \label{eq:FlatToCurve}
\tilde{\nu}_{\hat{\alpha}} = N \nu_{\hat{\alpha}} - \bm{\beta} \cdot {\bf \hat{n}}\,,
\eea 
while the eigenvector (characteristic mode) expressions are unchanged from their flat spacetime counterparts. The analysis of the characteristic structure is then essentially independent of the spacetime curvature. 

For the minimal evolution system, the right eigenvalues satisfying the equation 
\bea
(\hat{n}_i \tilde{A}^i)^{\alpha}{}_{\beta} e^{\beta}_{\hat{\alpha}} = \tilde{\nu}_{\hat{\alpha}} e^{\alpha}_{\hat{\alpha}}
\eea
are \cite{Harald} 
\bea
\tilde{\nu}_{\hat{1}} &=& -N - \bm{\beta} \cdot {\bf \hat{n}}\,, \label{eq:EigenValEx1}\\
\tilde{\nu}_{\hat{2}} &=& N - \bm{\beta} \cdot {\bf \hat{n}}\,, \label{eq:EigenValEx2}\\
\tilde{\nu}_{\hat{3}} &=& N(\nu - \omega)- \bm{\beta} \cdot {\bf \hat{n}}\,, \label{eq:EigenValEx3}\\
\tilde{\nu}_{\hat{4}} &=& N(\nu + \omega)- \bm{\beta} \cdot {\bf \hat{n}}\,, \label{eq:EigenValEx4}\\
\tilde{\nu}_{\hat{5}} &=& - \bm{\beta} \cdot {\bf \hat{n}}\,, \label{eq:EigenValEx5}\\
\tilde{\nu}_{\hat{6}} &=& - \bm{\beta} \cdot {\bf \hat{n}}\,, \label{eq:EigenValEx6}
\eea
where 
\bea
\nu &=& \frac{{\bf\hat{n}}\cdot ({\bf E}\times {\bf B})}{B^2}\,, \label{eq:EigenPara1} \\
\omega &=& \frac{1}{B^2}\sqrt{({\bf\hat{n}}\cdot {\bf B})^2(B^2-E^2)}\,. \label{eq:EigenPara2}
\eea 
The right eigenvectors are 
\bea
e^{\alpha}_{\hat{1}} &=& \left( -\mathbb{P}{\bf E} + {\bf \hat{n}} \times {\bf B}, \frac{}{}\quad \mathbb{P}{\bf B} + {\bf \hat{n}} \times {\bf E} \right)\,, \label{eq:Eig1} \\
e^{\alpha}_{\hat{2}} &=& \left(-\mathbb{P}{\bf E} - {\bf \hat{n}} \times {\bf B}, \frac{}{}\quad \mathbb{P}{\bf B} - {\bf \hat{n}} \times {\bf E} \right)\,, \label{eq:Eig2} \\
e^{\alpha}_{\hat{3},\hat{4}} &=& \left(-\mathbb{P}{\bf B} + \nu_{\hat{3},\hat{4}} {\bf \hat{n}} \times {\bf E} + (1-\nu^2_{\hat{3},\hat{4}}){\bf B}, \frac{}{} \right. \notag \\
&& \left. \quad -\mathbb{P}{\bf E} -\nu_{\hat{3},\hat{4}} {\bf \hat{n}} \times {\bf B} \right)\,, \label{eq:Eig34} \\
e^{\alpha}_{\hat{5}} &=& \left(0, \frac{}{}\quad {\bf \hat{n}} \right)\,,  \label{eq:Eig5}\\
e^{\alpha}_{\hat{6}} &=& \left(({\bf \hat{n}} \cdot {\bf B}) {\bf \hat{n}},\frac{}{} \quad \mathbb{P}{\bf E} \right)\,, \label{eq:Eig6}
\eea
where $\nu_{\hat{3},\hat{4}} = \nu \pm \omega$, and for a vector ${\bf A}$, $(\mathbb{P}{\bf A})$ is defined as ${\bf A} - ({\bf \hat{n}} \cdot {\bf A}) {\bf \hat{n}}$. The $e^{\alpha}_{\hat{1}}$ and $e^{\alpha}_{\hat{2}}$ are fast modes travelling at the speed of light. The terms $e^{\alpha}_{\hat{3}}$ and $e^{\alpha}_{\hat{4}}$ are the Alfv\'en modes, while 
$e^{\alpha}_{\hat{5}}$ and $e^{\alpha}_{\hat{6}}$ are the unphysical modes (there can only be four physical modes as constraints $\nabla \cdot {\bf B} =0$ and ${\bf E} \cdot {\bf B} =0$ reduce the number of independent degrees of freedom to four). 

Note that the characteristic speeds become complex when $E^2 > B^2$, and the evolution system will not be strongly hyperbolic. This is one issue that stems from the physical constraint of subluminal motion for the plasma particles. There is however another hyperbolicity related problem with the FFE evolution equations. Namely, even when all the constraints are satisfied, we do not have a complete set of eigenvectors when $({\bf \hat{n}} \cdot {\bf B})^2 = ({\bf \hat{n}}\times {\bf E})^2$ \cite{Harald}. For example, when ${\bf \hat{n}} \cdot {\bf B} = 0$ and ${\bf E} = E{\bf \hat{n}}$, we can choose coordinates such that ${\bf \hat{n}} ={\bf \hat{x}}$ and ${\bf B} = B{\bf \hat{y}}$; then for the minimal system, we have 
\bea \label{eq:DegMatrix}
{\bf \hat{n}} \cdot {\bf A} = \bma
0 & 0 & 0 & 0 & 0 & 0 \\
0 & 0 & 0 & 0 & 0 & 0 \\
\frac{E}{B} & 0 & 0 & 0 & -1 & 0 \\
0 & 0 & 0 & 0 & 0 & 0 \\
0 & 0 & -1 & 0 & 0 & 0 \\
0 & 1 & 0 & 0 & 0 & 0 \\
\ema,
\eea
whose characteristic equation is
\bea
{\rm det}({\bf \hat{n}} \cdot {\bf A} - \lambda \mathbb{1}) = \lambda^4(\lambda^2-1) =0,
\eea
so that we have four zero eigenvalues. In order to allow for four eigenvectors corresponding to these zero eigenvalues, matrix ${\bf \hat{n}} \cdot {\bf A}$ must have rank $6-4=2$. However, its actual rank is $3$, so we don't have a complete set of eigenvectors. As an aside, we mention that in some numerical implementations, a divergence cleaning scalar field is introduced into the evolution system \cite{Dedner2002,Palenzuela:2009hx,Alic:2012df}, which enlarges the space of evolved variables, and changes the characteristic structure of the evolution system. However, doing so does not cure this particular hyperbolicity problem (details are provided in Appendix \ref{sec:HyperCleaning}). 
Nevertheless, the directions for which we do not have a complete set of eigenvectors is a set of measure zero among all possible ${\bf \hat{n}}$ directions, and there are enough eigenvectors to represent the constraint-satisfying solutions even for these directions \cite{Harald}. However, for constraint violating solutions, the constraints may grow on arbitrarily short time scales (beyond the ability of our constraint damping additions to control) when the evolution system is not well posed. Thankfully, for the numerical studies in this paper, we do not encounter such a situation. 

We note, nevertheless, that it is possible to obtain strictly strongly-hyperbolic evolution equations by
augmenting them with terms that vanish for constraint-satisfying solutions, so that only \emph{unphysical} modes are altered.
Ref.~\cite{Harald} provides one such system, and we further improve upon it by bringing in more augmentation terms and proposing two systems with additional desirable properties. In particular, one system remains symmetric hyperbolic even when the constraint ${\bf E} \cdot {\bf B} =0$ is violated. The other system, although no longer symmetric hyperbolic when the constraints are violated, has a strongly hyperbolic set of constraint evolution equations. 
The details of these augmented systems are given in Appendix \ref{sec:AugSys}. 

\section{Well-posed FFE evolution systems \label{sec:AugSys}}
It is desirable for an evolution system to be strongly hyperbolic, as then it will be well posed. It has been shown in a recent paper \cite{Harald} that it is possible to augment the FFE evolution equations with terms containing the derivatives of the constraints, such that the resulting system is symmetric hyperbolic when the constraints are satisfied. We show in this appendix section that by considering additional augmentation terms, it is possible to make the evolution system retain its symmetric hyperbolicity even when the FFE constaint ${\bf E} \cdot {\bf B} =0$ is violated. 
As constraints are never exactly satisfied in numerical simulations, such nice off shell (off of the constraint surface) properties are obviously desirable. In addition, we also provide an alternative augmented evolution system, whose evolution equations do not remain symmetric hyperbolic off shell, but whose associated constraint evolution equations are strongly hyperbolic and particularly simple, so that the constraints evolve in a well-understood and controlled manner. 

\subsection{The main evolution equations}
The unaugmented evolution equations are not strictly hyperbolic even when the constraints are satisfied. Namely when $({\bf \hat{n}} \cdot {\bf B})^2=({\bf \hat{n}} \times {\bf E})^2$, the matrix $\hat{n}_i A_{\alpha}^{i \beta}$ does not possess a complete set of eigenvectors \cite{Harald}. This problem can be cured by adding constraints to the evolution equations. Such additions will not change the physical, constraint-satisfying solutions, but will modify the characteristic structure of Eq.~\eqref{eq:EvoEquGen} if the new terms contain derivatives. For our case, we consider seven possible additional terms that look similar to already existing ones. With them, the evolution equations become
\begin{widetext}
\bea
(\partial_t - \mathcal{L}_{\beta}) {\bf E} &=& N K {\bf E} + \nabla \times (N{\bf B}) - N\frac{{\bf B}}{B^2}\left( {\bf B} \cdot \nabla \times {\bf B} - {\bf E} \cdot \nabla \times {\bf E} - 2K_{ij}E^i B^j + 2K {\bf E} \cdot {\bf B} + \delta {\bf E} \cdot {\bf B}\right) \notag \\
&& - \frac{{\bf E} \times {\bf B}}{B^2} N \nabla \cdot {\bf E} -  \alpha_1 N \frac{{\bf E}}{B^2} \times \nabla({\bf E} \cdot {\bf B}) - \alpha_2 N \frac{{\bf E} \cdot {\bf B}}{B^2} \nabla \times {\bf E} - \alpha_3 N ({\bf E} \cdot {\bf B}) {\bf E} \times \nabla \frac{1}{B^2}  \label{eq:CruvedEvoE}\\
(\partial_t - \mathcal{L}_{\beta}) {\bf B} &=& N K {\bf B} - \nabla \times (N {\bf E}) - \alpha_4 N \frac{{\bf E} \times {\bf B}}{B^2} \nabla \cdot {\bf B} - \alpha_5 N \frac{{\bf B}}{B^2} \times \nabla({\bf E} \cdot {\bf B}) - \alpha_6 \frac{{\bf E} \cdot {\bf B}}{B^2} \nabla \times (N {\bf B}) \notag \\
&&- \alpha_7 N({\bf E} \cdot {\bf B}) {\bf B} \times \nabla \frac{1}{B^2}
\label{eq:CruvedEvoB}
\eea
\end{widetext}
where we have included the constraint damping term as well (it does nothing to the characteristic structure of the equations as it does not contain any derivatives). 
Note that in order to acquire additional desirable properties, we have considered a larger collection of possible augmentation terms than Ref.~\cite{Harald}, which included those terms whose coefficients are $\alpha_1$, $\alpha_4$ and $\alpha_5$, and with these coefficients fixed to $0$, $1$ and $1$, respectively. 
In other words, the minimally augmented system (abbreviated to AU) introduced in \cite{Harald} corresponds to 
\bea \label{eq:AU1Coe}
\text{AU}: {\bm \alpha} = (0,0,0,1,1,0,0).
\eea
in our notation.

We will propose two additional augmented systems AU2 and AU3 given by 
\bea \label{eq:AU2Coe}
\text{AU2}: {\bm \alpha} = (-1,0,-1/2,1,1,0,1/2)
\eea
and 
\bea \label{eq:AU3Coe}
\text{AU3}: {\bm \alpha} = (0,1,0,1,1,-1,1)
\eea
respectively. Both of these systems are symmetric hyperbolic (see Sec.~\ref{sec:MainOnConsSurf} of this appendix) just like the AU system, but possess additional desirable properties. The AU2 system retains its symmetric hyperbolicity when $\bf{E} \cdot \bf{B} \neq 0$ (see Sec.~\ref{sec:MainOffConsSurf}) and the AU3 system has a particularly simple constraint evolution system (see Sec.~\ref{sec:ConsEqsHyper}). 

\subsection{Hyperbolicity of the main evolution equations when constraints are satisfied \label{sec:MainOnConsSurf}}
The requirement of hyperbolicity of Eqs.~\ref{eq:CruvedEvoE} and \ref{eq:CruvedEvoB} 
will imply restrictions on the coefficients 
${\bm \alpha}$. To investigate these restrictions,
we first consider the case of ${\bf \hat{n}} \cdot {\bf B} = 0$ and ${\bf E} = E {\bf \hat{n}}$, which is a special case of the $({\bf\hat{n}} \cdot {\bf B})^2=({\bf\hat{n}} \times {\bf E})^2$ configurations. Because the curvature of the spacetime impacts the characteristic structure of the FFE equations trivially, we will use only flat spacetime expressions for the rest of this section, with the understanding that curved spacetime counterparts can be recovered using Eq.~\eqref{eq:FlatToCurve}. For the augmented systems, we have that 
\bea
{\bf \hat{n}} \cdot {\bf A} = \bma
0 & 0 & 0 & 0 & 0 & 0 \\
0 & 0 & 0 & 0 & 0 & 0 \\
E/B & 0 & 0 & 0 & -1 & 0 \\
0 & 0 & 0 & 0 & 0 & 0 \\
0 & 0 & -1 & 0 & 0 & 0 \\
0 & 1-\alpha_5 & 0 & (\alpha_4-\alpha_5)E/B & 0 & 0 \\
\ema\,,
\eea
whose characteristic equation is
\bea
{\rm det}({\bf \hat{n}} \cdot {\bf A} - \lambda \mathbb{1}) = \lambda^4(\lambda^2-1) =0\,,
\eea
so we have four zero eigenvalues. In order to allow for four eigenvectors corresponding to this zero eigenvalue, matrix ${\bf \hat{n}} \cdot {\bf A}$ must have rank two, which implies $1-\alpha_5=0$ and $\alpha_4-\alpha_5=0$, or $\alpha_4 = 1= \alpha_5$. 
We note that all three augmented systems given by Eqs.~\eqref{eq:AU1Coe}, \eqref{eq:AU2Coe} and \eqref{eq:AU3Coe} satisfy this requirement. 

We now turn to prove symmetric hyperbolicity of AU2 and AU3 for generic cases by explicitly calculating the symmetrizer ${\bf S}$ for them. We do so by writing down the most general symmetrizer when the constraint ${\bf E} \cdot {\bf B} =0$ is satisfied, with each term multiplied by a yet-to-be-determined coefficient. We then solve for these coefficients by ensuring $S^{\beta}{}_{\alpha}({\bf \hat{n}} \cdot {\bf A})^{\gamma}{}_{\beta}$ is symmetric for all ${\bf \hat{n}}$. This is a tedious but straightforward process. The condition of $\alpha_4 = 1= \alpha_5$ turns out to be necessary to ensure the positive definiteness of $S^{\beta}{}_{\alpha}$. The symmetrizer for AU2 is simply 
\bea \label{eq:AU2Sym}
\bf{S}= \bma B^2 g_{ij} + (\zeta -1) B_iB_j & -E_iB_j + \zeta B_i E_j \\ 
-B_iE_j + \zeta E_i B_j &
B^2 g_{ij} + (\zeta -1)E_iE_j
\ema\,,  
\eea 
where $\zeta$ is a free constant, and we require $\zeta>0$ for the positive definiteness of $S$. The symmetrizer for AU3 is 
\bea
\bf{S} = \bma 
\Delta g_{ij} + (\xi \Delta - 2\Delta)\frac{B_iB_j}{B^2} &
-\Delta \frac{E_iB_j}{B^2} + \xi \Delta \frac{B_iE_j}{B^2} \\
-\Delta \frac{B_iE_j}{B^2} + \xi \Delta \frac{E_iB_j}{B^2} &
\Delta g_{ij} + \xi \Delta \frac{E_iE_j}{B^2}
\ema\,,
\eea
where $\Delta = 1-E^2/B^2$, and we should pick a $\xi > 1/\Delta$ to ensure positive definiteness of $S$. 

\subsection{Hyperbolicity of the main evolution equations when ${\bf E} \cdot {\bf B} \neq 0$ \label{sec:MainOffConsSurf}}
The $\bm{\alpha}$ vector for AU2 in Eq.~\eqref{eq:AU2Coe} is chosen to ensure that the symmetrizer remains valid when ${\bf E} \cdot {\bf B} \neq 0$ (this property is not shared by the AU3 or AU evolution systems). Indeed, by picking $\zeta = 1/2$, it is straightforward to verify explicitly that 
$S^{\beta}{}_{\alpha}({\bf \hat{n}} \cdot {\bf A})^{\gamma}{}_{\beta}$ is symmetric. Namely,  if we break the greek indices into a pair of spatial indices and write $S^{\beta}{}_{\alpha}({\bf \hat{n}} \cdot {\bf A})^{\gamma}{}_{\beta}$ in a block form
\bea \label{eq:BlockSA}
\bma (SA_{EE})^j{}_{k} & (SA_{EB})^j{}_{k}  \\
(SA_{BE})^j{}_{k} & (SA_{BB})^j{}_{k}
\ema\,,
\eea
we then have 
\bea
(SA_{EE})_{ik}\epsilon^{jik} = 0 = (SA_{BB})_{ik}\epsilon^{jik}\,, \notag \\
(SA_{EB})_{ik} - (SA_{BE})_{ki} =0\,, \label{eq:SymTest}
\eea
regardless of the value of ${\bf E} \cdot {\bf B}$. 

In greater detail, the block form of ${\bf \hat{n}} \cdot {\bf A}$ is 
\bea \label{eq:BlockCharMatrix}
\bma (A_{EE}){}^j{}_k & (A_{EB}){}^j{}_k \\
(A_{BE}){}^j{}_k & (A_{BB}){}^j{}_k 
\ema\,,
\eea
where for generic $\bm{\alpha}$ choices
\begin{widetext}
\bea
(A_{EE}){}^j{}_k &=& -\frac{B^j}{B^2}({\bf E} \times {\bf \hat{n}})_k + \frac{({\bf E} \times {\bf B})^j}{B^2}\hat{n}_k + \alpha_1 \frac{({\bf E} \times {\bf \hat{n}})^j}{B^2}B_k + \alpha_2 \frac{{\bf E}\cdot {\bf B}}{B^2} \epsilon^{ji}{}_k {\hat{n}}_i\,, \label{eq:Block1}\\
(A_{EB}){}^j{}_k &=& -\epsilon^{ji}{}_k \hat{n}_i + \frac{B^j}{B^2}({\bf B} \times {\bf \hat{n}})_k + \alpha_1 \frac{({\bf E} \times {\bf \hat{n}})^j}{B^2} E_k - 2\alpha_3 \frac{{\bf E} \cdot {\bf B}}{B^4} ({\bf E} \times {\bf \hat{n}})^j B_k\,, \label{eq:Block2} \\
(A_{BE}){}^j{}_k &=& \epsilon^{ji}{}_k \hat{n}_i + \alpha_5 \frac{({\bf B} \times {\bf \hat{n}})^j}{B^2}B_k\,, \label{eq:Block3}\\
(A_{BB}){}^j{}_k &=& \alpha_4 \frac{({\bf E} \times {\bf B})^j}{B^2}\hat{n}_k + \alpha_5 \frac{({\bf B}\times {\bf \hat{n}})^j}{B^2}E_k+ \alpha_6 \frac{{\bf E}\cdot {\bf B}}{B^2} \epsilon^{ji}{}_k \hat{n}_i - 2\alpha_7 \frac{{\bf E} \cdot {\bf B}}{B^4} ({\bf B} \times {\bf \hat{n}})^j B_k\,. \label{eq:Block4}
\eea
Multiplying with symmetrizer $\bf{S}$ as given by Eq.~\eqref{eq:AU2Sym} with $\zeta=1/2$ and an extra overall factor of $2$ for convenience, we obtain the components in Eq.~\eqref{eq:BlockSA}:
\bea
(SA_{EE})_{ik} &=& -B_i ({\bf E} \times {\bf \hat{n}})_k + 2 ({\bf E}\times{\bf B})_i \hat{n}_k - 2 ({\bf E} \times {\bf \hat{n}})_i B_k + {\bf B} \cdot ({\bf E} \times {\bf \hat{n}}) B_i B_k + B_i ({\bf E}\times {\bf \hat{n}})_k \notag \\
&&- 2 E_i ({\bf B} \times {\bf \hat{n}})_k + \frac{{\bf E} \cdot ({\bf B}\times {\bf \hat{n}})}{B^2} B_i B_k\,, \\
(SA_{BB})_{ik} &=& -E_i({\bf B} \times {\bf \hat{n}})_k + 2 B_i ({\bf E} \times {\bf \hat{n}})_k + E_j ({\bf B} \times {\bf \hat{n}})_k - 2\frac{{\bf E} \cdot {\bf B}}{B^2} B_i ({\bf B}\times {\bf \hat{n}})_k - \frac{{\bf B} \cdot ({\bf E} \times {\bf \hat{n}})}{B^2}E_i E_k \notag\\
&&+ \frac{{\bf E}\cdot {\bf B}}{B^4} {\bf B} \cdot ({\bf E} \times {\bf \hat{n}})E_i B_k + 2 ({\bf E} \times {\bf B})_i n_k + 2({\bf B} \times {\bf \hat{n}})_i E_k - \frac{{\bf E} \cdot ({\bf B} \times {\bf \hat{n}})}{B^2}E_i E_k \notag\\
&&-2 \frac{{\bf E}\cdot{\bf B}}{B^2}({\bf B} \times {\bf \hat{n}})_i B_k + \frac{{\bf E} \cdot {\bf B}}{B^2} \frac{{\bf E}\cdot ({\bf B} \times {\bf \hat{n}})}{B^2} E_i B_k \,,\\
(SA_{EB})_{ik} &=& 2 B^2 \epsilon_{ikl} \hat{n}^l -2 ({\bf E}\times {\bf \hat{n}})_i E_k +2 \frac{{\bf E}\cdot{\bf B}}{B^2}({\bf E} \times {\bf \hat{n}})_i B_k + 2B_i ({\bf B} \times {\bf \hat{n}})_k \,,\\
(SA_{BE})_{ik} &=& -2 B^2 \epsilon_{ikl} \hat{n}^l -2 ({\bf E}\times {\bf \hat{n}})_k E_i +2 \frac{{\bf E}\cdot{\bf B}}{B^2}({\bf E} \times {\bf \hat{n}})_k B_i + 2B_k ({\bf B} \times {\bf \hat{n}})_i \,,\\
\eea 
where we have specialized to the AU2 system of Eq.~\eqref{eq:AU2Coe}. We then have 
$(SA_{EB})_{ik}-(SA_{BE})_{ki}=0$, and it is straightforward to show that the antisymmetric part of the diagonal blocks are
\bea
(SA_{EE})_{ik} \epsilon^{jik} = (SA_{BB})_{ik} \epsilon^{jik} = 2({\bf B} \times ({\bf E} \times {\bf \hat{n}}) + {\bf \hat{n}}\times ({\bf B} \times {\bf E}) + {\bf E} \times ({\bf \hat{n}} \times {\bf B}))^j = 0 
\eea
\end{widetext}

For completeness, we explicitly write out the characteristic modes and speeds for the AU2 system. The right eigenvalues satisfying the equation 
\bea
(n_i A^i)^{\alpha}{}_{\beta} e^{\beta}_{\hat{\alpha}} = \nu_{\hat{\alpha}} e^{\alpha}_{\hat{\alpha}}
\eea
are  
\bea
\nu_{\hat{1}} &=& -1\,, \label{eq:EigenValEx1}\\
\nu_{\hat{2}} &=& 1\,, \label{eq:EigenValEx2}\\
\nu_{\hat{3}} &=& \nu - \omega\,, \label{eq:EigenValEx3}\\
\nu_{\hat{4}} &=& \nu + \omega\,, \label{eq:EigenValEx4}\\
\nu_{\hat{5}} &=& \nu\,, \label{eq:EigenValEx5}\\
\nu_{\hat{6}} &=& 2\nu\,, \label{eq:EigenValEx6}
\eea
where $\nu$ and $\omega$ are given in Eqs.~\eqref{eq:EigenPara1} and \eqref{eq:EigenPara2}.
The right eigenvectors are 
\bea
e^{\alpha}_{\hat{1}} &=& \left( -\mathbb{P}{\bf E} + {\bf \hat{n}} \times {\bf B}, \quad \mathbb{P}{\bf B} + {\bf \hat{n}} \times {\bf E} \right)\,, \label{eq:AU2Eig1} \\
e^{\alpha}_{\hat{2}} &=& \left( -\mathbb{P}{\bf E} - {\bf \hat{n}} \times {\bf B}, \quad \mathbb{P}{\bf B} - {\bf \hat{n}} \times {\bf E} \right)\,, \label{eq:AU2Eig2} \\
e^{\alpha}_{\hat{3},\hat{4}} &=& \left( -\mathbb{P}{\bf B} + \nu_{\hat{3},\hat{4}} {\bf \hat{n}} \times {\bf E} + (1-\nu^2_{\hat{3},\hat{4}}){\bf B}, \right. \notag \\
&& \left. \quad -\mathbb{P}{\bf E} -\nu_{\hat{3},\hat{4}} {\bf \hat{n}} \times {\bf B} \right)\,, \label{eq:AU2Eig34} \\
e^{\alpha}_{\hat{5}} &=& \left( {\bf E}({\bf B} \cdot {\bf \hat{n}}) -{\bf B}({\bf E} \cdot {\bf \hat{n}}), 
\quad B^2 {\bf \hat{n}} - {\bf E} \times {\bf B} \nu \right)\,, \quad \quad \label{eq:AU2Eig5} \\
e^{\alpha}_{\hat{6}} &=& \left({\bf B}, 0\right)\,. \label{eq:AU2Eig6}
\eea
We note that the unphysical modes for the AU2 system as given by Eqs.~\eqref{eq:AU2Eig5} and \eqref{eq:AU2Eig6} are much less complicated than for the AU system as given in Ref.~\cite{Harald}. This is beneficial for inverting the characteristic modes in order to obtain the fundamental variables ${\bf E}$ and ${\bf B}$, which is necessary for some pseudospectral implementations such as ours, where both the internal (between the adjacent subdomains) \cite{Hesthaven1997,Hesthaven1999,Hesthaven2000,Gottlieb2001} and the external boundary conditions \cite{Rinne2006,Rinne2007,Lindblom:2007} are imposed on the characteristic modes, and so need to be translated into the fundamental evolution variables before they become useful. 

The left eigenvalues satisfying 
\bea
e_{\alpha}^{\hat{\alpha}} (n_i A^i)^{\alpha}{}_{\beta} = \nu^{\hat{\alpha}} e_{\beta}^{\hat{\alpha}}
\eea
are identical to the right eigenvalues in Eq.~\eqref{eq:EigenValEx1}-\eqref{eq:EigenValEx6}, while the left eigenvectors are 
\bea
e_{\alpha}^{\hat{1}} &=& \left( {\bf E} - {\bf \hat{n}} \times {\bf B}, \right. \notag \\ 
&& \left. \quad -{\bf B} - {\bf \hat{n}} \times {\bf E} + \frac{(B^2-E^2)}{1+\nu} \frac{{\bf \hat{n}} \cdot {\bf B}}{B^2} {\bf \hat{n}}\right)\,, \\
e_{\alpha}^{\hat{2}} &=& \left( {\bf E} + {\bf \hat{n}} \times {\bf B}, \right. \notag \\ 
&&\left. \quad -{\bf B} + {\bf \hat{n}} \times {\bf E} + \frac{(B^2-E^2)}{1-\nu} \frac{{\bf \hat{n}} \cdot {\bf B}}{B^2} {\bf \hat{n}}\right)\,, \\
e_{\alpha}^{\hat{3}} &=& \left( \frac{{\bf \hat{n}} \cdot {\bf B}}{B^2} {\bf E} \times {\bf B} + \frac{\omega}{{\bf \hat{n}} \cdot {\bf B}} B^2 \left( {\bf \hat{n}} - \frac{{\bf \hat{n}} \cdot {\bf B}}{B^2} {\bf B} \right), \right.\notag \\
&&\left. \quad -{\bf E}(\omega -\nu) + {\bf \hat{n}} \times {\bf B}   \right)\,, \\
e_{\alpha}^{\hat{4}} &=& \left( -\frac{{\bf \hat{n}} \cdot {\bf B}}{B^2} {\bf E} \times {\bf B} + \frac{\omega}{{\bf \hat{n}} \cdot {\bf B}} B^2 \left( {\bf \hat{n}} - \frac{{\bf \hat{n}} \cdot {\bf B}}{B^2} {\bf B} \right),\right. \notag \\
&&\left. \quad -{\bf E}(\omega -\nu) - {\bf \hat{n}} \times {\bf B}   \right)\,, \\
e_{\alpha}^{\hat{5}} &=& \left( 0, {\bf \hat{n}} \right)\,, \\
e_{\alpha}^{\hat{6}} &=& \left( {\bf B}, {\bf E} \right)\,. 
\eea

These eigenvectors are degenerate when ${\bf E} = \pm {\bf \hat{n}} \times {\bf B}$, in which case we can find alternative complete sets of eigenvectors. When ${\bf E} = {\bf \hat{n}} \times {\bf B}$, we have $\nu_{\hat{1}} = \nu_{\hat{3}} = -1$, and we can pick ${\bf \hat{q}} \bot {\bf \hat{n}}$ to construct
\bea
e^{\alpha}_{\hat{1}} &=& \left({\bf \hat{q}}, \quad -{\bf \hat{n}} \times {\bf \hat{q}} \right)\,, \\
e^{\alpha}_{\hat{3}} &=& \left({\bf \hat{n}} \times {\bf \hat{q}} , \quad {\bf \hat{q}} \right)\,, \\
e_{\alpha}^{\hat{1}} &=& \left({\bf B} \times ({\bf \hat{n}} \times {\bf \hat{q}}), \quad {\bf B} \times {\bf \hat{q}} \right)\,, \\
e_{\alpha}^{\hat{3}} &=& \left(-{\bf B} \times {\bf \hat{q}}, \quad {\bf B} \times ({\bf \hat{n}} \times {\bf \hat{q}}) \right)\,, 
\eea 
while the remaining eigenvectors are still valid. 
When ${\bf E} = -{\bf \hat{n}} \times {\bf B}$, we have $\nu_{\hat{2}} = \nu_{\hat{4}} = 1$, and can use the new vectors 
\bea
e^{\alpha}_{\hat{2}} &=& \left({\bf \hat{q}}, \quad {\bf \hat{n}} \times {\bf \hat{q}} \right)\,, \\
e^{\alpha}_{\hat{4}} &=& \left({\bf \hat{n}} \times {\bf \hat{q}} , \quad -{\bf \hat{q}} \right)\,, \\
e_{\alpha}^{\hat{2}} &=& \left(-{\bf B} \times ({\bf \hat{n}} \times {\bf \hat{q}}), \quad {\bf B} \times {\bf \hat{q}} \right)\,, \\
e_{\alpha}^{\hat{4}} &=& \left({\bf B} \times {\bf \hat{q}}, \quad {\bf B} \times ({\bf \hat{n}} \times {\bf \hat{q}}) \right)\,, 
\eea 
together with the remaining eigenvectors that are still valid.

Lastly, as an aside, we note that for the AU2 system, we can further use the identity 
\bea
- \frac{{\bf B}}{B} \times \nabla \frac{{\bf E} \cdot {\bf B}}{B} &=&
-\frac{{\bf B}}{B^2}\times \nabla({\bf E} \cdot {\bf B})  \notag \\
&&- \frac{1}{2}({\bf E} \cdot {\bf B}) {\bf B} \times \nabla \frac{1}{B^2} 
\eea
to combine terms in the evolution Eqs.~\eqref{eq:CruvedEvoE} and \eqref{eq:CruvedEvoB}. 

\subsection{The constraint evolution equations \label{sec:ConsEqsHyper}}
We can derive the evolution equations of the constraints
$\nabla \cdot \bf{B}$ and $\bf{E} \cdot \bf{B}$ from the main evolution Eqs.~\ref{eq:CruvedEvoE} and \ref{eq:CruvedEvoB}. The result is 
\bea
\partial_t \nabla\cdot \bf{B} &=& -\alpha_4 \nabla(\bf{v} \nabla \cdot \bf{B}) - \bf{a} \cdot \nabla(\bf{E} \cdot \bf{B}) \notag \\
&&+ \Psi \bf{E} \cdot \bf{B}\,, \label{eq:divBEvoEq}\\
\partial_t (\bf{E} \cdot \bf{B}) &=& -\alpha_5 \nabla \cdot (\bf{v} (\bf{E}\cdot \bf{B})) + \alpha_1 \bf{v}\cdot \nabla(\bf{E}\cdot\bf{B}) \notag \\
&&+ \Psi' (\bf{E} \cdot \bf{B})\,,
\eea
where 
\bea
\bf{v} &=& \frac{\bf{E} \times \bf{B}}{B^2}\,, \\ 
\bf{a} &=& (\alpha_5-\alpha_7) \nabla \times \frac{\bf{B}}{B^2} + (\alpha_6 + \alpha_7)\frac{1}{B^2} \nabla \times \bf{B} \,,\\
\Psi &=& -(\alpha_6+\alpha_7)\left(\nabla \frac{1}{B^2} \right) \cdot (\nabla \times \bf{B})\,, \\
\Psi' &=& (\alpha_5-\alpha_7+\alpha_3)(\bf{E} \times \bf{B}) \cdot \nabla \frac{1}{B^2} \notag \\
&&+ (\alpha_5-\alpha_2)\frac{\bf{B}\cdot \nabla \times \bf{E}}{B^2} \notag \\
&&-(\alpha_6+ \alpha_5) \frac{\bf{E}\cdot \nabla \times \bf{B}}{B^2}\,.
\eea
It is desirable for the constraint evolution equations to be strongly hyperbolic, so that the constraints evolve in a predictable and controlled manner. Such a property is especially useful when the main evolution equations are not symmetric hyperbolic off shell, as a well posed constraint evolution system would signal that at least some good behaviors are retained off shell. The ${\bf \hat{n}}\cdot {\bf A}$ matrix for the constraint evolution system is simply 
\bea
\bma 
\alpha_4 {\bm v}\cdot {\bm \hat{n}} &  {\bm a}\cdot {\bm \hat{n}} \\
0 & (\alpha_5-\alpha_1) {\bm v}\cdot {\bm \hat{n}}
\ema\,,
\eea
which does not have a complete set of eigenvectors when ${\bm v}\cdot {\bm \hat{n}}=0$ unless ${\bm a}\cdot {\bm \hat{n}}=0$, which can be achieved by setting $\alpha_5=-\alpha_6=\alpha_7$ as in the AU3 system. Note that the AU and AU2 systems do not satisfy this condition, so their constraint evolution equations are not strongly hyperbolic. 

The $\partial_t(\bf{E}\cdot \bf{B})$ equation simplifies further when $\alpha_1 =0$ and $\Psi'=0$. So under the coefficient choice $\alpha_5=-\alpha_6=\alpha_7$, $\alpha_2=\alpha_5$, and $\alpha_1 = 0 = \alpha_3$, the evolution equations reduce to a pair of decoupled advection equations 
\bea
\partial_t (\nabla \cdot \bf{B}) &=& -\nabla (\alpha_4 \bf{v} \nabla \cdot \bf{B})\,, \label{eq:AU3divBEvo}\\
\partial_t (\bf{E} \cdot \bf{B}) &=& -\nabla (\alpha_5 \bf{v} \bf{E} \cdot \bf{B})\,. 
\eea
When combined with the $\alpha_4 = \alpha_5 =1$ condition for the hyperbolicity of the main evolution equations, we obtain the $\bm{\alpha}$ coefficients for the AU3 system as given by Eq.~\eqref{eq:AU3Coe}. 

We note that Eq.~\eqref{eq:divBEvoEq} contains a second derivative of ${\bf B}$ whenever $\alpha_4\neq 0$.  Since hyperbolicity requires $\alpha_4=1$, this is always the case (shared by all of the AU, AU2 and AU3 systems ), with both positive and negative implications: A disadvantage of this term is that the second derivative increases the sensitivity to high-frequency noise in ${\bf B}$.  However, the $\alpha_4$-term in Eq.~\eqref{eq:divBEvoEq} will cause any constraint violations $\nabla\cdot {\bf B}\neq 0$ to propagate along ${\bf v}$, thus allowing it to propagate away.

\section{Hyperbolicity of an evolution system with a divergence cleaning scheme \label{sec:HyperCleaning}}
In our FFE code implementation, it turns out that just as observed in Ref.~\cite{Parfrey:2011ta}, the intrinsic accuracy of the pseudospectral code is sufficient to keep the $\nabla \cdot {\bf B} =0$ constraint under control (provided we remove it from the initial data of course), and there is no need for any additional constraint cleaning procedures. 
We note, however, that Refs.~\cite{Dedner2002,Palenzuela:2009hx} used a dynamical divergence cleaning scheme.
Namely, one   
adds a $-N \nabla \phi$ term to $\partial_t {\bf B}$, where $\phi$ is a scalar field satisfying the evolution equation 
\bea \label{eq:ConsDamping}
(\partial_t - \mathcal{L}_{\bm{\beta}}) \phi  = -N \nabla \cdot {\bf B} - N \sigma_2 \phi \,. 
\eea
The $\phi$ field then damps the $\nabla \cdot {\bf B}$ constraint with a time scale of $\sigma_2^{-1}$ \cite{Palenzuela:2009hx}. 

The introduction of a new field like $\phi$ enlarges the space of evolution variables, and would generally alter the characteristic structure of the evolution system. Unfortunately this does not remove the hyperbolicity problem when $({\bf \hat{n}} \cdot {\bf B})^2 = ({\bf \hat{n}}\times{\bf E})^2$. The characteristic matrix for the now enlarged system is 
\bea \label{eq:CleaningEigenMatrix}
{\bf \hat{n}} \cdot {\bf A} = \bma (A_{EE}){}^j{}_k & (A_{EB}){}^j{}_k & 0 \\
(A_{BE}){}^j{}_k & (A_{BB}){}^j{}_k & \hat{n}^j \\
0 & \hat{n}_k & 0 
\ema,
\eea
where the four sub-blocks in the top left are the same as the minimal system. 
Under the same assumptions that lead to Eq.~\eqref{eq:DegMatrix}, we have a $7\times 7$ matrix whose characteristic equation is $\lambda^3(\lambda^2-1)^2=0$, so there are three vanishing characteristic speeds. The matrix needs to have a rank of $4$ for there to be three independent characteristic modes of vanishing speed, but the matrix actually has a rank of $5$. Therefore, the enlarged evolution system is still not strictly strongly hyperbolic.

\section{The null solutions in Kerr-Schild coordinates \label{sec:NullSolKerrSchild}}
In our simulations, we specify the metric using the Kerr-Schild slicing and coordinates, and the transformations between the Kerr-Schild coordinates $(\tilde{t},x,y,z)$ and the ingoing Kerr coordinates $(\nu,r,\theta,\psi)$ are simply (we have kept $a$ in case readers need it elsewhere, for this paper, one should set $a=0$ in all these expressions)
\bea
\tilde{t} = \nu - r, \quad x+iy = (r+ia) e^{i\psi}\sin\theta, \quad z = r\cos\theta, 
\eea
with the inverse transformations being 
\bea
r &=& \sqrt{\frac{1}{2}\left(b+ \sqrt{b^2 + 4 a^2 z^2} \right)}\,, \\
b &\equiv& x^2+y^2+z^2-a^2\,, \\
\theta &=& \text{arccos}\left( \frac{z}{r}\right)\,,  \quad \nu = \tilde{t} + r\,, \\
\psi &=& \text{arctan}\left(\frac{y}{x}\right) + \frac{\pi}{2}\left(1 - \frac{x}{|x|}\right) -\text{arctan}\left(\frac{a}{r}\right)\,.
\eea
   
Aside from the null solution, we will also add in a monopole contribution according to Eq.~\eqref{eq:MagMonopole} to ensure magnetic dominance, whereby the Faraday tensor becomes
\bea
F_{ab} &=& 4\Re\left(\phi_0 \bar{m}_{[a}n_{b]} + \phi_1 (m_{[a}\bar{m}_{b]} + n_{[a}l_{b]})\right) \,.
\eea
Note that the second term for $\phi_1$ in the expression above vanishes when $\phi_1$ is given by Eq.~\eqref{eq:MagMonopole}, which is purely imaginary, and so the above expression reduces to Eq.~\eqref{eq:FieldTensorWithMono}. 

We also note that the  
Kinnersley tetrad in the Cartesian (vector components are in the Cartesian basis) Kerr-Schild coordinates is given by 
\bea
l^0 &=& 2 \frac{r^2+a^2}{\Delta} -1\,, \\
l^1 + i l^2 &=& e^{i\psi}\sin\theta\left(\frac{2ai}{\Delta}(r+ia) +1\right)\,, \\
l^3 &=& \cos\theta \,, \\[0.2em]
\hline \notag \\[0.em]
n^0 &=& \frac{\Delta}{2\Sigma}\,, \\
n^1 + i n^2 &=& -\frac{\Delta}{2\Sigma}  e^{i\psi}\sin\theta\,,\\
n^3 &=& -\frac{\Delta}{2\Sigma} \cos\theta\,, \\[0.2em]
\hline \notag \\[0.em]
m^0 &=& -\frac{\bar{\rho}}{\sqrt{2}}ia\sin\theta\,, \\
m^1 &=& -\frac{\bar{\rho}}{\sqrt{2}}\left( \Re A \cos\theta - i \Im A \right)\,,\\
m^2 &=& -\frac{\bar{\rho}}{\sqrt{2}}\left( \Im A \cos\theta + i \Re A \right)\,,\\
m^3 &=& \frac{\bar{\rho}}{\sqrt{2}} r \sin\theta\,,  
\eea
where $A = (r+ia)\exp(i\psi)$. 

For use in Sec.~\ref{sec:PertPropDir}, we also summarize the relationship between the Boyer-Lindquist $(t,r,\theta,\phi)$
and Kerr-Schild $(\tilde{t},x,y,z)$ spatial coordinates. The transformations between them are
\bea
\tilde{t} &=& t + 2M {\rm ln}\left|
\frac{r}{2M} -1 \right|, \\
x+iy &=& (r+ia){\rm exp}(i\tilde{\phi}) \sin\theta \\
z &=& r\cos\theta
\eea
where 
\bea
\tilde{\phi} = \phi + \frac{a}{r^+-r^-} {\rm ln} \left| \frac{r-r^+}{r-r^-} \right|
\eea
which in the $a=0$ limit gives the Jacobian
\bea
\frac{\partial {\bf x}_{\rm KS}}{\partial {\bf x}_{\rm BL}} =
\left(
\begin{array}{cccc}
1 & \frac{2M}{r-2M} & 0 & 0 \\ 
0 & \cos\phi\sin\theta & r\cos\phi\cos\theta & -r\sin\phi\sin\theta \\
0 & \sin\phi\sin\theta & r\sin\phi\cos\theta & r\cos\phi\sin\theta \\
0 & \cos\theta & -r \sin\theta & 0 
\end{array}\right). \notag
\eea

\section{The spinor formalism \label{sec:SpinorIntro}}
For a comprehensive
introduction to spinors, please consult, for example,  Refs.~\cite{Wald,Penrose1992,Penrose1986V2}. 
Here for the sake of completeness we give a brief summary.

When spinor bundles can be defined on a spacetime (see Ref.~\cite{Wald}), we have a two-dimensional complex vector space $W$, as well as its complex conjugation $W'$, over each spacetime location. 
The elements of $W'$ are written with an overhead bar (signifying complex conjugation) and primed indices (e.g. $\bar{\xi}^{A'}\in W'$), while elements of $W$ have unprimed indices 
and no special overhead symbols (e.g. $\xi^A \in W$). We then map between $W$ ($W'$) and its dual space $W^*$ ($W'^*$) using an antisymmetric spinor $\epsilon_{AB}$ (and $\epsilon_{A'B'}$, 
where it is customary to leave out the overhead bar on $\epsilon_{A'B'}$). 
In other words, we raise and lower spinor indices 
with an antisymmetric second rank spinor as
\bea 
\xi^B \epsilon_{BA} =  \xi_A \in W^*\,, \quad \epsilon^{AB} \xi_B = \xi^A \in W\,,
\eea
rather than with a symmetric one.
The result is that spinor self-contractions vanish, i.e. 
\begin{equation} \label{eq:SpinorSelfContraction}
\xi^A\xi_A = \epsilon_{[BA]}\xi^{(A} \xi^{B)} = 0. 
\end{equation}
In fact, we have the stronger result (Proposition (2.5.56) in Ref.~\cite{Penrose1992}) that: 

\noindent \emph{$\alpha_A \beta^A = 0$ iff $\alpha_A$ and $\beta_A$ are (complex) scalar multiples of each other.}

\noindent When this happens, we will call $\alpha_A$ and $\beta_A$ coincident.  

We also have a map between the tangent space of the spacetime and the tensor product space $W\otimes W'$ given by
\bea
v^{AA'} = v^a \sigma_{a}^{AA'}\, \quad v^a = v^{AA'}\sigma^a_{AA'}\,
\eea
where $\bm{\sigma}$ are the soldering forms satisfying 
\bea \label{eq:SolderForm}
\sigma_a^{AA'}\sigma_{AA'}^b = -\delta_a{}^b\,, \quad \sigma^a_{AA'}\sigma_a^{BB'} = -\epsilon_A{}^{B} \epsilon_{A'}{}^{B'}\,,
\eea
where the minus signs are due to our metric signature choice of $(-+++)$ instead of the customary $(+---)$ for spinor calculations. The consequence is that whenever we translate a contraction between a pair of spacetime indices into the contraction between the corresponding pair of double (one primed and the other unprimed) spinor indices, and vice versa, we should add an extra minus sign. This step would not be necessary had we adopted the $(+---)$ signature as Refs.~\cite{Wald,Penrose1992,Penrose1986V2} did. 
We note that the spacetime null vectors map to particularly nice factorized spinor forms   
\bea \label{eq:NullVecVsSpinor}
l^{a} = \sigma^{a}_{AA'}\xi^{A} \bar{\xi}^{A'}\,.
\eea
With the relationship \eqref{eq:NullVecVsSpinor} and after applying Eq. \eqref{eq:SolderForm}, we see that Eq. \eqref{eq:SpinorSelfContraction} is equivalent to null vectors
having zero norms under the Lorentzian metric. 

Using the soldering forms, we can also transfer higher rank spacetime tensors into their corresponding multispinors.
For example the metric maps to
\begin{equation} \label{eq:SpinorMetric}
g^{ab} = \sigma^{a}_{AA'} \sigma^{b}_{BB'} g^{AA'BB'}, 
\end{equation}
where 
\begin{equation} 
g^{AA'BB'} = -\epsilon^{AB} \epsilon^{A'B'}.
\end{equation}
More importantly for us, the Weyl tensor also has a spinor counterpart $\Psi_{ABCD}$ defined by 
\bea
C_{abcd} &=& \Psi_{ABCD}\epsilon_{A'B'}\epsilon_{C'D'}\sigma^{AA'}_a \sigma^{BB'}_b 
	     \sigma^{CC'}_c \sigma^{DD'}_d \notag \\
    &&+ \bar{\Psi}_{A'B'C'D'}\epsilon_{AB}\epsilon_{CD}\sigma^{AA'}_a
\sigma^{BB'}_b \sigma^{CC'}_c \sigma^{DD'}_d,
\eea
where $\Psi_{ABCD}$ has a much more straight-forward relationship with its 
principal spinors than the original tensor version $C_{abcd}$ had with its principal null vectors. Specifically, the algebraic closedness of the complex numbers field
underlying spinors ensures that we always have the factorization
\begin{equation} \label{eq:GPNDBreakdown}
\Psi_{ABCD} = \alpha^{(1)}_{(A}\alpha^{(2)}_{B}\alpha^{(3)}_{C}\alpha^{(4)}_{D)}
\,,
\end{equation}
where the $\alpha^{(\cdot)}$s are the principal spinors, and then the GPNDs of $C_{abcd}$ are simply given by
\bea
\sigma_a^{AA'}\alpha^{(a)}_{A}\bar\alpha^{(a)}_{A'}\,,\quad (a) \in \{(1),(2),(3),(4)\}.
\eea

Many other quantities take on more transparent forms under the spinor formalism. For example, let a spinor dyad $(\o,\iota)$ be defined such that it relates to a Newman-Penrose null tetrad by 
\begin{equation} \label{eq:TetradVSpinor}
\begin{split}
l^a &= \sigma^a_{AA'} \o^{A}\bar{\o}^{A'}\,, \quad n^a = \sigma^a_{AA'} \iota^{A}\bar\iota^{A'}\,, \\ 
m^a &= \sigma^a_{AA'}
\o^{A}\bar\iota^{A'}\,, \quad \bar m^a = \sigma^a_{AA'} \iota^{A}\bar{\o}^{A'}\,. 
\end{split}
\end{equation}
Then the definitions for the Newman-Penrose scalars under that tetrad
\begin{subequations} \label{eq:PsisTensor}
\begin{align}
\Psi_0 &= C_{abcd} l^a m^b l^c m^d\,, \\
\Psi_1 &= C_{abcd} l^a n^b l^c m^d\,, \\
\Psi_2 &= C_{abcd} l^a m^b \bar{m}^c n^d\,, \\
\Psi_3 &= C_{abcd} l^a n^b \bar{m}^c n^d\,, \\
\Psi_4 &= C_{abcd} n^a \bar{m}^b n^c \bar{m}^d\,,
\end{align}
\end{subequations}
can be rewritten as 
\begin{subequations} \label{eq:PsisSpinor}
\begin{align}
\Psi_0 &= \Psi_{ABCD} \o^A \o^B \o^C \o^D\,, \\
\Psi_1 &= \Psi_{ABCD} \o^A \o^B \o^C \iota^D\,, \\
\Psi_2 &= \Psi_{ABCD} \o^A \o^B \iota^C \iota^D\,, \\
\Psi_3 &= \Psi_{ABCD} \o^A \iota^B \iota^C \iota^D\,, \\
\Psi_4 &= \Psi_{ABCD} \iota^A \iota^B \iota^C \iota^D\,,
\end{align}
\end{subequations}
whereby the decreasing number of times that $\o^A$ appears in the definitions establishes the hierarchy of decay rates of these scalars under the peeling theorem \cite{Penrose1986V2,Zhang:2012ky}, a feature not as visible in the tensorial expressions in Eq.~\eqref{eq:PsisTensor}. Lastly, to carry out some calculations in the main text, we note the fact that under the dyad basis, $\epsilon_{AB}$ can be written as 
\bea \label{eq:SympDyad}
\epsilon_{AB} = \o_A \iota_B - \iota_A \o_B\,.
\eea

\vfil

%\bibliography{References} 

\bibliography{paperv2.bbl} 
\end{document}